\documentclass[prl,aps,amssymb,amsmath,twocolumn]{revtex4}
\begin{document}
\draft
\title{Equality of Inertial and Gravitational Masses for Quantum Particle}
\author{Jaros{\l}aw Wawrzycki\footnote{Electronic address: waw@alf.ifj.edu.pl}}
\address{Institute of Nuclear Physics, ul. Radzikowskiego 152, 
31-342 Krak\'ow, Poland}
\newcommand{\ud}{\mathrm{d}}
\begin{abstract}
We investigate the interaction of the gravitational field with a quantum particle. We prove two results.
First, we give the proof of the equality of the inertial and the gravitational mass for the nonrelativistic
quantum particle, independently of the equivalence principle.
Second, we show that the macroscopic body cannot be described by the many-particle Quantum Mechanics.
As an important tool we generalize the Bargmann's theory of ray representations and 
explain the connection with the state vector reduction problem. The Penrose's hypothesis is discussed,
\emph{i.e.} the hypothesis that the gravitational field may influence the state 
vector reduction. 
   
\end{abstract}
\maketitle

\vspace*{-1cm}
\narrowtext
\section{Introduction}

In this paper we investigate the interaction of the gravitational field with a quantum particle.
In comparison with the electromagnetic interactions, we have little to say about the question.
Because all attempts to quantize the relativistic gravitational interactions have encountered 
insurmountable difficulties, we consider here the nonrelativistic gravitational interactions.

In the paper \cite{AHJW} we have considered the question if the Einstein equivalence principle
makes any sense in the (nonrelativistic) Quantum Mechanics. We have argued there, that in some
sense the Einstein equivalence principle is meaningful and can be at least operationally applied
to the local wave equation. Compare also similar arguments in \cite{Brevie}. As was shown
in \cite{AHJW}, this principle determines the wave equation uniquely, which provides in that sense
the full analogue of the Einstein equivalence principle in classical physics -- where the principle
determines the motion of test particles uniquely. In particular, this principle applied to the wave equation
implies the equality of the inertial $m_{i}$ and gravitational mass $m_{g}$ of the nonrelativistic
quantum particle: $m_{i} = m_{g}$. In this paper we derive the equality $m_{i} = m_{g}$ for
a spinless nonrelativistic quantum particle, but our assumptions are weakened in comparison
to \cite{AHJW}. Our assumptions are as follows ($X$ denotes all spacetime coordinates, 
we use the nonrotating Cartesian coordinates)
\begin{enumerate}
\item[(A)] The quantum particle, when its kinetic energy is small in comparison to its 
rest energy $mc^2$, does not exert any influence on the spacetime structure.
\item[(B)] The Born interpretation for the wave function $\psi$ is valid, and 
the transition probabilities in the Newton-Cartan spacetime are equal to the integral 
\begin{displaymath} 
\vert(\psi\vert \varphi)\vert^{2}=\Big\vert\int_{t(X)=t} \psi^*\varphi\, {\ud}^3x\Big\vert^{2}
\end{displaymath}
over a three dimensional simultaneity hyperplane and are preserved under the coordinate transformations.
\item[(C)] The wave equation is linear, of second order at most, generally covariant, and
can be built in a local way with the help of the geometrical objects describing the 
spacetime structure (other quantities must not be used to build it).
\item[(D)] The transition probabilities -- as defined in (B) -- vary continously under the continuous 
variation of the coordinate transformation.
\end{enumerate}
The assumption (D) will be described in detail in the Appendix {\bf A}.

The covariance condition in Quantum Mechanics is so strong that it determines (together with assumptions 
(A) $\div$ (D)) almost uniquely the wave equation. Moreover, the equality $m_{i} = m_{g}$ has to hold
as a consequence of the covariance condition. This is in a complete contradistinction to the equations of motion 
of the classical particle. There are many possibilities for the generally covariant classical theory of classical
particle in the gravitational field with $m_{i} \neq m_{g}$. The Theorem that $m_{i} = m_{g}$ for nonrelativistic 
quantum particle results from (A) $\div$ (D) seems to be strange. We analyse some consequences of this Theorem
in this paper. The first consequence is that we are able to prove the equality $m_{i} = m_{g}$ for  
nonrelativistic quantum system -- consisting of arbitrary many particles, when a very broad assumptions
are fulfilled by the potential which describes the mutual interactions of the constituent particles. This is 
almost unacceptable, however, because it means that we are able to prove the equality 
$m_{i} = m_{g}$ for macroscopic systems if only we assume that the macroscopic body 
can be desribed by the many-particle Quantum Mechanics,
and that the classical body does not influence the spacetime. There are two main
solutions. 1) If we assume that the equality $m_{i} = m_{g}$ for macroscopic body cannot be derived
from many-particle Quantum Mechanics, we have to conclude that the macroscopic body cannot be 
described by the many-particle Quantum Mechanics. 2) The macroscopic body, which we describe 
as a many-particle quantum system, does influence the spacetime geometry and the 
influence is not negligible. But, the many-particle system interacting with the gravitational field 
cannot be described by any quantum system of infinitely many degrees of freedom. 
That is, the commonly applied second quantization of the gravity field fails also in the nonrelativistic theory. 

As we will see, the analysis of the possibility 2) can be carried out in both the nonrelativistic
and the relativistic theories. In this sense the present paper can be treated as an introduction 
to the investigation of the quantum relativistic gravitational interactions. 
However, the equality $m_{i} = m_{g}$ cannot be analysed in 
the relativistic theory, because of the mass renormalization process. 

As we have seen, we use the notion of covariance of the wave equation. It should be stressed here, that
we distinguish between the covariance condition and the invariance (symmetry) condition in accordance with 
\cite{And}. There are no substantial difficulties with the notion of covariance as considered for the 
wave equation. On the other hand not all principles of Quantum Mechanics are contained in 
the Schr\"odinger equation. The Hilbert space and transition probabilities are of fundamental 
importance. Therefore we are forced to make a deeper analysis of the covariance condition
in Quantum Mechanics. Especially we investigate the representation $T_{r}$ of a covariance group
$G$ from a very general point of view. Because the group $G$ is not any symmetry group but only 
a covariance group it does not act in the ordinary Hilbert space. As we will see, if the wave equation
possesses in addition a (say) time dependent gauge freedom, then the exponent $\xi(r,s,t)$ in the
formula
\begin{displaymath}
T_{r}T_{s} = e^{i\xi(r,s,t)}T_{rs},
\end{displaymath}  
depends on the time $t$. On the other hand we will show that the transformation $T_{r}$
for the spinless particle has the general form
\begin{displaymath}
T_{r}\psi(X)=\psi'(X)= e^{-i\theta(r,X)}\psi(r^{-1}X),
\end{displaymath}  
and that the classification of all $\theta(r,X)$ in this formula is equivalent to the classification
of all possible $\xi(r,s,t)$. So, we are forced to generalize the Bargmann's \cite{Bar} classification 
theory of exponents $\xi$ to the case, when $\xi(r,s) = \xi(r,s,t)$ depends on the time $t$. 
That is, we get in this way all possible $T_{r}$ for spinless nonrelativistic particles. 
Next, we insert this $T_{r}$ to the covariance condition of the wave equation, 
and determine the wave equation almost uniquely with $m_{i} = m_{g}$. 

It is a peculiar property of the gravitational field, that the symmetry group $G$ of the system 
particles + field with the back reaction to the field allowed for, that $G$ is greater then the 
maximal symmetry group of the spacetime. For such a group $G$ when $G$ is a covariance group
the exponent $\xi(r,s,t)$ depends on the time $t$ (in the nonrelativistic theory, in the relativistic
one the spacetime dependence is possible) in general. This by no means can be reconciled 
with the fact that the exponent $\xi(r,s)$ of the group $G$ when $G$ is a symmetry group must not
depend on the time $t$. This is the origin of all difficulties of the possibility marked above
as the possibility 2).  
 
The Newtonian gravity is described in the geometrical fashion compatibly with the equivalence 
principle, given first by Cartan \cite{Car}. 

In section {\bf II} we give the strict definitions of covariance and invariance for classical 
equations, which are valid for the Schr\"odinger equation. In section {\bf III} we describe 
in short the Newton-Cartan spacetime. In section {\bf IV} the covariance condition
for the wave equation is given in the explicit form. In section {\bf V} the representation
$T_{r}$ of a covariance group is investigated in detail. Subsection {\bf V.A} contains
the general analysis of the representation $T_{r}$ of a covariance group as compared to 
a symmetry group.  The subsection {\bf V.B} contains the classification of the representations
$T_{r}$ (of the form given above for the spinless particle)
of the Galilean group as a covariance group. In the subsection
{\bf V.C} the classification of $T_{r}$ (for the spinless particle) of the Milne group
-- the covariance group relevant for us --  is given. Section {\bf VI}  
contains the analysis of some consequences of the equality $m_{i} = m_{g}$ (for
the nonrelativistic quantum particle) proved independently of the equivalence principle.
This paper contains the Summary. In Appendix {\bf A} the generalization of the 
Bargmann's theory is presented. In Appendix {\bf B} a general comments concernig 
the representation $T_{r}$ of a covariance group of a theory with spacetime dependent 
gauge freedom is presented.

\section{Covariance and invariance}

In general the notions of \emph{covariance} and \emph{invariance} are very 
important and decisive if one looks for a possible dynamical equation describing a
physical system. Because we make use of them, the general and strict definitions of
those notions should be given. We provide such a definitions according to \cite{And},
the only difference in comparison to \cite{And} is the use of the 
\emph{geometric object} notion. The motivation for the use of the notion is that in our
case (and in general case) the quantities we are dealing with do not in general form 
any representation of the covariance group (as in Anderson's definition) but they are
always \emph{geometrical objects}. For example, the wave function $\psi$ does not 
form any representation but only a ray representation. 
\\ Let us consider a spacetime $M$ and the group $G$ 
(or pseudogroup) of transformations of $M$. 
\vspace{1ex}
\\  DEFINITION 1. There is given a geometrical object $y$ in $M$ with $m$ 
components if for each point $x$ there exists a neighbourhood such that to each 
point of the neighbourhood correspond uniquely $m$ numbers $y$. The correspondence
is such that the components $y'$ at each point $x$ in a new coordinate system
$u'$ depend only on the components $y$ in the old system $u$ and the  
transformation $T$ of $G$, $T: u \to u'$, \emph{i.e.} $y'=F(y,T)$.(Compare e.g 
\cite{Acz}).
\vspace{1ex}
\\ Let us consider a physical system, which by assumption is completely described by 
a  geometrical object $y$ (in the case of the Maxwell theory $y$ is the spacetime metric and the antisymmetric 
tensor field -- possibly the solution of the Maxwell equations). The set of all possible values of $y$ -- not 
necessarily realizable physically by the system -- will be called to be the set of 
\emph{kinematically possible trajectories (kpt)} (in the case of the Maxwell theory
the set of \emph{kpt} consists of all antisymmetric tensor fields, however, not 
necessarily solutions of the Maxwell equations). If the $y$ can be physically realized, then it will
be called \emph{dynamically possible trajectory (dpt)} (in the case of the Maxwell theory 
the set of \emph{dpt} consists of all solutions to the Maxwell equations). 
\vspace{1ex}
\\ DEFINITION 2. A group $G$ will be called covariance group of a theory if
\begin{enumerate}
\item[1)] The set of all \emph{kpt} constitutes a geometric object under the action
of $G$.
\item[2)] The action in 1) is such that it associates \emph{dpt} with \emph{dpt}.
\end{enumerate}
\vspace{1ex}
That is, if a real state is seen by an observer as $y$ in the reference frame $u$,
(so, it is a solution to the dynamical equations of the theory in question),
then the same physical state described by $y'$ by an another observer in 
the reference frame $u'$ should be a real state of course (and should fulfil the equations of motion of the theory)
Note, however, that the equations of motion have different form in different reference frames $u$ and $u'$
in general.   

If a theory possesses a covariance group $G$, then one can divide the set of 
\emph{dpt} into equivalence classes of a given \emph{dpt}. Two \emph{dpt} are defined
to be the members of the same class if they are associated by an element of $G$. The equivalence
class represents the same physical state of a system but in various reference 
frames. Mathematicans call the equivalence classes \emph{transitive fibres}.
\vspace{1ex} 
\\ DEFINITION 3. In general it is possible to divide $y$-s in two parts, dynamical 
$y_D$ and absolute $y_A$ in such a way that:
\begin{enumerate}
\item[1)] The part $y_D$ is that which distinguishes between various equivalence
classes.
\item[2)] $y_D$ constitute a geometrical object under the action of $G$. 
\item[3)] $y_A$ constitute a geometrical object under the action of $G$.
\item[4)] Any $y_A$ that satisfies the equations of motion of the theory appears,
together with all its transforms under $G$, in every equivalence class of 
\emph{dpt}.
\end{enumerate}
\vspace{1ex}
Exceptionally, when there exists only one equivalence class, then $y$ is wholly 
absolute. 
\vspace{1ex}
\\ DEFINITION 4. The subgroup of the covariance group $G$, which is the invariance 
(symmetry) group of all absolute objects $y_A$ is said to be the \emph{invariance}
(\emph{symmetry}) group of a theory.
\vspace*{1ex}
\\ It has to be clear what the invariance of absolute object means.

Let us illustrate those rather abstract definitions by an instructive example. In the example 
\emph{kpt} is described by a scalar field $ \Phi $ in the Minkowski spacetime. Let the dynamical law
in an inertial frame has the form of the heat equation:
\begin{equation}\label{6}
\vec{\nabla}^2\Phi - \kappa \partial_t\Phi = 0.
\end{equation}
If we do not introduce any other objects (beside the scalar field $ \Phi $) the covariance  
group $G_{\mathfrak{c}}$ of this theory consists of the inhomogeneous rotations and time translations group 
and is identical to the invariance group $G_{\mathfrak{i}}$ in this case. But one can use (in addition to the
scalar field) explicitely the Minkowskian metric $ g_{\mu \nu} $ and covariantly constant timelike unit vector field 
$ n^{\nu} $. With the help of these two additional objects one can write the above dynamical law (\ref{6}) in the
following form 
\begin{equation}\label{7}
h^{\nu \mu} \widetilde{\nabla}_{\nu} \widetilde{\nabla}_{\mu} \Phi - \kappa n^{\nu} \partial_{\nu} \Phi = 0,
\end{equation}
where $ h^{\nu \mu} \equiv g^{\nu \mu} - n^{\nu} n^{\mu} $ is the metric on the spacelike hyperplane
with the normal vector $ n^{\nu} $ induced by spacetime metric, $ \widetilde{\nabla}_{\nu} $ is the covariant derivative
due to this induced metric on the hyperplane.  The equation (\ref{7}) is generally covariant, \emph{i.e.} its 
covariance group $G_{\mathfrak{c}}$ consists of "all" spacetime mappings, this time however
$G_{\mathfrak{i}} \ne G_{\mathfrak{c}}$. Moreover, it is always possibe to 
make a theory generally covariant if one introduces the suitable objects to the theory. It is a known result obtained by 
Kretschmann \cite{Kret}. Generally covariant description of a theory reveals the additional \emph{i.e.} nondynamical
structures, which the theory really uses. In practice $ y_A $ is a part of $ y $ which can be consistently (independently
of reference frame) defined as that part of $ y $, which does not enter into the equations of motion for the remaining part
$ y_D $ of $ y $. So, $ y_A $ can be eliminated from the theory in question without altering the structure of its 
equivalence classes. After such elimination $ y $ is equal to its dynamical part $ y_D $ -- $ y_A $ does not enter
into the theory explicitly. The elimination makes the covariance group smaller then the initial one 
(before the elimination), but the group is only apparently smaller. In the above example the first formulation is 
given after such eliminaton and in this formulation the covariance group is the same as the invariance (symmetry) 
group (but only apparently). Note also that after such elimination $G_{\mathfrak{i}} = G_{\mathfrak{c}}$ and moreover
$\Phi$ and $\Phi'$ -- the transformed $\Phi$ by an element of $G_{\mathfrak{i}} = G_{\mathfrak{c}}$ --  both are
solutions to (\ref{6}) which is now identical for all admissible reference frames. Note for the later use, \emph{that this is 
always the case if one eliminates the absolute elements $y_A$ as above, i.e.
the invariance group $G_{\mathfrak{i}}$ acts in the space of solutions to the dynamical equations}. 
 Generally covariant formulation makes clear what are the absolute elements and what 
is the symmetry group. In the above example given by Eqs (\ref{6}), (\ref{7}) the spacetime metric $g_{\mu\nu}$
and the vector field $n^{\mu}$ form the absolute object $y_{A}$.
\\ So, even if one can always make a theory covariant, the covariance condition is not empty, because the price
for such covariant formulation is in general high: one has to introduce \emph{ad hoc} new objects.     

\section{Newton-Cartan spacetime}

Contrary to the spacetime of General Relativity, the Newton-Cartan spacetime is 
described by three independent objects. The first is the absolute time $t$ or 
equivalently the gradient of absolute time $t_\mu \equiv \partial_\mu t$, defining
the Euclidean simultaneity hyperlanes $t=constant$. The second is the contravariant 
symmetric tensor field $g^{\mu\nu}$, which is degenerate: $g^{\mu\nu}t_\nu=0$ and its
rank is three. The third is the connection $\Gamma_{\nu\rho}^\mu$, such that
$\nabla_\mu g^{\nu\rho}=0$ and $\nabla_\mu t_\nu=0$. The last two conditions do not 
determine the connection because $g^{\mu\nu}$ is degenerate. Compare e.g. 
\cite{Tra,Dau}.

It is extremely inconvenient to rewrite the Schr\"odinger equation
in a generally covariant form in terms of the connection $\Gamma_{\nu\rho}^\mu$
(and the remaining two objects $t_\mu$ and $g^{\mu\nu}$), compare e.g. \cite{DK,Waw}. 
To do this, it is convenient to describe the Newton-Cartan spacetime in a "metric-like 
way", as was shown in \cite{Waw}. Namely, with the help of the connection 
$\Gamma_{\nu\rho}^\mu$ a covariant tensor $g_{\mu\nu}$ and contravariant vector $u^\mu$ 
can be defined. $g_{\mu\nu}$ and $u^\mu$ \emph{replace completely the connection}. 
$\Gamma_{\nu\rho}^\mu$ is determined by motions of the freely falling particles, 
\emph{i.e.} geodesics. Geodesic is a solution of the Lagrange-Euler equations for a 
free particle Lagrange function $L$. Rewriting $L$ in a generally covariant form,
one gets (compare \cite{Waw})
\begin{displaymath}
L=\frac{m}{2}\frac{a_{\mu\nu}\dot{x^\mu}\dot{x^\nu}}{\dot{t_\sigma x^\sigma}},
\end{displaymath}
where $a_{\mu\nu}$ is some covariant field and the particle's trajectory is 
$x^\mu=x^\mu(\tau)$, dot denotes derivative with respect to the parameter $\tau$. 
We define $g_{\mu\nu}$ as equal to $a_{\mu\nu}$ in the Lagrange function $L$.      
So, we have 
\begin{displaymath}
L=\frac{m}{2}\frac{g_{\mu\nu}\dot{x^\mu}\dot{x^\nu}}{\dot{t_\sigma x^\sigma}}.
\end{displaymath}
$u^\mu$ is defined in the following way
\begin{displaymath}
g_{\mu\nu}g^{\nu\sigma}=\delta_\mu^\sigma-u^\sigma t_\mu, \quad u^\mu t_\mu=1.
\end{displaymath}
Because $L$ is determined up to a full parameter derivative $L\rightarrow L+
m\frac{df}{d\tau}$, $g_{\mu\nu}$ is determined up to the gauge transformation
$g_{\mu\nu}\rightarrow g_{\mu\nu} + t_\mu\partial_\nu f+t_\nu\partial_\mu f$, and in consequence
$u^\mu$ also is determined up to the gauge transformation $u^\mu\rightarrow u^\mu-g^{\mu\nu}
\partial_\nu f$.
\\The Lagrange-Euler equations for $L$ give the geodetic equation with the connection
\begin{equation}\label{kon}
\Gamma_{\nu\rho}^\mu=u^\mu \partial_\nu t_\rho + \frac{1}{2}g^{\mu\sigma}
\{\partial_\nu g_{\rho\sigma} + \partial_\rho g_{\nu\sigma} - 
\partial_\sigma g_{\nu\rho}\}.
\end{equation}
The connection given by the formula (\ref{kon}) is gauge independent, as immediately 
follows from the fact that the solution of the Lagrange-Euler equations does not 
change after adding a full differential term to the Lagrange function $L$. But, of 
course, one can check it directly. The quantities $g_{\mu\nu}, u^\nu$ as well as 
the Eq. (\ref{kon}) were introduced first by Da\u utcourt \cite{Dau} but in 
a different way. He considered in \cite{Dau} the Newtonian limit of General Relativity.
Now an important moment comes. Namely, what is the role of the gauge transformations
introduced here? The full transformation group of $g_{\mu\nu}$ and $u^\mu$ consists of coordinate 
transformations, but not only. Beside the coordinate freedom there exists the gauge freedom.
So, the full transformations are 
\begin{equation}\label{trg}
g_{\mu\nu} \rightarrow \frac{\partial x^{\mu}}{\partial x^{\mu'}}
\frac{\partial x^{\nu}}{\partial x^{\nu'}}(g_{\mu\nu}+t_{\nu} \partial_{\mu} f  + 
t_{\mu} \partial_{\nu} f),
\end{equation}
\begin{equation}\label{tru}
u^{\mu} \rightarrow \frac{\partial x^{\mu'}}{\partial x^{\mu}} (u^{\mu} - g^{\mu\nu}
\partial_{\nu} f).
\end{equation}
\emph{One should not think of coordinate and gauge transformations as of separate 
transformations}, for they mix with each other.

Consider first the Galilean spacetime, \emph{i.e.} the Riemann curvature
of $\Gamma_{\nu\rho}^\mu$ is zero. Now we ask the question: can we find an appropriate
connection between $f$ and coordinate transformation such that $g_{\mu\nu}$ and $u^\mu$
become invariant for the Galilean group of transformations? The answer is -- and should 
be -- positive. Moreover, such $f$ connected with coordinate transformation is unique 
and non-trivial, that is, \emph{f cannot be identically equal to zero}. This question is 
equivalent to the following problem: what is the invariant form of the Lagrange function 
$L$ of a freely falling particle? The problem is not well defined even in the Landau and 
Lifshitz \cite{Lan} very good course of theoretical physics. It can be shown, that accordingly to 
their definition of invariance, there are infinitely many invariant Lagrange functions in Galilean 
spacetime, see Ref. \cite{Waw}. 

Consider now the Newton-Cartan spacetime. This spacetime as a whole does not possess
any symmetry in general and one could suppose the gauge and the coordinate transformations to be 
completely independent in this case. But the Newton-Cartan spacetime possesses the absolute 
elements in addition to the dynamical ones (contrary to the General Relativity), namely $t_\mu,
g^{\mu\nu}, u^\mu$ and $\widetilde{g}_{\mu\nu} \equiv g_{\mu\nu}-g_{\sigma\rho}u^\sigma 
u^\rho t_\mu t_\nu$ are absolute ones in the general sense given above. The Newtonian 
potential $\phi=-\frac{1}{2}g_{\mu\nu}u^\mu u^\nu$ is dynamical one. The absolute 
elements have the symmetry group of transformations -- the so called Milne group, 
which in the Cartesian non-rotating coordinates have the following form
\begin{equation}\label{tra}
t'=t+b, \qquad \vec x'=R\vec x + \vec A(t),
\end{equation}
where $R$ is an orthogonal martix and $\vec A(t)$ is "any" function of time and $b$ is 
any constant. Then we see that (\ref{tra}) connect the Cartesian nonrotating frame with the
Cartesian nonrotating frame in which $t_\mu, g^{\mu\nu}, u^\mu$ and 
$\widetilde{g}_{\mu\nu}$ have particulary simple form. The gauge function
$f$ in Eqs (\ref{trg}), (\ref{tru}) is  uniquely determined by 
the invariance condition of $t_\mu, g^{\mu\nu}, u^{\mu}$ and $\widetilde{g}_{\mu\nu}$
with respect to the group (\ref{tra}). 
Namely, we ask the question: can we find such a connection of $f$ with coordinate 
transformation which brings $t_\mu, g^{\mu\nu}, u^\mu$ and $\widetilde{g}_{\mu\nu}$ 
into a form invariant with respect to the group (\ref{tra})? Again the answer is positive,
 \emph{i.e.} $f$ is determined by this condition up to an arbitrary function of time, see
\cite{Waw} for the proof. 

Now, one can write the Schr\"odinger equation in the Galilean spacetime in a generally covariant form and 
then obtain the Schr\"odinger equation in Newton-Cartan spacetime 
\begin{equation}\label{Sch}
i\hbar u^\mu\partial_\mu \psi=-\frac{\hbar^2}{2m} g^{\mu\nu}\nabla_{\mu}\nabla_{\nu}\psi-
\frac{m}{2} g_{\mu\nu}u^{\mu}u^{\nu}\psi-\frac{i\hbar}{2} \nabla_{\mu}u^{\mu}\psi,
\end{equation}
with the following transformation law for $\psi$
\begin{equation}\label{tpsi}
\psi'(X')=e^{i\frac{m}{\hslash}f}\psi(X),
\end{equation}
using the minimal coupling procedure, where $f$ in (\ref{tpsi}) is exactly the same as in (\ref{trg}) 
and (\ref{tru}), see  \cite{Waw}. 
However, we are interested (in this paper) in the most general form of the wave equation, 
which is covariant and built up with the use of the spacetime objects and is in agreement 
with the principles of quantum mechanics, linear, local and of second order at most. It can 
\emph{a priori} be different from (\ref{Sch}). Moreover, the gravitational mass can be 
different from the inertial one, contrary to (\ref{Sch}). (Eq. (\ref{Sch}) is equivalent to 
the generally covariant wave equation which was found 
in \cite{Kuch} and \cite{DK}, see \cite{Waw} for the proof.)

\section{Covariance condition}

In this Chapter we derive the general form of the wave equation in the Newton-Cartan spacetime 
assuming (A) $\div$ (D). It is a crucial point for us that there exist absolute elements in this spacetime. 
There exist the privileged, \emph{i.e.} nonrotating Cartesian coordinates (which we call after Kucha\v r \cite{Kuch}
the \emph{galilean} coordinates),
in which the absolute elements take on a \emph{canonical} particulary simple form:
\begin{displaymath}
(t_{\mu})=(1, 0, 0, 0),  (g^{\mu \nu})= \left( \begin{array}{cccc}
0  & & {} & 0\\
& {}  1 & & {}\\
& & {} 1 & {} \\
0 &  & {} & 1
\end{array} \right),
\end{displaymath}
\begin{displaymath}
(u^{\mu})=\left( \begin{array}{c}
1\\
0\\
0\\
0
\end{array} \right), 
 (\tilde{g}_{\mu \nu}) = \left( \begin{array}{cccc}
0 & & {} & 0\\
& {}  1 & & {}\\
& & {} 1 & {} \\
0 &  & {} & 1
\end{array} \right).
\end{displaymath}
It largely simplifies the investigation of such a problem as general covariance. The simplifacation comes, 
because the absolute elements are invariant under the transformation group (\ref{tra}) and have the same
\emph{canonical} form in all \emph{galilean} coordinate frames. So, the  potential $\phi$, is the only object, which 
describes the geometry and does not trivialy simplify to 
a constant equal to 0 or 1, in those coordinates. From the fact that the wave equation is generally covariant follows,
of course, that -- written in \emph{galilean} coordinates -- is covariant under the Milne group (\ref{tra}). 
The group (\ref{tra}) is sufficiently rich to determine the wave equation as the covariant equation under 
the group (\ref{tra}) fulfilling the assumtion (C) (and making use of the classification of $\theta$-s in the 
formula (\ref{trapsi}), which itself uses (A) $\div$ (D)). We confine ourselves then, to the \emph{galilean} coordinates. 

First, we will show that $T_{r}\psi$,
for the transformation $r$ of the Milne group (\ref{tra}) has the form 
\begin{equation}\label{trapsi}
\psi'(X) = T_{r}\psi(X) = e^{-i\theta(r,X)}\psi(r^{-1}X).
\end{equation}
Everything, which we need now to know about the quantum interpretation of $\psi$, is that 
\begin{displaymath}
\int_{{\mathbf{R}}^3}\psi^*(\vec{x},t)f(\vec{x},t)\psi(\vec{x},t) \, {\ud}^3 x
\end{displaymath}
is the average $\overline{f}$ of the $f(X)$ in the state described by the wave function $\psi(X)$. 
The second fact which we need is that the spacetime natural measure $\mu$ depends on the absolute objects
(of the Newton-Cartan spacetime) only: 
\begin{displaymath}
\mu({\mathbf{D}}) = \int_{{\mathbf{D}}}\nu(X) \, {\ud}^{4 }X, 
\end{displaymath}
where 
\begin{displaymath}
\begin{array}{c}

\vspace{0.3cm}

\nu(X)= \sqrt{\det[g_{\mu \nu}+(1-g_{\alpha \beta}u^{\alpha}u^{\beta})t_{\mu}t_{\nu}]} \equiv \\
\equiv \sqrt{\det[\widetilde{g}_{\mu \nu} +t_{\mu}t_{\nu}]}, 
\end{array}
\end{displaymath} 
see \cite{Waw} for the proof. This means that $\mu$ is invariant under the group (\ref{tra}):
$\mu({\mathbf{D}})=\mu(r {\mathbf{D}})$ or 
\begin{displaymath}
\int_{{\mathbf{D}}}g(X) \, \nu(X) \, {\ud}^{4}X=\int_{r^{-1}{\mathbf{D}}}g(rX)\nu(X) \, {\ud}^{4}X.
\end{displaymath}
Recall that $\nu(X)\equiv 1$ in \emph{galilean} coordinates. Consider two \emph{galilean} coordinate systems
$(X)$ and $(X')=(rX)$ 
connected by a transformation $r$ of the form (\ref{tra}). Then make the experiment in which the average  of a
quantity $f(X)$ in the two reference frames is measured. Let $\overline{f}'$ and $\overline{f}$ be the result of the 
experiment in reference frames $(X')$ and $(X)$ respectively,
 or in states $\psi'=T_{r}\psi$ and $\psi$. Then we have: $\overline{f(r^{-1}X)}'=
\overline{f(X)}$, \emph{i.e.} the average of the quantity in the moved frame has to be the same as that 
of the correspondingly moved quantity in the initial
frame. Consider  $f(X)$ with compact support. So we have \cite{Herdegen}
\begin{displaymath}
\int_{{\mathbf{R}}^{3}}\psi^{'*}(\vec{x},t)f(r^{-1}(\vec{x},t))\psi'(\vec{x},t) \, {\ud}^{3}x   \equiv  \overline{f(r^{-1}X)}' = 
\end{displaymath}

\vspace{-0.5cm}

\begin{displaymath}
= \overline{f(X)} \equiv \int_{{\mathbf{R}}^{3}}\psi^*(\vec{x},t)f(\vec{x},t)\psi(\vec{x},t) \, {\ud}^{3}x.
\end{displaymath}
Now we integrate both sides of this equation over $t$: $\int_{t_{1}}^{t_{2}} \, {\ud} t \ldots$ with $t_{1}$ and $t_{2}$
chosen in such a way that the four dimensional domain of the resulting integration contains supports of
$f(X)$ and $f(r^{-1}X)$. We obtain in this way the equality 
\begin{displaymath}
\int_{{\mathbf{R}}^4}\vert \psi'(X)\vert^{2}f(r^{-1}X) \, {\ud}^{4}X=\int_{{\mathbf{R}}^4}\vert \psi(X) \vert^{2}f(X) \, {\ud}^{4}X,
\end{displaymath}
for "all" $\psi$ and $f(X)$. By invariance property of those integrals we easily obtain
\begin{equation}\label{integration}
\int_{{\mathbf{R}}^4} [\vert \psi'(rX) \vert ^{2}-\vert \psi(X) \vert^{2}]f(X) \, {\ud}^{4}X=0
\end{equation}
for "all" $\psi$ and $f(x)$ with compact support. Because $\psi$ fulfils the differential wave equation
it has to be continuous. Then, $g(X) \equiv \vert \psi'(rX) \vert^{2} - \vert \psi(X) \vert^{2} = 0$ and 
$\psi' \equiv T_{r}\psi$ is of the form (\ref{trapsi}), or equivalently the probability density $\rho$
is a scalar field. Indeed, suppose that there is a point $X_{o}$ in which 
$g(X_{o}) \ne 0$. Because of the continuity of $g(X)$ there exists a neighborhood of $X_{o}$ with 
a compact closure $\mathcal{C}$ on which $g(X)>0$. Then, because there exists a differentiable function 
$f(X) \geqslant 0$  with the support $\mathcal{C}$, the integral in (\ref{integration}) is greater then 0, 
contrary to (\ref{integration}). 

 Recall, that the existence of the invariant measure is important in the proof. This is
a peculiar property of the Newton-Cartan spacetime (we have in mind the invariance as defined at the 
begining of this chapter, not the 'invariance' frequently used in differential geometry where it
means that the scalar density $\nu$ of the measure is covariantly constant). In the Einsteinian
spacetime the measure has no invariance group when the spacetime has no symmetry. 
Recall also that $\psi$ is differentiable up to second order being a solution of second order wave equation,
by this $\theta(r,X)$ also is differentiable up to second order.
  
In accodance to the assumption (C) the wave equation has the form 
\begin{equation}\label{eq}
[a\partial_{t}^{2} + b^{i}\partial_{i}\partial_{t} + c^{ij}\partial_{i}\partial_{j} + d\partial_{t} + f^{i}\partial_{i} + g]\psi(X) = 0,
\end{equation}
where $a(X), b^{i}(X), \ldots g(X)$ are fields which can be built up in a local way with the 
help of the geometrical objects describing the spacetime structure.
(Note that $a(X), \ldots, g(X)$ as functions of spacetime coordinates $X$ are not the same 
in each \emph{galilean} system.) But in accordance with our assumption they are functions 
of spacetime geometric objects the same in all systems, built in a local way from those objects. 
However, the only non-trivial object, which describes the geometry, and does not trivially simplify to a constant
equal to 0 or 1 in the \emph{galilean} coordinates, is the Newtonian potential $\phi$. 
Then, $a = a(X), \ldots, g(X)$ are algebraic functions of $\phi$ and its derivatives identical 
in all \emph{galilean} systems,
\emph{i.e.}
\begin{displaymath}
\begin{array}{c}

\vspace{0.3cm}

a=a(\phi, \partial_{i}\phi, \partial_{i}\partial_{j}\phi, \partial_{t}\phi, \partial_{t}^{2}\phi, \ldots, \partial_{t}^{n}\phi), \\ 

\vspace{0.3cm}

\ldots \ldots \ldots \ldots \ldots \ldots \ldots \ldots \ldots \ldots \ldots \ldots \\
g=g(\phi, \partial_{i}\phi, \partial_{i}\partial_{j}\phi, \partial_{t}\phi, \partial_{t}^{2}\phi, \ldots, \partial_{t}^{n}\phi). \\
\end{array}
\end{displaymath}
In the mathematical terminology this means that $a, \ldots, g$ are \emph{differential concomitants} 
of the potential, see \cite{Acz}.  
We assume in addition that $n=2$. We do not lose any generality by this assumption, beside this
the whole reasoning could be applied for any $n$. But the case with $n>2$ would not be physically
interesting. Namely, it is \emph{a priori} possible that the derivatives of second order are discontinuous, 
such that the derivatives of order $n>2$ do not exist, at least the classical geometry does allow such a
situation. On the other hand there does not exist any physical obstruction for a discontinuity of the wave
equation coefficients, take for example the wave equation with the "step-like" potential. Then the assumtions
about the existence of higher oder derivatives -- not necessary for the spacetime geometry -- confines
our resoning rather then generalizes it. 

Because $a, \ldots, g$ do not depend explicitly on the spacetime coordinates $X$, then our 
reasoning is apparently not general, but only apparently. Indeed, suppose they do depend explicitly
on $X$, so they are built with the help of an additional function of $X$. Then, according to our assumtion
(C) this function would have to be build of the potential and its derivatives, and $a, \ldots, g$ would be
algebraic functions of the potential and its derivatives. 
The situation with $\theta$ is however different. Namely, $\theta$ does not itself contribute to the transformed
equation (because the wave equation is linear the exponent $\theta$ of (\ref{trapsi}) cancells in the transformed equation),
but its first and second derivatives with respect to $X$ do. So, $\theta$ can explicitly depend on $X$. 

By $a(X), ... \,$ we always mean $a(\phi(X), ... \, \partial_{t}^{2}\phi(X)), ... \, $ and for simplicity we do not introduce
different notation for the two different sets of functions, but no confusion should arise from it. 
 
The covariance condition for (\ref{eq}) under the group (\ref{tra}) in the \emph{galilean} systems
gives the following transformation laws of $a, b^{i}, c^{i j}$
\begin{equation}\label{trb}
{b'}^{i}(X')=R^i_j b^{j}(X)+2a(X) \dot{A}^{i},
\end{equation}
\begin{equation}\label{a}
a'(X')=a(X),
\end{equation}
\begin{equation}\label{trc}
{c'}^{ij}(X')=R^i_s R^j_k c^{sk}(X)+a(X) \dot{A}^{i} \dot{A}^{j}+b^{k}R^i_k \dot{A}^{j}.
\end{equation}
From (\ref{a}) it follows that $a$ is a \emph{scalar comitant} of the potential, or 
equivalently 
\begin{displaymath}
\begin{array}{c}

\vspace{0.3cm}

a(\phi'(X'), \partial'_{i}\phi'(X'), \ldots, {\partial'}_t^2\phi'(X'))= \\
=a(\phi(X), \partial_{i}\phi(X), \partial_{i}\partial_{j}\phi(X), \partial_{t}\phi(X), \partial_{i}\partial_{t}\phi(X), 
\partial_{t}^{2}\phi(X))
\end{array} 
\end{displaymath}
and $a$ is constant  along each \emph{transitive fibre} of the object 
\begin{displaymath}
\Omega = (\phi, \partial_{i}\phi, \partial_{i}\partial_{j}\phi, \partial_{t}\phi, \partial_{i}\partial_{t}\phi, \partial_{t}^{2}\phi),
\end{displaymath}
as a function of $\phi$.
It can be shown that this determinates $a$ as a function of the Kronecker's invariants of the matrix 
$(\partial_{i}\partial_{j}\phi)$, using the well known Theorems of the \emph{Theory of Geomtric Objects}, 
compare e.g. \cite{Zaj}. However, we will proceed in a different way begining with the concomitant
$b^{i}$ which transforms as in (\ref{trb}). If one started the whole anlysis with $a$ and (\ref{a}) then
one would have to analyse (\ref{trb}) repeating many steps. On the other hand already from (\ref{trb}) it will
follow that $a=0$ and independent analysis of (\ref{a}) will not be necessary. Because the analysis
is rather long, we break it in the steps in which $a, \ldots, g$ are computed respectively.

\subsection{ $b^{i}, a$}

The situation with $b^{i}$ is more complicated because it is not a scalar concomitant, but has a more 
complicated transformation law. But we gradually simplify the situation.       
That means, we find subgroups of (\ref{tra}) which define such transitive fibers of $\Omega$ on 
which $b^{i}(\Omega)$ has to be constant. We will show it step by step. 

To simplify the reading we write the explicit form of the transformation laws for $\phi$ and its derivatives.
\begin{displaymath}
\phi'(X')=\phi(X)-\ddot{A}^{i}x_{i}, 
\end{displaymath}

\vspace{-0.5cm}

\begin{displaymath}
\partial'_{j}\phi'(X')={R^{-1}}^{i}_{j}\partial_{j}\phi(X)-{R^{-1}}^{i}_{j}\ddot{A}_{i}, 
\end{displaymath}

\vspace{-0.5cm}

\begin{displaymath}
\partial'_{i}\partial'_{j}\phi'(X')={R^{-1}}^{k}_{i}{R^{-1}}^{s}_{j}\partial_{k}\partial_{s}\phi(X), 
\end{displaymath}

\vspace{-0.5cm}

\begin{displaymath}
\partial'_{t}\phi'(X')=\partial_{t}\phi(X)-\dddot{A}^{i}x_{i}-{R^{-1}}^{i}_{k}\dot{A}^{k}\partial_{i}\phi(X)+
{R^{-1}}^{i}_{k}\dot{A}^{k}\ddot{A}_{i}, 
\end{displaymath}

\vspace{-0.5cm}

\begin{displaymath}
\partial'_{j}\partial'_{t}\phi'(X')={R^{-1}}^{i}_{j}\partial_{i}\partial_{t}\phi(X)- 
\end{displaymath}

\vspace{-0.5cm}

\begin{displaymath}
- {R^{-1}}^{r}_{k}{R^{-1}}^{i}_{k}\dot{A}^{k}\partial_{r}\partial_{i}\phi(X) - {R^{-1}}^{r}_{j}\dddot{A}_{r}, 
\end{displaymath}

\vspace{-0.5cm}

\begin{displaymath}
{\partial'_{t}}^{2}\phi'(X')= {\partial_{t}}^{2}\phi(X)-2{R^{-1}}^{i}_{k}\dot{A}^{k}\partial_{i}\partial_{t}\phi(X)+ 
\end{displaymath}

\vspace{-0.5cm}

\begin{displaymath}
+ {R^{-1}}^{j}_{k}{R^{-1}}^{i}_{s}\dot{A}^{k}\dot{A}^{s}\partial_{j}\partial_{i}\phi(X)- 
\end{displaymath}

\vspace{-0.5cm}

\begin{displaymath}
- \ddddot{A}_{i}{x}^{i}-{R^{-1}}^{i}_{k}\ddot{a}^{k}\partial_{k}\phi(X)+{R^{-1}}^{i}_{k}\ddot{A}^{k}\ddot{A}_{i}+
2{R^{-1}}^{i}_{k}\dot{A}^{k}\dddot{A}_{i}. 
\end{displaymath}
At the beginning we will show, that the concomitant $\vec{b}$ does not depend on the time derivatives
and the space-time derivatives of the potential. To do this we use the transformations 
(\ref{tra}) with $R={\bf 1}$ and $b=0$. The formula (\ref{trb}) is valid in each \emph{galilean} 
system and for any potential, and implicitly at any spacetime point. Let us take then, such 
a system and let $X_{o}$ be any (but fixed) spacetime point. At last, we define $\vec{A}(t)$ in (\ref{tra}) 
as follows: $\vec{A}(t)=A(t)\vec{n}$, where $A(t)=\lambda (t-t_{o})^4$
and $\vec{n}$ is a constant (in space and time) spacelike unit vector, $\lambda$ is any constant. 
Then, all derivatives of $\vec{A}$ are zero in $t_{o}$ except the fourth order derivative.  The formula 
(\ref{trb}) with this transformation gives  
\begin{displaymath}
b^{k}(\phi, \partial_{i}\phi, \partial_{i}\partial_{j}\phi, \partial_{t}\phi, \partial_{j}\partial_{t}\phi, \partial_{t}^{2}\phi-
4!\lambda n_{s}x_{o}^{s})=
\end{displaymath}

\vspace{-0.7cm}
 
\begin{displaymath}
= b^{k}(\phi, \partial_{i}\phi, \partial_{i}\partial_{j}\phi, \partial_{t}\phi, \partial_{j}\partial_{t}\phi, \partial_{t}^{2}\phi),
\end{displaymath}
at the point $X_{o}$.
Now one can infer that $b^{k}$ does not depend on $\partial_{t}^{2}\phi$ at the point $X_{o}$, because this 
formula is fulfilled for all $\lambda$-s. Of course with the exception of $\vec{x}_{o}=0$. But one can choose
the point  $X_{o}$ in an arbitrary way and $b^{k}$ does not depend on $\partial_{t}^{2}\phi(\vec{x}, t)$ with the
exception of the point $(0,t)$, \emph{i.e.} $b^{k}$ can \emph{a priori} depend on $\partial_{t}\phi(0,t)$. This 
dependence, however, is irrelevant, because the potential is determined up to any additive function of time
$F(t)$, \emph{i.e.} the potential $\phi$ is determined  up to the gauge freedom. 
We choose $F(t) \equiv -\phi(0, t)+\alpha t + \beta$ and eliminate 
this irrelevant dependence (in each \emph{galilean} system the gauge is chosen separately, of course). 
The two constants $\alpha$ and $\beta$ in $F(t)$ is the only potential gauge freedom which remains at our 
disposal now.

We apply now a similar transformation, but this time $A(t)=\lambda (t-t_{o})^{3}$ and only the third order 
derivative of $\vec{A}$ at $t_{o}$ is not equal to zero. The formula (\ref{trb}) for such a transformation reads
\begin{displaymath}
b^{k}(\phi, \partial_{i}\phi, \partial_{i}\partial_{j}\phi, \partial_{t}\phi-3!\lambda n_{s}x_{o}^{s}, \partial_{i}\partial_{t}\phi
-3!\lambda n_{i}) = 
\end{displaymath}

\vspace{-0.7cm}

\begin{equation}\label{rb1}
=b^{k}(\phi, \partial_{i}\phi, \partial_{i}\partial_{j}\phi, \partial_{t}\phi, \partial_{i}\partial_{t}\phi),
\end{equation}
for all $\lambda$. It means that $b^{k}$ is constant along those two directions of $\vec{\partial}\partial_{t}\phi(X_{o})$ 
which are perpendicular to $\vec{x}_{o}$. Because the point $X_{o}$ and the \emph{galilean} frame was arbitrary, it is 
true for any \emph{galilean} system and at any spacetime point. Because $b^{k}$ in the chosen frame
is identical function (for each $k$)
of $\vec{\partial}\phi$ as in the translated reference frame, $b^{k}$ is constant along any direction of 
$\vec{\partial}\partial_{t}\phi$,
so, $b^{k}$ does not depend on $\vec{\partial}\partial_{t}\phi$. (For, one may reach any direction of $\vec{x'}_{o}$ by 
the appropriate translation $X \to X'$, and by this one may aquire any direction of constancy of 
$b^{k}$ in the space of variables 
$\vec{\partial}\phi'(X'_{o})$.)

Applying the same transformation as in the last step to the formula (\ref{trb}) one gets
\begin{displaymath}
b^{k}(\phi, \partial_{i}\phi, \partial_{i}\partial_{j}\phi, \partial_{t}\phi - 3!\lambda n_{s}x_{o}^{s})= 
\end{displaymath}

\vspace{-0.7cm}

\begin{equation}\label{rb3}
=b^{k}(\phi, \partial_{i}\phi, \partial_{i}\partial_{j}\phi, \partial_{t}\phi) 
\end{equation}
at $X_{o}$. So, in a similar way as before one can see that $b^{k}$ does not depend on $\partial_{t}\phi(X)$ except 
that it can depend on $\partial_{t}\phi(\vec{x}=0,t)$. But again this dependence is irrelevant and can be eliminated 
by an appropriate gauge. Namely, choosing $\alpha=0$ in $F(t)$ we eliminate this dependence, so that
$b^{k}=b^{k}(\phi, \partial_{i}\phi, \partial_{i}\partial_{j}\phi)$.

At last, we make use of the transformation of the same type as before with $A(t)= \lambda (t-t_{o})^{2}$, (\ref{trb})
for this transformation gives at $X_{o}$
\begin{displaymath}
b^{k}(\phi - 2!\lambda n_{s}x_{o}^{s}, \partial_{i} -2!\lambda n_{i}, \partial_{i}\partial_{j}\phi)=
b^{k}(\phi, \partial_{i}\phi, \partial_{i}\partial_{j}\phi),
\end{displaymath}
and the situation with the variables $\phi$ and $\partial_{i}\phi$  is exactly the same as was with 
$\partial_{t}\phi$ and $\partial_{i}\partial_{t}\phi$. Then, in the analogous way one gets
\begin{displaymath}
b^{k}(\vec{x},t)=b^{k}(\partial_{i}\partial_{j}\phi(\vec{x}, t)), \, \vec{x} \ne 0, 
\end{displaymath}

\vspace{-0.7cm}

\begin{displaymath}
b^{k}(0,t)=b^{k}(\partial_{i}\partial{j}\phi(0,t)), \, \vec{x}=0,
\end{displaymath}
so, we have
\begin{displaymath}
b^{k}(X)=b^{k}(\partial_{i}\partial_{j}\phi(X)).
\end{displaymath}
It means, that $b^{k}$ is a vector concomitant (at least under rotations, spatial inversion and spatial reflections) 
of a tensor $\partial_{i}\partial_{j}\phi$ of valence 2. As is well known from the theory of geometric objects, see
e.g. \cite{Kucharzewski}, such a vector concomitant has to be zero. The argumentation is as follows. Take any 
\emph{galilean} system and any point $X_{o}$. Apply now the space inversion with the origin in $X_{o}$,
\emph{i.e.} $R=-{\bf 1}$ and $ \vec{A}=2\vec{x}_{o}, \, b=0$. Then, (\ref{trb}) at $X_{o}$ with this inversion gives 
the equation: $\vec{b}(X_{o})=-\vec{b}(X_{o})$ because the valence of $\partial_{i}\partial_{j}\phi$ is even 
and $\partial_{i}\partial_{j}\phi$ does not change the sign under the inversion. Because the point $X_{o}$ 
and the \emph{galilean} reference frame can be chosen in an arbitrary way the concomitant $\vec{b} = 0$. 

From (\ref{trb}) immediately follows, that also $a=0$.

\vspace{1ex}

REMARK. If we were not specify the gauge function $F(t)$ (to be equal $-\phi(0,t) + \beta$), then $b^{k}(X)$ 
would be equal to zero with the possible exception of $b^{k}(0,t)$, which could be different from zero. For 
the later use, however,  we will not specify $F(t)$, and we will analyse the quantities $a(X), b^{k}(X), ... , g(X)$
at those points $X$ whose space coordinates are not equal to zero. We will not indicate it explicitly.
Then, in particular
\begin{displaymath}
b^{k}(x^{i},t) = 0, \, a(x^{i},t) = 0, \, \textrm{if}  \, \,  x^{i} \ne 0.
\end{displaymath}

\subsection{$\vec{f}, c^{ij}$}

We have reduced our equation (\ref{eq}) to the following form
\begin{displaymath}
[c^{ij}\partial_{i}\partial_{j}+d\partial_{t}+f^{i}\partial_{i}+g]\phi=0.
\end{displaymath}
Covariance condition of the equation under the Milne group of coordinate transformations (\ref{tra}) gives
the following transformation law for $f^{j}$
\begin{equation}\label{22}
{f'}^{j}(X')=R^{j}_{i}f^{i}(X) - d\dot{A}^{j}-2ic^{ij}\partial_{i}\theta,
\end{equation}
where $\theta$ is the exponent in the transformation law of $\psi$ given by the formula (\ref{trapsi}).
Now, we make use of the classification of $\theta$-s presented in the next section and in the Appendix {\bf A}.
According to this classification
(see in particular the subsection {\bf V.C})
\begin{equation}\label{23}
\theta=-\gamma_{1}\frac{d}{dt}A_{j}x^{j}- \ldots -\gamma_{n}\frac{d^{n}}{{dt}^{n}}A_{j}x^{j} 
+\widetilde{\theta}(A^{k}, t),
\end{equation}
for any polynomial Milne transformation $A_{j}(t)$, where $n$ is $\geq$ then the maximal degree
of the polynomials $A_{j}(t)$ in the Milne transformation, $\gamma_{j}$ are constants.
This is the very important point. The whole meaning of the analysis of the section 
{\bf V} and the Appendix {\bf A} for the derivation of the equality 
$m_{i} = m_{g}$ lies in the above formula (\ref{23}) for $\theta$.   
Note, that $\theta$ is determined up to any function of time and the function
$\widetilde{\theta}$ is not determined by the classification (according to the fact that the wave function 
is determined up to a time dependent phase factor, see the next chapter).  

First of all let us take notice of the fact that $\gamma_{k}=0$ for $k>4$. Indeed, let $X_{o}=
(\vec{x_{o}},t_{o})$ be any point. We apply now a transformation (\ref{tra}) for which all derivatives of 
$\vec{A}(t)$ disappear at $t_{o}$ with the exception of the $k$-th order derivative (for example we can choose
such a transformation as in the preceding subsection with $A(t)=(t-t_{o})^{k}$). Then, we insert the transformation to the
law (\ref{22}). Because the derivatives of the order higher then the 4-th do not appear in the transformation laws for
$\phi, \partial_{i}\phi, \ldots, \partial_{t}^{2}\phi$, then (\ref{22}) at $X_{o}$ implies that $\gamma_{k}=0$.  
Note, that $f^{j}$ is local, \emph{i.e.} it can be built from the potential in a local way, that is, in an algebraic way
from the potential and its finite order derivatives with the order $\leqslant$ then say $m$. 
This is essential for the finitnees of $n$ which is $\leqslant m+2$. 

Note, that the covariance condition under the polynomial accelerations $\vec{A}(t)$ 
is sufficient for us.

Let $X_{o}$ be any spacetime point. We define the following object
\begin{displaymath}
\widetilde{f}^{j} \equiv f^{j}  +2i\gamma_{2}c^{ij}\partial_{i}\phi + 2i\gamma_{3}c^{ij}\partial_{t}\partial_{i}\phi.
\end{displaymath}
It has the transformation law 
\begin{displaymath}
{\widetilde{f}}^{\, ' i}(X')=R^{i}_{s}\widetilde{f}^{s}(X)-(d -2i\gamma_{1}c^{sj}R^{i}_{s} -
\end{displaymath}

\vspace{-0.7cm}

\begin{equation}\label{trf'}
-2i\gamma_{3}{R^{-1}}^{i}_{s}{R^{-1}}^{q}_{p}c^{sk}\partial_{q}\partial_{k}\phi\delta^{pj})\dot{A}_{j} 
-2i\gamma_{4}c^{sj}R^{i}_{s}\ddddot{A}_{j}.
\end{equation} 
The inhomogeneous part of the law for the transformation (\ref{tra}) with $R={\bf 1}$  takes on the following form
\begin{displaymath}
d^{ik}\dot{A}_{k}-2i\gamma_{4}c^{jk}\ddddot{A}_{k},
\end{displaymath}
where $d^{ik}$ is defined as follows
\begin{displaymath}
d^{ij} \equiv d\delta^{ij} - 2i\gamma_{1}c^{ij} -2i\gamma_{3}c^{ik}\partial_{k}\partial_{p}\phi\delta^{pj}.
\end{displaymath}
Now, we make use of the transformations such as before with $A(t)=\lambda (t-t_{o})^{3}$. Inserting the 
transformation to (\ref{trf'}) we get at $X_{o}$ the equation (\ref{rb1}) but with $\widetilde{f}^{k}$ instead of $b^{k}$. 
 In the same way, then, as for $b^{k}$ we infer that $\widetilde{f}^{k}$ does not depend on 
$\partial_{t}\partial_{j}\phi$. Applying the same transformation to (\ref{trf'}) we get (\ref{rb3}) for $\widetilde{f}^{k}$ 
(instead of $b^{k}$). Concluding: $\widetilde{f}^{k}=
\widetilde{f}^{k}(\phi, \partial_{i}\phi, \partial_{i}\partial_{j}\phi, \partial_{t}^{2}\phi)$.  

Denote the values of $d^{ij}(X)$ and $c^{ij}(X)$ at $X_{o}$ by  $d_{o}^{ij} $ and $c_{o}^{ij}$. Note, that if 
$c_{o}^{ij}=0$, the analysis for $f^{k}$ reduces to the case such as with $b^{k}$ and $f^{k}=0$. So, we
assume that $c_{o}^{ij} \ne 0$. Suppose that $d_{o}^{ij}$ is degenerate and possesses a zero direction 
$m^{k} \ne 0$: $d_{o}^{ij}m_{i}=0$. There exist a vector $v^{k}$ for which 
\begin{displaymath}
\sigma \equiv m^{i}m^{j}\partial_{i}\partial_{j}\phi(X_{o})+2m^{i}v_{i}
\end{displaymath}
has non zero value, which can be negative or positive. Denote the value of 
\begin{displaymath}
-2m^{i}\partial_{i}\partial_{t}\phi(X_{o})
\end{displaymath}   
by $\omega$. Consider the transformation (\ref{tra}) with $R={\bf 1}$ and $\vec{A}(t)$ such that $\dot{\vec{A}}(t_{o})=
\lambda \vec{m}, \, \ddot{\vec{A}}(t_{o})=0, \, \dddot{\vec{A}}(t_{o})=\lambda \vec{v}, \, \ddddot{\vec{A}}(t_{o})=0$, and
apply it to (\ref{trf'}) obtaing at $X_{o}$
\begin{displaymath}
\widetilde{f}^{k}(\phi, \partial_{i}\phi, \partial_{i}\partial_{j}\phi, \partial_{t}^{2}\phi +\omega \lambda + 
\sigma \lambda^{2})=
\end{displaymath}

\vspace{-0.7cm}
 
\begin{displaymath}
\widetilde{f}^{k}(\phi, \partial_{i}\phi, \partial_{i}\partial_{j}\phi, \partial_{t}^{2}\phi),
\end{displaymath}
for all $\lambda$. So, in the case when $\det{(d_{o}^{ij})}=0$ $\widetilde{f}^{k}$ does not depend on 
$\partial_{t}^{2}\phi$.

Suppose now that $\det{(d_{o}^{ij})} \ne 0$. The same analysis, but with $m^{i}$ such that $d_{o}^{ij}m_{i}$ is
perpendicular to the (for example) first  axis of the coordinate system shows that $\widetilde{f}^{1}$ does not
depend on $\partial_{t}^{2}\phi$. On the same footing, this is also true for the remaining coordinates of 
$\widetilde{f}^{k}$. 

In the further analysis identical as for $b^{k}$ we show that $\widetilde{f}^{k}=0$, or equivalently
\begin{displaymath}
f^{k}=-2i\gamma_{2}c^{ij}\partial_{j}\phi-2i\gamma_{3}c^{ij}\partial_{j}\partial_{t}\phi.
\end{displaymath}
But this is possible only if $\gamma_{2}=\gamma_{3}=0$ or equivalently, only if $f^{k}=0$. Indeed,
applying the transformation laws for $\partial_{i}\phi$ and $\partial_{i}\partial_{t}\phi$ to the above 
formula one gets the transformation law for $f^{k}$
\begin{displaymath}
{f'}^{i}(X')=R^{i}_{s}f^{s}(X)-2i\gamma_{3}{R^{-1}}^{i}_{s}{R^{-1}}^{q}_{p}c^{sk}\partial_{k}\partial_{q}\phi\dot{A}^{p}-
\end{displaymath}

\vspace{-0.5cm}

\begin{displaymath}
-2i\gamma_{2}R^{i}_{s}c^{sk}\ddot{A}_{k}-2i\gamma_{3}R^{i}_{s}c^{sk}\dddot{A}_{k}.
\end{displaymath}
Comparing it with (\ref{22}) at $X_{o}$ for the transformation (\ref{tra}) with $R \ne {\bf 1}$
and $A^{i}(t)=(t-t_{o})^{2}n^{i}$ such that $v^{j}\equiv c_{o}^{ij}n_{i} \ne 0$ (this is possible because $c_{o}^{ij} \ne 0$)
one gets
\begin{displaymath}
\gamma_{2}R^{j}_{i}v^{i}=\gamma_{2}v^{j},
\end{displaymath}
for all orthogonal $R$ and $\vec{v} \ne 0$, which means that $\gamma_{2}=0$. In the similar way, but with
$A^{i}=(t-t_{o})^{3}n^{i}$, one shows that $\gamma_{3}=0$. Summing up $f^{k}=0$. Now, looking back to the 
transformation law (\ref{22}) we realize that 
\begin{displaymath}
2ic^{ij}\partial_{j}\theta=-d\dot{A}^{i}
\end{displaymath}
for all polynomial $\vec{A}(t)$. Comparing the above formula with (\ref{23}) one can see that 
\begin{equation}\label{theta}
\partial_{j}\theta=-\gamma_{1}\dot{A}_{j},
\end{equation}

\vspace{-0.5cm}

\begin{displaymath}
c^{ij}=c\delta^{ij},
\end{displaymath}
where 
\begin{displaymath}
c \equiv \frac{d}{2i\gamma_{1}}
\end{displaymath}
is a scalar field: $c'(X')=c(X)$ (which follows from the fact that $c^{ij}$ is a tensor field, 
compare (\ref{trc}) and recall that $b^{k}=0$ as well as $a=0$, or immediatelly from the fact that $d$ is a scalar,
which is a consequence of the covariance of the wave equation). Note that $\gamma_{1}$ is the inertial
mass of the particle in question and by this $\gamma_{1} \ne 0$, see the next section.

\subsection{$g,\, d$}

We have simplified our equation (\ref{eq}) to the following form 
\begin{displaymath}
\Big[\frac{k}{2\gamma_{1}}\delta^{ij}\partial_{i}\partial_{j}+ik\partial_{t}+g \Big]\psi=0,
\end{displaymath}
where we introduce $ik \equiv d$.
The covariance condition of the equation gives the following transformation law of $g$
\begin{displaymath}
g'(X')=g(X)-\frac{k\gamma_{1}}{2}\dot{A}_{i}\dot{A}^{i}+k\gamma_{1}\dot{A}_{i}\dot{A}^{i}-
\end{displaymath}

\vspace{-0.5cm}

\begin{displaymath}
-k\partial_{t}\widetilde{\theta}(A^{k},t)-k\gamma_{1}\ddot{A}_{i}x^{i}.
\end{displaymath}

Let us define a new object 
\begin{displaymath}
\Lambda(X)=g(X)+\gamma_{1}k(X)\phi(X).
\end{displaymath}
It is clear that the transformation law of $\Lambda$ is as follows
\begin{displaymath}
\Lambda'(X') = \Lambda(X) + \frac{k\gamma_{1}}{2} \dot{A}_{i}\dot{A}^{i} - k\partial_{t}\widetilde{\theta}(A^{k}, t).
\end{displaymath}
Both $\widetilde{\theta}$ and $\Lambda$ taken separately are not uniquely defined. This is because the potential
$\phi$ is determined up to a time dependent additive term (the gauge freedom). 
So, one can assume any form for $\widetilde{\theta}$ by an appropriate (gauge) redefinition of $\phi$ 
(changing the first one by $G(t)$ and the second one by
$(1/\gamma_{1})\dot{G}(t)$). This is the place at which we need the gauge freedom $F(t) = (1/\gamma_{1}) \dot{G}$, 
see the Remark at the end of the subsection {\bf IV.A}. 
 Assume then, that $\widetilde{\theta}$ is chosen in such a way that 
\begin{displaymath}
\partial_{t}\widetilde{\theta}=\frac{\gamma_{1}}{2}\dot{A}_{i}\dot{A}^{i}.
\end{displaymath} 
After this the above transformation law for $\Lambda$ takes on the following form
\begin{displaymath}
\Lambda'(X')=\Lambda(X)
\end{displaymath}
and $\Lambda$ is a scalar field. In the identical way as for $b^{k}$ we
show that 
\begin{displaymath}
\Lambda=\Lambda(\partial_{i}\partial_{j}\phi).
\end{displaymath}

Now, we come back to the equation and easily show that it can be covariant if and only if $k$ is a constant.
We get, then, the wave equation 
\begin{equation}\label{eqm}
\Big[\frac{k}{2\gamma_{1}}\delta^{ij}\partial_{i}\partial_{j}+ik\partial_{t}-k\gamma_{1}\phi+\Lambda \Big]\psi,
\end{equation}
valid for each \emph{galilean} system at each point $X$ whose space coordinates are not equal to zero.
But we may uniquely extend the equation and it has the form (\ref{eqm}) for all spacetime points.  
If the gravitational field goes to zero, then the equation takes on the standard form (or has to be equivalent to 
the standard equation). But this is possible if and only if there exist a non zero number such that if one 
multiplies the equation (\ref{eqm}) by this number one gets: $k=\hslash$ and $k/2\gamma_{1}=\hslash^{2}/2m$, 
where $m$ is the inertial mass. So, we get
\begin{displaymath}
m \equiv k\gamma_{1}=m_{g},
\end{displaymath}
where $m_{g}$ is the gravitational mass. We have just interpreted the parameter at the gravitational potential as
the gravitational mass $m_{g}$. 

Concluding, the inertial and gravitational masses are equal and $\psi$ fulfils the equation
\begin{equation}\label{ec1}
\Big[\frac{\hslash^{2}}{2m}\delta^{ij}\partial_{i}\partial_{j}+i\hslash\partial_{t}-m\phi
+\Lambda(\partial_{a}\partial_{b}\phi) \Big] \psi=0,
\end{equation}
with the transformation law of $\psi$ given by the representation $T_{r}$ of the form (\ref{trapsi}) and
the $\theta$ in $T_{r}$ given by
\begin{equation}\label{ec2}
\theta=\frac{m}{2\hslash} \int_{0}^{t} \dot{\vec{A}}^{2}(\tau) \, {\ud{\tau}} +\frac{m}{\hslash}\dot{A}_{i}x^{i}.
\end{equation}  
Note that we have used the classification theory of $\theta$-s, presented in the next section and in 
the Appendix, to derive the formulas (\ref{ec1}) and (\ref{ec2}). The assumption (C) is too weak to get 
the Eqs (\ref{ec1}) and (\ref{ec2}). The covariance condition in the Quantum Mechanics is so strong 
because it imposes the conditions on the representation $T_{r}$ independently of the covariance 
condition of the wave equation. In the next section we will find the covariance conditions for $T_{r}$.

As we have shown in \cite{AHJW} the Eqs (\ref{ec1}) and (\ref{ec2}) with the scalar term $\Lambda = 0$
result uniquely from the Einstein equivalence principle applied to the local wave equation.
Of course, beside the state vector evolution law given by the wave equation, there are other
independent structures of the theory. The Hilbert space and the transition probabilities are
not contained in the Schr\"odinger equation. Moreover, the state vector reduction process
during the measurement is a nonlocal process \cite{EPR}. So, the status of such a local
principle as the Einstein equivalence principle is not clear in the Quantum Mechanics. 
By the notion of the Einstein equivalence principle in the Quantum 
Mechanics we always mean implicitly the application of this principle to the local wave equation.

There were done the experiments checking the equality $m_{i} = m_{g}$ for quantum objects.
The earliest one is the so called COW experiment \cite{COW} for the neutron.
The most precise experiment was done by M. Kasevich and S. Chu \cite{Kasevich},
for the sodium atom in which the E\"otv\"os parameter was estimated to be $\leq 10^{-6}$.
The recent experiment \cite{Nesvi} in which the bound states are formed by the gravitational
potential is a qualitative one.

\section{Generalization of Bargmann's Theory. Classification of $\theta$-s}

\subsection{Preliminary Remarks}

In this subsection we carry out the general analysis of the representation $T_{r}$ of a covariance
group and compare it with the representation of a symmetry group. 
We describe also the correspondence between the space of wave functions $\psi(\vec{x},t)$ 
and the Hilbert space. Before we give the general description, it will be instructive to ivestigate the problem 
for the free particle in the flat Galilean spacetime. 

The set of solutions $\psi$ of the Schr\"odinger equation which are admissible in Quantum Mechanics is 
precisely given by 
\begin{displaymath}
\psi(\vec{x},t)=(2\pi)^{-3/2} \int \varphi(\vec{k})e^{-i\frac{t}{2m}\vec{k}\centerdot \vec{k}+i\vec{k}\centerdot\vec{x}}
\, {\ud}^{3}{k},
\end{displaymath}
where $p=\hslash k$ is the linear momentum and $\varphi(\vec{k})$ is any square integrable function. The 
functions $\varphi$ (wave functions in the "Heisenberg picture") form a Hilbert space 
$\mathcal{H}$ with the inner product
\begin{displaymath}
(\varphi_{1}, \varphi_{2})=\int \varphi_{1}^{*}(\vec{k})\varphi_{2}(\vec{k}) \, {\ud}^{3}{k}.
\end{displaymath}
The correspondence between $\psi$ and $\varphi$ is one-to-one. The above construction fails in the curved 
Newton-Cartan spacetime, because in this spacetime we do not have plane wave, see \cite{Waw}. So, there
does not exist any natural counterpart of the Fourier transform. However, we need not to use the 
Fourier transform. What is the role of the Schr\"odinger equation in the above construction of $\mathcal{H}$?
Note, that 
\begin{displaymath}
\Vert \psi \Vert^{2} \equiv
\int \psi^{*}(\vec{x},0)\psi(\vec{x},0) \, {\ud}^{3}{x} = (\varphi,\varphi) = 
\end{displaymath}

\vspace{-0.5cm}

\begin{displaymath}
=\int \psi^{*}(\vec{x},t)\psi(\vec{x},t) \, {\ud}^{3}{x}.
\end{displaymath}
This is in accordance with the Born interpretation of $\psi$. Namely, if $\psi^{*}\psi(\vec{x},t)$ is the
probability density, then
\begin{displaymath}
\int \psi^{*}\psi \, {\ud}^{3}{x}
\end{displaymath}
has to be preserved in time. In the above construction the Hilbert space $\mathcal{H}$ is isomorfic to the 
space of square integrable functions $\varphi(\vec{x})\equiv \psi(\vec{x},0)$ -- the set of square integrable
space of initial data for the Schr\"odinger equation, see e.g. \cite{Giulini}. The connection between $\psi$ and
$\varphi$ is given by the time evolution $U(0,t)$ operator (by the Schr\"odinger equation):
\begin{displaymath}
U(0,t)\varphi=\psi.
\end{displaymath}
The correspondence between $\varphi$ and $\psi$ has all formal properties such as in the above Fourier 
construction. Of course, the initial data for the Schr\"odinger equation do not cover the whole Hilbert space 
$\mathcal{H}$ of square integrable functions, but the time evolution given by the Schr\"odinger equation can be 
uniquely extended on the whole Hilbert space $\mathcal{H}$ by the unitary evolution operator $U$. 

The construction can be applied to the particle in the Newton-Cartan spacetime. As we implicitly assumed in (B) and (C), 
the wave equation is such that the set of its admissible initial data is dense in the space of square integrable functions
(we need it for the uniquenees of the extension). Because of the Born interpretation the integral
\begin{displaymath}
\int \psi^{*}\psi \, {\ud}^{3}{x}
\end{displaymath}
has to be preserved in time. Denote the space of the initial square integrable data $\varphi$ on the hyperplane
$t(X)=t$ by ${\mathcal{H}}_{t}$. The evolution is, then, an isometry between ${\mathcal{H}}_{0}$ and 
${\mathcal{H}}_{t}$. But such an isometry has to be a unitary operator, and the construction is well defined,
\emph{i.e.} the inner product of two states corresponding to the wave functions $\psi_{1}$ and $\psi_{2}$
does not depend on the choice of ${\mathcal{H}}_{t}$. 
Let us mention, that the assumptions (B) and (C) are not independent, that is, the wave equation has 
to be linear in accordance with the Born interpretation of $\psi$ (any unitary operator is linear,
so, the time evolution operator is linear). The space of wave functions $\psi(\vec{x},t) = U(0,t)\varphi(\vec{x})$
isomorphic to the Hilbert space ${\mathcal{H}}_{0}$ of $\varphi$ is called in the common "jargon" the 
"Schr\"odinger picture".  

However, the connection between $\varphi(\vec{x})$ and $\psi(\vec{x},t)$ is not unique in general,
if the wave equation possesses a gauge freedom. Namely, consider 
the two states $\varphi_{1}$ and $\varphi_{2}$ and ask the question: when the two states are equivalent and 
by this indistinguishable? The answer is as follows: they are equivalent if 
\begin{displaymath}
\vert(\varphi_{1},\varphi)\vert \equiv \Big\vert\int \psi_{1}^{*}(\vec{x},t)\psi(\vec{x},t) \, {\ud}^{3}{x}\Big\vert =
\vert(\varphi_{2},\varphi)\vert \equiv
\end{displaymath}

\vspace{-0.5cm}

\begin{equation}\label{row}
\equiv \Big\vert\int \psi_{2}^{*}(\vec{x},t)\psi(\vec{x},t) \, {\ud}^{3}{x}\Big\vert,
\end{equation}
for any state $\varphi$ from $\mathcal{H}$, or for all $\psi=U\varphi$ ($\psi_{i}$ are defined to be = 
$U(0,t)\varphi_{i}$). Substituting $\varphi_{1}$ and then 
$\varphi_{2}$ for $\varphi$ and making use of the Schwarz's inequality one gets: $\varphi_{2}=e^{i\alpha}\varphi_{1}$,
where $\alpha$ is any constant \cite{Weyl}. The situation for $\psi_{1}$ and $\psi_{2}$ is however different. 
In general the condition (\ref{row}) is fulfilled if 
\begin{displaymath}
\psi_{2}=e^{i\xi(t)}\psi_{1}
\end{displaymath}
and the phase factor can depend on time. Of course it has to be consistent with the wave equation, that is, together
with a solution $\psi$ to the wave equation the wave function $e^{i\xi(t)}\psi$ also is a solution
to the appropriately gauged wave equation. \emph{A priori}
one can not exclude the existence of such a consistent time evolution. This is not a new observation, it was 
noticed by John von Neumann \cite{Neumann}, but it seems that it has never been deeply investigated
(probably because the ordinary nonrelativistic Schr\"odinger equation has a gauge symmetry 
with constant $\xi$). Note, that $e^{i\xi(t)}\psi$ has to be a solution to 
the \emph{gauged} wave equation. The equation has to be gauged 
together with $\psi$. Indeed, suppose that both $\psi_{1}$ and $\psi_{2}=e^{i\xi(t)}\psi_{1}$ are the solutions of the 
identical wave equation and by this both $\psi_{1}$ and $\psi_{2}$ belong to the same "copy" of
the "Schr\"odinger picture". Then, the time evolution induced by the wave equation would not be 
any unitary operator between ${\mathcal{H}}_{0}$ and ${\mathcal{H}}_{t}$ because 
\begin{displaymath}
\int \psi_{1}^{*}\psi_{2} \, {\ud}^{3}{x} =e^{i\xi(t)} \Vert \psi_{1} \Vert^{2} = e^{i\xi(t)}const.
\end{displaymath}
would be time dependent. This is the main difference in comparison to the space of $\varphi$. The two 
states $\psi_{1}$ and $\psi_{2}$ belong then to two (equivalent) "gauge copies" of the "Schr\"odinger
picture", the "gauge copies" having no elements in common beside the trivial one: $\psi=0$. 
Note, that such a "Schr\"odinger picture" corresponds to a fixed observer (reference frame) 
and in general the two "pictures" corresponding to two different observers are different in general.    
It will be useful, however, to consider an enlarged linear space $\mathfrak{S}$ consisting of all  the above
"Schr\"odinger pictures" corresponding to all (in our case \emph{galilean}) observers and all "gauge copies" of them. 
That is, $\mathfrak{S}$ is the smallest linear space 
which together with any element $\psi$ of any "Schr\"odinger picture" contains $e^{i\xi(t)}\psi'$ for any differentiable 
$\xi(t)$ and any transformation $\psi' = T_{r}\psi$ of $\psi$ to any (\emph{galilean}) reference frame, $r$ being any 
element of the group $G$ in question (the Milne group in our case). In other words, if we fix a "Schr\"odinger picture" and denote it by ${\mathcal{H}}_{0}$,
then $\mathfrak{S}$ is the smallest linear space which contains the set of elements $e^{i\xi(t)}T_{r}\psi$,
where $r$ is any element of the group $G$, $\xi(t)$ is any real differentiable function, and 
$\psi \in {\mathcal{H}}_{0}$. So, $\mathfrak{S}$ is the smallest linear space containing a "Schr\"odinger picture"
on the whole, in which the representation $T_{r}$ acts.   

But in the space $\mathfrak{S}$ the integral 
\begin{displaymath}
(\psi_{1},\psi_{2})_{t} \equiv \int \psi_{1}^{*}\psi_{2} \, {\ud}^{3}{x}
\end{displaymath} 
is time dependent, which will be indicated by the subscript $t$. Even the modulus 
$\vert(\psi_{1},\psi_{2})_{t}\vert$ of this expression is time dependent in
general. The expression $\vert (\psi, \chi)_{t}\vert$ does not depend on the time $t$ if 
$\psi$ belongs to a "copy" of the "Schr\"odinger picture" which corresponds to an
observer and $\chi$ belonging to a (possibily
different) "gauge copy" but corresponding exactly to the same observer. 
Indeed, we have $\chi=e^{i\tau(t)}\psi_{1}$ for some function $\tau(t)$ and $\psi_{1}$ belonging
to the same "copy" of the "Schr\"odinger picture" as the wave function $\psi$
(the "copies" both are connected with a fixed observer). So, we have
\begin{equation}\label{innt}
(\psi,\chi)_{t} \equiv  e^{i\tau} \int \psi^{*}\psi_{1} \, {\ud}^{3}{x} = 
e^{i\tau (t)}\vert (\varphi,\varphi_{1}) \vert
\end{equation}
for $U(0,t)\varphi=\psi$ and $U(0,t)\varphi_{1} = \psi_{1}$. 
The modulus $\vert (\psi, \chi)_{t} \vert = \vert (\varphi_{1}, \varphi_{2}) \vert $ 
of the expression is constant then. In the remaining situations, however, even the modulus does depend 
on the time $t$. Namely, consider three waves $\psi, \psi_{1}$ and $\psi_{2}$ belonging to the same 
"copy" of the "Schr\"odinger picture" and the wave function $\chi = \psi_{1}+ e^{i\sigma(t)}\psi_{2}$. Then,
$\chi$ cannot belong to any "Schr\"odinger picture" (see the above remarks) and moreover 
$\vert (\psi, \chi)_{t} \vert = \vert (\varphi, \varphi_{1}) + e^{i\sigma(t)}(\varphi, \varphi_{2}) \vert$ has to be
time dependent in general where $\varphi, \varphi_{i}$ are defined as above. 
The same is true in general for $\psi(X)$ and $\chi(X)=\psi'(X)$ equal to a transform
of $\psi$ to a different frame.   

Note also, that only the wave functions $\psi=e^{i\xi(t)}\psi_{1}$ with $\psi_{1}$ belonging to a "Schr\"odinger picture"
correspond to the physical states (\emph{dynamically possible trajectories dpt}, compare section {\bf II}), 
the remaining being unphysical. Such sets $\boldsymbol{\psi}=\{\sigma(t) \psi_{1}, \vert \sigma \vert =1\}
\subset \mathfrak{S}$ of wave functions will be called sets of \emph{rays},
more precisely, we confine ourselves to the \emph{unit rays}, \emph{i.e} with $\Vert \psi \Vert =1$ (this is meaningful
definition because this time the norm is constant), and differentiable $\sigma(t)$, which is natural according to the 
fact that $\psi$ fulfils a differential equation -- the wave equation. Any $\psi \in \boldsymbol{\psi}$ will be called
a \emph{representative} of $\boldsymbol{\psi}$. More precisely, we assume $\sigma(t)$ to be differentiable
up to any order, see the further discussion for the justification (we will always use the word 'diffrentiable' in this sense
in this paper, unless the order of differentiability is specified).       

Let $G$ be a group consistent with the simultaneity structure of the Galilean (or the Newton-Cartan spacetime), 
\emph{i.e.} any transformation $r$ of the group acts on the time coordinate $t \mapsto rt$ in such a way that $rt$ 
is again a function of the time $t$ only. Consider the two cases: $G$ is an invariance (symmetry) group and the 
second one when $G$ is a covariance group. 

First, let $G$ be an invariance group. As we know (see section {\bf II}, 
the comment to the Definitions 1 $\div$ 4) one can always eliminate
the absolute elements $y_{A}$.  As is well known -- in the Quantum Mechanical 
description --  the group $G$ (or its ray representation $U_{r}$) acts in the Hilbert 
space $\mathcal{H}$. Then, $U_{r}U_{s}\varphi$ has to be equivalent to $U_{rs}\varphi$ and we have
\begin{displaymath}
U_{r}U_{s}\varphi=e^{i\alpha(r,s)}U_{rs}\varphi
\end{displaymath}
with $\alpha$ depending on $r$ and $s$ only. Consider the action of $G$ in the space $\mathfrak{S}$ of wave functions
$\psi$. The representation $U_{r}$ induces a representation $T_{r}$ acting in the space $\mathfrak{S}$
of wave functions $\psi$. But this time $T_{r}$ acts in the space of solutions to the wave equation and $T_{r}\psi$ 
and $\psi$ belongs to the same "copy" of the Schr\"odinger picture".  
This is because we have eliminated the absolute elements $y_{A}$, so that the symmetry group $G$ 
becomes identical to the covariance group and the group $G$ acts in the space of solutions to the dynamical equations
(see the section {\bf II}, the comment to the Definitions 1 $\div$ 4). So, $T_{r}T_{s}\psi$ and $T_{rs}\psi$ belong 
to the same "Schr\"odinger picture" and can differ by a constant phase factor $\alpha$ at most:
\begin{displaymath}
T_{r}T_{s}\psi = e^{i\alpha(r,s)}T_{rs}\psi,
\end{displaymath}
which should be the same as in the Hilbert space $\mathcal{H}$ of states $\varphi$. So, we have a 

\vspace{1ex}

THEOREM 1. \emph{If $G$ is a symmetry group, then the phase factor $\alpha$ should be time independent 
(or equivalent to a time independent one)}.

\vspace{1ex}

Second, let $G$ be a covariance group. This time $G$ does not act in the Hilbert space of the system in 
question. Namely, $\psi_{l}$  and its transform $T(l,l')\psi_{l} = \psi_{l'}$ under $r \in G, \, r: l \mapsto l' =rl$, 
belong to the two different  "Schr\"odinger pictures" corresponding to the two different  observers  
in the two reference frames $l$ and $l' =rl$. 
So, $T_{r} = T(l,l') = T(l,rl)$ does not act in a fixed 
"Schr\"odinger picture" but in a space of $\psi_{l}$ for all \emph{galilean}
observers $l$ \cite{Wigner}. However, because in general the gauge freedom cannot  \emph{a priori} be excluded the 
representation $T_{r}$ of $G$ acts in the space $\mathfrak{S}$. But after this, the equivalence of 
$T_{r}T_{s}\psi$ and $T_{rs}\psi$ means that (we omit the subscript $l$ at $\psi_{l}$) 
\begin{equation}\label{rowt}
T_{r}T_{s}\psi=e^{i\xi(r,s,t)}T_{rs}\psi
\end{equation}
and $\xi$ depends on $r$, $s$ and in addition on the time $t$. It has to be consistent with the wave equation. 
Namely, consider a transformation $g = rs$ between the two reference frames $l$ and $l'$, $rs: l \mapsto l'$, and its 
action on the wave equation (and on the wave function). It can be realized in a different way, that is, in the two steps
$r: l\mapsto l"$ and $s: l" \mapsto l'$. Then, the result should differ by an appropriate gauge transformation in such
a way that if $\psi_{1}$ is a solution to the first equation then the wave function $e^{-i\xi(r,s,t)}\psi_{1}$ is a solution
to the gauged equation obtained as the second result, by transforming the equation in the two steps. Again,
one can not \emph{a priori} exclude the existence of such a consistent wave equation. Note that because the wave 
equation is a differential equation (in our case possibly of second order in accordance with the assumption (C))
then, the exponent $\xi(r,s,t)$ in (\ref{rowt}) has to be a differentiable function of $t$ (in our case up to the second
order). But we assume in the sequel, that $\xi(r,s,t)$ in (\ref{rowt}) is differentiable in $t$ up to any order. 
This assumption is of a technical character only. The second order continuous differentiability of $\xi$ with respect to
the time $t$ is sufficient for us. We use this assumption for the sake of simplicity. 

Note that $T_{r}$ being a representation of a covariance group transforms \emph{rays} into \emph{rays}
(or \emph{dpt}-s into \emph{dpt}-s in accordance with our general definition, see section {\bf II}). In
other words a physical ("real") state observed by an observer cannot be seen as an unphysical ("unreal")
one by any other observer, it would be a nonsense to allow such a situation.  

A natural question arises then: why the phase factor $e^{i\xi}$ in (\ref{rowt})
 is time independent for the Galilean group even when
the Galilean group is considered as a covariance group? This is the case for the Schr\"odinger equation
with a potential which does not possess the Galilean group as a symmetry group -- as in the hydrogen
atom, for example. The explanation of the paradox is as follows. 
The Galilean covariance group $G$ induces the representation
$T_{r}$ in the space $\mathfrak{S}$ of wave functions fulfilling (\ref{rowt}). But, as we will show later on, 
the structure of $G$ is such that there always exists a function $\zeta(r,t)$ continuous in $r$ and 
differentiable in $t$ with the help of which one can define a new equivalent representation 
$T'_{r}=e^{i\zeta(r,t)}T_{r}$ fulfilling
\begin{displaymath}
T'_{r}T'_{s}=e^{i\alpha(r,s)}T'_{rs} 
\end{displaymath}
with a time independent $\alpha$. The representations $T_{r}$ and $T'_{r}$ are equivalent 
because $T'_{r}\psi$ and $T_{r}\psi$ are equivalent for all $r$ and $\psi$. 

Of course, the exponent $\xi$ has to be constant if the group $G$ in question is a symmetry group as follows
from the above argumentation. In the case of the Galilean group one can show it in a completely different way. 
Indeed, the Schr\"odinger equation possesses a gauge freedom
$\psi\mapsto e^{if(X)}\psi$ even in the flat Galilean spacetime, compare \cite{DK}, \cite{Waw} or section {\bf III}, 
and together with $\psi$ the appropriately gauged wave equation possesses the solution $e^{if(X)}\psi$. 
In particular, if it has a solution $\psi$, then $e^{if(t)}\psi$ is a solution to the appropriately gauged equation.
However, if one imposes the Galilean invariance on the wave equation 
(in addition to the covariance condition) then the gauge freedom is eliminated and the gauge function $f$ has to 
be reduced to a constant, see \cite{Waw}. 

Consider now two \emph{galilean} coordinate systems $l$ and $l'=rl$ in the 
Newton-Cartan spacetime, $r$ being the Milne transformation (\ref{tra}). 
Let $\psi_{l}$ and $\psi_{l'} = T(l,l')\psi_{l}$ be the
wave functions describing the same physical state in the two systems respectively. 
The covariance Milne group (\ref{tra}) induces a representation $T(l,l') =T(r)\equiv T_{r}$ 
which acts in the space $\mathfrak{S}$ of wave functions $\psi$. In general it fulfills
the condition (\ref{rowt}). As we will see there does not exist in this case any equivalent representation $T'_{r}$
for which $\xi$ is time independent. As is well known there exists an elegant theory which classifies all $\xi$-s
in (\ref{rowt}), providing that $\xi$ does not depend on the time $t$, given by Bargmann \cite{Bar}. Then, a natural
problem arises to generalise the theory to the case with time dependent $\xi$, which we do in the Appendix {\bf A}. 
As we will see the correspondence between equivalence classes of $\xi$-s in (\ref{rowt}) and 
$\theta$-s in (\ref{trapsi}) is biunique and by this
we are able to classify all $\theta$-s. In other words one can reproduce the most general transformation law of the 
wave function under the Milne transformations given by Eq. (\ref{tra}). Having given $\xi$ one also has the 
commutation rules for the representation $T_{r}$. The Bargmann's  theory (as well as its generalisation) 
is based on the fact that the exponents $\xi$ in ray representation are determined by the group associative law. Namely,
consider the transformation between the two reference frames $l$ and $l'$: $l\rightarrow l'$,
and its action on the wave function. It can be realized in a different way, 
\emph{i.e.} in the three steps: $l\mapsto l''\mapsto l''' \mapsto l'$. The three-step transformation
can be realized in the following two ways. First one applies $l \mapsto l'' \mapsto l'''$ and then $l''' \mapsto l'$, 
or first $l \mapsto l''$ is applied and then $l'' \mapsto l''' \mapsto l'$.  
The result should be identical. But not all exponents are compatible in this sense, and by this 
the above requirement is very strong. Two problems will arise here. The first is connected with the time dependence 
of the phase factor and the second follows from the fact that the Milne group (\ref{tra}) is not any 
classical Lie group (Bargmann's theory largely concerns classical Lie groups). 
But we need not the whole Milne group in our analysis. It is sufficient for us to consider a finite dimensional Lie 
subgroup of the Milne group -- the polynomial Milne transformations group with the
appropriate  polynomial degree, compare subsection {\bf IV}. 
However, we construct the most general transformation (\ref{trapsi}) for the Milne group,
to make the analysis more complete. Namely, we built a sequence of Lie groups $G(1)\subset G(2) \subset 
\ldots \subset G(m)\subset \ldots$ which is dense in the group $G$ of Milne transformations  (the 
former $G(m-1)$ being a subgroup of the later one $G(m)$). Then we extend $\theta$ from $G(m)$ on the whole 
group $G$, see subsection {\bf V.C}. 

Before we proceed further on we make a general comment concerning the relation (\ref{rowt}). There is a physical 
motivation to investigate representations $T_{r}$ fulfilling (\ref{rowt}) with $\xi$ depending on all 
spacetime coordinates $X$:
\begin{equation}\label{rowX}
T_{r}T_{s}=e^{i\xi(r,s,X)}T_{rs}.
\end{equation}
Namely, in the Quanum Field Theory the spacetime coordinates $X$ play the role of parameters such 
as the time plays in the nonrelativistic theory (recall that, for example, the wave functions
$\Psi$ of the Fock space of the quantum electromagnetic field are functions of the Fourier components of the field,
the spacetime coordinates playing the role of parameters like the time $t$ in the nonrelativistic Quantum Mechanics). 
However, we are not interested here in it, and will present such an application elsewhere. There is another less 
interesting motivation for (\ref{rowX}), in the nonrelativistic context presented here. Suppose the assumption (B) 
is weakened, in such a way that the transition probabilities are given by integrals over the hyperplanes 
$h(X)=constant$ from a family covering the whole spacetime
\begin{displaymath}
\vert(\psi_{1},\psi_{2})\vert^{2} 
= \Big\vert \int_{H(X)=h} \psi_{1}^{*}\psi_{2}(h,y^{k}) \, \sigma_{h} (y^{k}) {\ud}^{3}{y}\Big\vert^{2},
\end{displaymath}
where $\sigma_{h}(y^{k})$ is the density of the measure induced on the hyperplane $H(X)=h$
by the spacetime mesure $\mu$ (note that we can show that $T_{r}$ has the form (\ref{trapsi})
in the same way as at the begining of {\bf II}). 
According to the new definition of transition probability the conception of equivalence is changed:
two waves being equivalent if they differ by a phase factor constant along the hyperplanes $H(X)=const.$,
\emph{i.e.}
\begin{displaymath}
T_{r}T_{s}=e^{i\xi(r,s,H(X))}T_{rs}.
\end{displaymath}
But it can be shown that any representation $T_{r}$ of the group of transformations (\ref{tra}) fulfilling (\ref{rowX})
is equivalent to a representation $T'_{r}$ for which $\xi = \xi(t)$, and the above weakened assumption
simplifies to the assumption (B).   

In most proofs there is no essential difference between the time dependent $\xi$ in (\ref{rowt}) and the
spacetime dependent $\xi$ in (\ref{rowX}). However, we are most interested here in (\ref{rowt}) and confine
ourselves to this case, but we mark the place at which important difference arises between 
the two cases.

\subsection{Galilean group as a covariance group}

As was mentioned in the subsection {\bf V.A}  the wave equation 
possessses the gauge freedom $\psi \to e^{if(X)}\psi$ even in the flat Galilean spacetime. 
So, in the situation when the Galilean group $G$ is only a covariance group and 
cannot be considered as a symmetry group (as for the electron wave function 
in the hydrogen atom) one should \emph{a priori} investigate such representations
$G$ which fulfil the Eq. (\ref{rowt}), with $\xi$ depending on the time. 
The following paradox, then, arises. Why the transformation law $T_{r}$ under 
the Galilean group has time-independent $\xi$ in (\ref{rowt}) independently of 
the fact if it is a covariance group or a symmetry group? We will solve the 
paradox in this subsection. Namely, we will show that any representation
of the Galilean group fulfilling (\ref{rowt}) is equivalent to a representation
fulfilling (\ref{rowt}) with time-independent $\xi$. This is a rather peculiar property
of the Galilean group not valid in general. For example, this is not true for
the group of transformations (\ref{tra}), and this fact will be used in section {\bf VI}.

According to {\bf V.A} and the Appendix {\bf A} we shall determine all equivalence 
classes of infinitesimal exponents $\Xi$ of the Lie algebra $\mathfrak{G}$ of $G$
to classify all $\xi$ of $G$. The commutation relations for the Galilean group are as follows
\begin{equation}\label{26a}
[a_{ij},a_{kl}] = \delta_{jk}a_{il} - \delta_{ik}a_{jl}+\delta_{il}a_{jk}-\delta_{jl}a_{ik}, 
\end{equation}

\vspace{-0.5cm}

\begin{equation}\label{26b}
[a_{ij},b_{k}] = \delta_{jk}b_{i} - \delta_{ik}b_{j}, \, [b_{i},b_{j}] = 0,
\end{equation}

\vspace{-0.5cm}

\begin{equation}\label{26c}
[a_{ij.d_{k}}] = \delta_{jk}d_{i} - \delta_{ik}d_{j}, \, [d_{i},d_{j}] = 0, [b_{i},d_{j}] = 0,
\end{equation}

\vspace{-0.5cm}

\begin{equation}\label{26d}
[a_{ij},\tau] = 0, [b_{k},\tau] = 0, [d_{k},\tau] = b_{k},
\end{equation}
where $b_{i},d_{i}$ and $\tau$ stand for the generators of space translations, the proper Galilean 
transformations and time translation respectively and $a_{ij} = - a_{ji}$ are rotation generators. 
Note, that the Jacobi identity (\ref{Jac2}) is identical to the Jacobi identity in the ordinary
Bargmann's Theory of time-independent exponents (see \cite{Bar}, Eqs (4.24) and (4.24a)). 
So, using (\ref{26a}) -- (\ref{26c}) we can proceed exactly after Bargmann (see \cite{Bar}, pages 39,40)
and show that any infinitesimal exponent 
defined on the subgroup generated by $b_{i}, d_{i}, a_{ij}$ is equivalent to an exponent
equal to zero with the possible exception of $\Xi(b_{i},d_{k},t) = \gamma \delta_{ik}$, where 
$\gamma = \gamma(t)$. So, the only components of $\Xi$ defined on the whole algebra $\mathfrak{G}$
which can a priori be not equal to zero are: $\Xi(b_{i},d_{k},t) = \gamma \delta_{ik}, \, \Xi(a_{ij}, \tau,t), \,
\Xi(b_{i}, \tau,t)$ and $\Xi(d_{k},\tau,t)$. We compute first the function $\gamma(t)$. Substituting
$a= \tau, \, a' = b_{i}, a'' = d_{k}$ to (\ref{Jac1}) we get ${\ud}\gamma/{\ud}t = 0$, so that $\gamma$
is a constant, we denote the constant value of $\gamma$ by $m$. Inserting $a = \tau, \, a' = a_{i}^{s},
\, a'' = a_{sj}$ to (\ref{Jac1}) and summing up with respect to $s$ we get $\Xi(a_{ij},\tau,t) = 0$. 
In the same way, but with the substitution $a = \tau, a' = a_{i}^{s}, a'' = b_{s}$, one shows that
$\Xi(b_{i},\tau, t) = 0$. At last the substitution $a=\tau, a' = a_{i}^{s}, a''=d_{s}$ to (\ref{Jac1})
and summation with respect to $s$ gives $\Xi(d_{i},\tau,t) = 0$. 
We have proved in this way that any time depending $\Xi$ on $\mathfrak{G}$ is equivalent 
to a time-independent one. In other words, we get a one-parameter family of possible $\Xi$, 
with the parameter equal to the inertial mass $m$ of the system in question. 
Any infinitesimal time-dependent exponent  of the Galilean group is equivalent
to the above time-independent exponent $\Xi$ with some value of the parameter $m$; and any two infinitesimal
exponents with different values of $m$ are inequivalent.   
As was argued in Appendix {\bf A} (Theorems 3 $\div$ 5) the classification of $\Xi$ gives the full classification of $\xi$.
Moreover, the classification of $\xi$ 
is equivalent to the classification of possible $\theta$-s in (\ref{trapsi}), see Appendix {\bf A}. On the other hand, the 
exponent $\xi(r,s,t)$ of (\ref{trapsi}) can be easily computed to be equal $\theta(rs,X) - \theta(r,X) - \theta(s,r^{-1}X)$,
and the infinitesimal exponent belonging to $\theta$ defined as 
$\theta(r,X) = -m\vec{v}\centerdot \vec{x} + \frac{m}{2}\vec{v}^{2}t$,
covers the whole one-parameter family of the classification (its infinitesimal exponent is equal to that 
infinitesimal exponent $\Xi$,
which has been found above). So, the standard $\theta(r,X) = - m\vec{v} \centerdot \vec{x} + \frac{m}{2}\vec{v}^{2}t$, 
covers the full classification of possible $\theta$-s in (\ref{trapsi}) for the Galilean group. 
Inserting the standard
form for $\theta$ we see that $\xi$ does not depend on $X$ but only on $r$ and $s$. By this, any time-depending
$\xi$ on $G$ is equivalent to a time-independent one. 

This result can be obtained in the other way. Namely, using now the Eq. (\ref{raycom})
we get the commutation relations for the \emph{ray} representation $T_{r}$ of the Galilean group 
\begin{displaymath}
[A_{ij},A_{kl}] = \delta_{jk}A_{il} - \delta_{ik}A_{jl} - \delta_{jl}A_{ik},  
\end{displaymath}

\vspace{-0.5cm}

\begin{displaymath}
[A_{ij},B_{k}] = \delta_{jk}B_{i} - \delta_{ik}B_{j}, \, [B_{i},B_{j}] = 0,
\end{displaymath}

\vspace{-0.5cm}

\begin{displaymath}
[A_{ij}, D_{k}] = \delta_{jk}D_{i} - \delta_{ik}D_{j}, 
\end{displaymath}

\vspace{-0.5cm}

\begin{displaymath}
[D_{i},D_{j}] =0, \, [B_{i},D_{j}] = m\delta_{ij},
\end{displaymath}

\vspace{-0.5cm}

\begin{displaymath}
[A_{ij}, T] = 0, \, [B_{k},T] = 0, \, [D_{k}, T] = B_{k},
\end{displaymath} 
where the generators $A_{ij}, \ldots $ which correspond to the generators $a_{ij}, \ldots$ of the one-parameter 
subgroups $r(\sigma) = \exp(\sigma a_{ij}), \ldots$ are defined in the following way \cite{Stone}
\begin{displaymath}
A_{ij}\psi(X) = \lim_{\sigma \to 0} \frac{(T_{r(\sigma)} - \boldsymbol{1})\psi(X)}{\sigma}.
\end{displaymath}
$A_{ij}$ is well defined for any differentiable $\psi(X)$.
So, we get the standard commutation relations such as in the case when the Galilean group is a symmetry group.
The above standard commutation relations for the transformation
$T_{r}$ of the form (\ref{trapsi}) gives a differential equations for $\theta$. It it easy to show, that they can be solved
uniquely (up to an irrelevant function $f(t)$ of time and the group parameters) and the solution has the standard form
$\theta(r,X) = - m\vec{v}\centerdot \vec{x} + f(t)$.

Note, that to any $\xi$ (or $\Xi$) there exists a coresponding $\theta$ (and by the result of Appendix {\bf A} 
such a $\theta$ is unique up to a trivial equivalence relation). As we will see this is not the case for the 
Milne group, where such $\Xi$ do exist which cannot be realized by any $\theta$.

\subsection{Milne group as a covariance group}

In this subsection we apply the theory of Appendix {\bf A} to the Milne group of transformations (\ref{tra}). We proceed 
like with the Galilean group in the preceding subsection {\bf V.B}. The Milne group $G$ does not form
any Lie group, which complicates the situation. We will go on according to the following plan. First, 1) we define
the topology in the Milne group. Second, 2) we define the sequence $G(1) \subset \ldots \subset G(m) \subset 
\ldots$ of subgroups of the Milne group $G$ dense in $G$. 3) Then we compute the infinitesimal exponents
and exponents for each $G(m)$, $m = 1,2, \ldots $, and by this the $\theta$ in (\ref{trapsi}) for $G(m)$.
4) As we have proved in Appendix {\bf A} the (strong) continuity of the exponent and $\theta(r,X)$ in the group variables
$r \in G$ follows as a consequence of the Theorem 2. By this, $\theta(r,X)$ defined for $r \in G(m), m = 1, 2, \ldots $
can be uniquely extended on the whole group $G$. As we will see, this can be done effectively thanks to
the assumption that the wave equation is local (assumption (C)), that is, the coefficients 
$a, b^{k}, \ldots $ in the wave equation (\ref{eq}) are built in a local way from the gravitational potential.

Before we go further on we make an important remark. The Milne group $G$ is an infinite dimensional group and 
there are infinitely many ways in which a topology can be introduced in $G$. On the other hand the physical contents
of the assumption (D) depends effectively on the topology in $G$. By this the assumption (D) is in some sense empty.
True, but it is important to stress here, that the whole analysis of this paper rests on the Lie subgroup $G(m)$ 
(see the further text for the definition of $G(m)$) for a sufficiently large $m$, and not on the whole $G$. Indeed,
the covariance condition with respect
to $G(m)$ instead of $G$ is sufficient for us, see {\bf IV} where the 4-th degree polynomial transformations are sufficient. 
Similarly, it is sufficient to consider $G(m)$ in the considerations of the section {\bf VI}. By this, there are no 
ambiguities in the assumption (D). The topology in $G$ is not important from the physical point of view, and 
the extension of the formula (\ref{trapsi}) from $G(m)$ to the whole group is of secondary importance. 
However, we construct such an extension to make our considerations more complete, living the opinion
about the "naturality" of this extension to the reader.    

1) Up to now the Milne group has not been strictly defined. The extent of arbitrariness of the function $\vec{A}(t)$ in
(\ref{tra}) has been left open up to now. The topology depends on the degree of this arbitrariness. 
It is natural to assume the function $\vec{A}(t)$ in (\ref{tra}) to be differentiable up to any order. Consider
the subgroups $G_{1}$ and $G_{2}$ of the Milne group which consist of the transformations:
$(\vec{x},t) \to (\vec{x}+\vec{A}(t),t)$ and $(\vec{x},t) \to (R\vec{x}, t + b)$ respectively, where $R$ is an orthogonal
matrix, and $b$ is constant. Then the Milne group $G$ is equal to the semidirect product $G_{1}\centerdot G_{2}$,
where $G_{1}$ is the normal factor of $G$. It is sufficient to introduce a topology in $G_{1}$ and then define
the topology in $G$ as the semi-cartesian product topology, where it is clear what is the topology in the Lie group
$G_{2}$. We introduce a linear topology in the linear group $G_{1}$ which makes it a Fr\'echet space,
in which the time derivation operator $\frac{{\ud}}{{\ud}t}: \vec{A} \to \frac{{\ud}\vec{A}}{{\ud}t}$ becomes  
a continuous operator. Let $K_{N}, N = 1, 2, \ldots$ be such a sequence of compact sets of $\mathcal{R}$, that
\begin{displaymath}
K_{1} \subset K_{2} \subset \ldots \, \, \, \textrm{and} \, \, \, \bigcup_{N} K_{N} = {\mathcal{R}}.
\end{displaymath} 
Then we define a separable family of seminorms
\begin{displaymath} 
p_{N}(\vec{A}) = \max  \big\{ \big\vert \vec{A}^{(n)}(t) \big\vert, t \in K_{N}, n \leq N \big\},    
\end{displaymath}
where $\vec{A}^{(n)}$ denotes the  $n$-th order time derivative of $\vec{A}$. Those seminorms define 
on $G_{1}$ a localy convex metrizable topology. For example, the metric
\begin{displaymath}
d(\vec{A_{1}},\vec{A_{2}}) = \max_{N \in {\mathcal{N}}} 
\frac{2^{-N}p_{N}(\vec{A_{2}} - \vec{A_{1}})}{1 + p_{N}(\vec{A_{2}} - \vec{A_{1}})}
\end{displaymath}
defines the topology. 

2) It is convenient to rewrite the Milne transformations (\ref{tra}) in the following form
\begin{displaymath}
\vec{x'} = R\vec{x} + A(t) \vec{v}, \, \, \, t' = t + b,
\end{displaymath}
where $\vec{v}$ is a constant vector, which does not depend on the time $t$. We define
the subgroup $G(m)$ of $G$ as the group of the following transformations
\begin{displaymath}
\vec{x'} = R\vec{x} + \vec{v}_{(0)} + t\vec{v}_{(1)} + \frac{t^{2}}{2!}\vec{v}_{(2)} + \ldots + \frac{t^{m}}{m!}\vec{v}_{(m)},
\, \, \, t' = t + b, 
\end{displaymath}
where $R = (R_{a}^{b}), v_{(n)}^{k}$ are the group parameters -- in particular the group $G(m)$ has the dimension
equal to $3m + 7$. 

3) Now we investigate the group $G(m)$, that is, we classify their infinitesimal exponents. The
commutation relations of $G(m)$ are as follows
\begin{equation}\label{Milne1}
[a_{ij},a_{kl}] = \delta_{jk}a_{il} - \delta_{ik}a_{jl} + \delta_{il}a_{jk} - \delta_{il}a_{ik},
\end{equation}

\vspace{-0.5cm}

\begin{equation}\label{Milne2}
[a_{ij},d_{k}^{(n)}] = \delta_{jk}d_{i}^{(n)} - \delta_{ik}d_{j}^{(n)}, \, [d_{i}^{(n)},d_{j}^{(k)}] = 0, 
\end{equation}

\vspace{-0.5cm}

\begin{equation}\label{Milne3}
 [a_{ij},\tau] = 0, \,  [d_{i}^{(0)},\tau]=0, \, [d_{i}^{(n)},\tau] = d_{i}^{(n-1)}, 
\end{equation}
where $d_{i}^{(n)}$ is the generator of the transformation $r(v_{(n)}^{i})$:
\begin{displaymath}
{x'}^{i} = x^{i} + \frac{t^{n}}{n!}v_{(n)}^{i},
\end{displaymath}
which will be called the $n$-acceleration, especially 0-acceleration is the ordinary space translation. 
All the relations (\ref{Milne1}) and (\ref{Milne2}) are identical with (\ref{26a}) $\div$ (\ref{26c}) with
the $n$-acceleration instead of the Galilean transformation. So, the same argumentation as that 
used for the Galilean group gives: $\Xi(a_{ij}, a_{kl})= 0$, $\Xi(a_{ij},d_{k}^{(n)}) = 0$, and 
$\Xi(d_{i}^{(n)}, d_{j}^{(n)}) = 0$. Substituting $a_{i}^{h}, a_{hi}, \tau$ for $a,a',a''$ into the Eq. (\ref{Jac1}),
making use of the commutation relations and summing up with respect to $h$ we get 
$\Xi(a_{ij},\tau) = 0$. Substituting $a_{i}^{h}, d_{h}^{(l)}, d_{k}^{(n)}$ for $a,a',a''$ into the Eq. (\ref{Jac2})
we get in the analogous way $\Xi(d_{i}^{(l)}, d_{k}^{(n)}) = \frac{1}{3}\Xi(d^{(l)h},d_{h}^{(n)}) \, \delta_{ik}$.
Substituting $a_{i}^{h}, d_{h}^{(n)}, \tau$ for $a,a',a''$ into the Eq. (\ref{Jac1}), making use of commutation relations,
and summing up with respect to $h$, we get $\Xi(d_{i}^{(n)}, \tau) = 0$. Now, we substitute 
$d_{k}^{(n)}, d_{i}^{(0)}, \tau$ for $a,a',a''$ in (\ref{Jac1}), and proceed recurrently with respect to $n$,
we obtain in this way $\Xi(d_{i}^{(0)},d_{k}^{(n)}) = P^{(0,n)}(t)\delta_{ik}$, where $P^{(0,n)}(t)$ is a polynomial
of degree $n-1$ -- the time derivation of $P^{(0,n)}(t)$ has to be equal to $P^{(0,n-1)}(t)$, and $P^{(0,0)}(t) = 0$. 
Substituting $d_{k}^{(n)}, d_{i}^{(l)}, \tau$ to (\ref{Jac1}) we get in the same way 
$\Xi(d_{k}^{(l)}, d_{i}^{(n)}) = P^{(l,n)}(t)\delta_{ki}$, where $\frac{{\ud}}{{\ud}t}P^{(l,n)} = P^{(l-1,n)} + P^{(l,n-1)}$.
This allows us to determine all $P^{(l,n)}$ by the recurrent integration process. Note that $P^{(0,0)} = 0$, and 
$P^{(l,n)} = - P^{(n,l)}$, so we can compute all $P^{(1,n)}$ having given the $P^{(0,n)}$. Indeed, we have 
$P^{(1,0)} = - P^{(0,1)}, P^{(1,1)} = 0, {\ud}P^{(1,2)}/{\ud}t = P^{(0,2)} + P^{(1,1)}, {\ud}P^{(1,3)}/{\ud}t = P^{(0,3)} + P^{(1,2)},
\ldots$ and after $m-1$ integrations we compute all $P^{(1,n)}$. Each elementary integration introduces a new
independent parameter (the arbitrary additive integration constant). Exactly in the same way we can compute
all $P^{(2,n)}$ having given all $P^{(1,n)}$ after the $m-2$ elementary integration processes. In general
the $P^{(l-1,n)}$ allows us to compute all $P^{(l,n)}$ after the $m-l$ integrations. So, $P^{(l,n)}(t)$ are 
$l+n - 1$-degree polynomial functions of $t$, and all are determined by $m(m+1)/2$ integration constants. 
Because $d[\Lambda](d_{i}^{(n)}, d_{k}^{(l)}) = 0$, the exponents $\Xi$ defined by different polynomials $P^{(l,n)}$
are inequivalent. By this the space of inequivalent classess of $\Xi$ is $m(m+1)/2$-dimensional.

However, not all $\Xi$ can be realized by the transformation $T_{r}$ of the form (\ref{trapsi}). All
the above integration constants have to be equal to zero with the exception of those in $P^{(0,n)}(t)$. 
By this, all exponents of $G(m)$, which can be realized by the transformations $T_{r}$ of the form
(\ref{trapsi}) are determined by the polynomial $P^{(0,m)}$, that is, by $m$ constants. We show it first
for the group $G(2)$ , because the case is the simplest one and it suffices to explain
the principle of all computations for all $G(m)$. From the above analysis 
we have $P^{(0,1)} = \gamma_{1}, P^{(0,2)} = \gamma_{1}t + \gamma_{2},
P^{(1,2)} = \frac{1}{2}\gamma_{1}t^{2} + \gamma_{2}t + \gamma_{(1,2)}$, where $\gamma_{i}, \gamma_{(1,2)}$
are the integration constants. We will show that $\gamma_{(1,2)} = 0$. A simple computation gives the following
formula $\xi(r,s) = \theta(rs,X) - \theta(r,X) - \theta(s,r^{-1}X)$ for the exponent of the representation 
$T_{r}$ of the form (\ref{trapsi}). Inserting this $\xi$ to the Eq. (\ref{20''})
and performing a rather straightforward computation we get the following formula 
\begin{displaymath}
 \Xi(d_{i}^{(k)}, d_{j}^{(n)}) = \frac{t^{n}}{n!}\frac{\partial^{2}\theta}{\partial x^{j}\partial v_{(k)}^{i}} - 
\frac{t^{k}}{k!}\frac{\partial^{2}\theta}{\partial x^{i}\partial v_{(n)}^{j}}, 
\end{displaymath}
for the infinitesimal exponent $\Xi$ of the representation $T_{r}$ given by (\ref{trapsi}),
where the derivation with respect to $v_{(p)}^{q}$ is taken at $v_{(p)}^{q} = 0$.  
Comparing this $\Xi(d_{i}^{(k)}, d_{j}^{(n)})$ with $P^{(k,n)}\delta_{ij}$ we get the equations
\begin{equation}\label{thetaXi}
\frac{t^{n}}{n!}\frac{\partial^{2}\theta}{\partial x^{j}\partial v_{(k)}^{i}} - 
\frac{t^{k}}{k!}\frac{\partial^{2}\theta}{\partial x^{i}\partial v_{(n)}^{j}} = P^{(k,n)} \delta_{ij}.
\end{equation}
Because of the linearity of the problem,
we can consider the three cases $1^{o}$. $\gamma_{(2)} = \gamma_{(1,2)} = 0$, 
$2^{o}$. $\gamma_{1} = \gamma_{(1,2)} = 0$ and $3^{o}$. $\gamma_{1} = \gamma_{2} = 0$, separately.
In the case $1^{o}$. we have the solution 
\begin{displaymath} 
\theta(r,X) = \gamma_{1}\frac{{\ud}\vec{A}}{{\ud}t}\centerdot \vec{x} + \widetilde{\theta}(t),
\end{displaymath}
where $\widetilde{\theta}(t)$ is an arbitrary function of time and the group parameters, and $\vec{A}(t) \in G(2)$. 
In the case $2^{o}$ we have
\begin{displaymath}
\theta(r,X) = \gamma_{2}\frac{{\ud}^{2}\vec{A}}{{\ud}t^{2}} \centerdot \vec{x} + \widetilde{\theta}(t), 
\end{displaymath}
with arbitrary function $\widetilde{\theta}(t)$ of time.
Consider at last the case $3^{o}$. From (\ref{thetaXi}) we have (corresponding to $(k,n) = (0,1), (0,2))$ 
and $(1,2)$ respectively) 
\begin{equation}\label{theta1}
t\frac{\partial^{2}\theta}{\partial x^{j}\partial v_{(0)}^{i}} - \frac{\partial^{2}\theta}{\partial x^{i}\partial v_{(1)}^{j}} = 0, 
\end{equation}

\vspace{-0.5cm}

\begin{equation}\label{theta2}
\frac{t^{2}}{2}\frac{\partial^{2}\theta}{\partial x^{j}\partial v_{(0)}^{i}} - 
\frac{\partial^{2}\theta}{\partial x^{i}\partial v_{(2)}^{j}} = 0,
\end{equation}

\vspace{-0.5cm}

\begin{equation}\label{theta3}
\frac{t^{2}}{2}\frac{\partial^{2}\theta}{\partial x^{j}\partial v_{(1)}^{i}} -
t \frac{\partial^{2}\theta}{\partial x^{i}\partial v_{(2)}^{j}} = \gamma_{(1,2)}\delta_{ij}.
\end{equation}
From (\ref{theta3}) and (\ref{theta2}) we get
\begin{equation}\label{theta4}
\frac{t^{2}}{2} \big\{\frac{\partial^{2}\theta}{\partial x^{j}\partial v_{(1)}^{i}}
 - t \frac{\partial^{2}\theta}{\partial x^{j}\partial v_{(0)}^{i}} \big\} = \gamma_{(1,2)} \delta_{ij}.
\end{equation}
But $\Xi(d_{i}^{(0)}, d_{j}^{(0)}) = 0 = \partial^{2}\theta/\partial x^{j}\partial v_{(0)}^{i}
- \partial^{2}\theta/\partial x^{i}\partial v_{(0)}^{j}$, so, from (\ref{theta4}) and (\ref{theta1}) we get
\begin{displaymath}
0 = \frac{\partial^{2}\theta}{\partial x^{i}\partial v_{(0)}^{j}}\big\{\frac{t^{3}}{2} - \frac{t^{3}}{2}\big\}
= \gamma_{(1,2)} \delta_{ij},
\end{displaymath}
and $\gamma_{(1,2)} = 0$.

The following   
\begin{equation}\label{thetaG(2)}
\theta(r,X) = \gamma_{1}\frac{{\ud}\vec{A}}{{\ud}t} \centerdot \vec{x} + 
\gamma_{2}\frac{{\ud}^{2}\vec{A}}{{\ud}t^{2}} \centerdot \vec{x} + \widetilde{\theta}(t)
\end{equation}
fulfils all Eqs. (\ref{thetaXi}) with $k,n \leq 2$ and its local exponents cover the full classification
of $\Xi$'s for $G(2)$ which can be realized by $T_{r}$ of the form (\ref{trapsi}), that is, all $\Xi$'s with
$\gamma_{(1,2)} = 0$. Then, according to Appendix {\bf A}, the formula (\ref{thetaG(2)}) gives the most general 
$\theta$ in (\ref{trapsi}) for $r \in G(2)$.  This is because the classification of $\Xi$'s covers
the classification of all possible $\theta$'s. 

It can be immediately seen that any integration constant $\gamma_{(l,q)}$ of the polynomial
$P^{(l,q)}(t)$ has to be equal to zero if $l,q \neq 0$, provided the exponent $\Xi$ belongs to the
representation $T_{r}$ of the form (\ref{trapsi}). The argument is essentially the same
as that for $\gamma_{(1,2)}$. 
It is sufficient to consider (\ref{thetaXi}) for the four cases: $(k,n) = (l-1,q-1), (l-1,q), (l,q)$
and $(l,q-1)$ respectively. Becuse of the linearity of the considered problem, it is sufficient to consider
the situation with the integration constants in $P^{(k,n)}$ equal to zero with the possible exception of the 
integration constant $\gamma_{(l,q)}$. We get in this way the equations (\ref{thetaXi}) corresponding to $(l-1,q-1),(l-1,q),
(l,q)$ and $(l, q-1)$ with the right hand sides equal to zero with the exception of the right hand side of the 
equations corresponding to $(k,n) = (l,q)$, which is equal to $\gamma_{(l,q)} \delta_{ij}$. 
From the equations (\ref{thetaXi}) corresponding to $(k,n) = (l,q)$ and $(l-1,q)$ we get   
\begin{displaymath}
\frac{t^{q}}{q!}\frac{\partial^{2}\theta}{\partial x^{j}\partial v_{(l-1)}^{i}} - 
\frac{t^{q+1}}{q!l}\frac{\partial^{2}\theta}{\partial x^{j}\partial v_{(l-1)}^{i}} = \gamma_{(l,q)}\delta_{ij}.
\end{displaymath}
From this and the equations (\ref{thetaXi}) corresponding to $(k,n) = (q-1,l-1)$ we get
\begin{displaymath}
\frac{t^{q}}{q!}\frac{\partial^{2}\theta}{\partial x^{j}\partial v_{(l)}^{i}} -
\frac{t^{l+1}}{l!q}\frac{\partial^{2}\theta}{\partial x^{i} \partial v_{(q-1)}^{j}} = \gamma_{(l,q)}\delta_{ij}.
\end{displaymath}
From this and the equations (\ref{thetaXi}) corresponding to $(k,n) = (l,q-1)$ one gets
\begin{displaymath}
0 = \frac{t^{l+1}}{l!q}\frac{\partial^{2}\theta}{\partial x^{i}\partial v_{(q-1)}^{j}} -
\frac{t^{l+1}}{l!q}\frac{\partial^{2}\theta}{\partial x^{i}\partial v_{(q-1)}^{j}} = \gamma_{(l,q)} \delta_{ij},
\end{displaymath}
which gives the result that $\gamma_{(l,q)} = 0$. 

Consider the $\theta$, given by the formula
\begin{equation}\label{thetaG(m)}
\theta(r,X) = \gamma_{1}\frac{{\ud}\vec{A}}{{\ud}t} + \gamma_{2}\frac{{\ud}^{2}\vec{A}}{{\ud}t^{2}} + \ldots + 
\gamma_{m}\frac{{\ud}^{m}\vec{A}}{{\ud}t^{m}} + \widetilde{\theta}(t),
\end{equation}
for $r \in G(m)$, where $\gamma_{i}$ are the integration constants which define the polynomial 
$P^{(0,m)} = \gamma_{1}\frac{t^{m-1}}{(m-1)!} + \gamma_{2}\frac{t^{(m-2)}}{(m-2)!} + \ldots + \gamma_{m}$, and
$\widetilde{\theta}(t)$ is any function of the time $t$ and eventually of the group parameters .
 A rather simple computation shows that this $\theta$ fulfils all (\ref{thetaXi}) for $k,n \leq m$ and that it covers 
all possible $\Xi$ which can be realized by (\ref{trapsi}). That is, the infinitesimal exponents
corresponding to the $\theta$ given by (\ref{thetaG(m)}) give all possible $\Xi$ with 
all integration constants $\gamma_{(k,n)} = 0$, for $k,n \neq 0$. 
So, according to Appendix {\bf A} the most
general $\theta(r,X)$ defined for $r \in G(m)$ is given by (\ref{thetaG(m)}). 

At this place we make use of the assumption (C), or more precisely the assumption that
the wave equation is local. Recall that we assumed the coefficients $a,b^{i}, \ldots , g$ in the wave
equation (\ref{eq}) to be local functions of the gravitational potential $\phi$. That is, they are
an algebraic functions of the potential and its derivatives up to the finite order -- in our case up to the second order.
Recall also, that we have proved from this assumption (begining of the Appendix {\bf A})
that the $\theta(r,X)$ can be a function of the 4-th order derivative of $\vec{A}(t)$ at most,
the higher derivatives cannot enter into $\theta$. By this, the most general $\theta(r,X)$ defined
for $r \in G(m)$ has the following form 
\begin{equation}\label{thetaG}    
\theta(r,X) = \gamma_{1}\frac{{\ud}\vec{A}}{{\ud}t} + \ldots + \gamma_{4}\frac{{\ud}^{4}\vec{A}}{{\ud}t^{4}} 
+ \widetilde{\theta}(t),
\end{equation}

4) Now, we extend the formula (\ref{thetaG}) on the whole Milne group $G$. It is a known fact that the 
time derivative operator ${\ud}/{\ud}t: \vec{A} \to {\ud}\vec{A}/{\ud}t$ is a continuous operator on $G$
in the topology introduced in 1), see e.g. \cite{Rudin}. It remains to show that the sequence $G(m), m \in {\mathcal{N}}$
is dense in $G$. The proof of this presents no difficulties \cite{dense}.
By this the function $\theta(r,X)$ can be uniquely extended on the whole group $r \in G$
\begin{displaymath}
\theta(r,X) = \gamma_{1}\frac{{\ud}\vec{A}}{{\ud}t} + \ldots + \gamma_{4}\frac{{\ud}^{4}\vec{A}}{{\ud}t^{4}} 
+ \widetilde{\theta}(t). 
\end{displaymath}
It should be stressed here that not only the topology in $G$ is needed to derive the formula, but
also the locality assumption is very important. If the coefficients $a, b^{i}, \ldots , g$ in the wave
equation were admitted to be nonlocal, then an infinite number of other solutions for $\theta$ in $G$ 
would exist.

\section{Connection with the reduction problem}

It is rather surprising that we are able to prove the equality of 
inertial and gravitational mass for \emph{one} quantum particle 
considered as a test body. We consider here physical consequences of this theorem.
As we have seen, the equivalence principle does not follow from this equality at the quantum level. 
Quantum Mechanical (QM) laws are non-local in general and the meaning
of the equivalence principle in QM is not clear of course. The QM laws are twofold
in some sense: 1) those concerning the unitary evolution for the state vector and
2) those concerning the Reduction Process. The equivalence principle 
can be applied to the wave equation which determines the evolution 
of the wave function and is local as a differential equation. The meaning
of equivalence principle is unclear when it is applied to the nonlocal
laws 2). So, even if the equivalence principle was true for 1) it would
be still an open question if it is true for QM. 
We have shown that the gravitational and inertial masses are equal for 
a system consisting of one quantum particle. But the equivalence principle for the wave equation
is not fulfilled in general because of the additional scalar term $\Lambda(\partial_{a}\partial_{b}\phi)$ depending on
the second order space derivatives of the potential. 
Consider now the classical equation of motion for the test body.
If we assume in addition that the equation of motion has the folowing form
\begin{equation}\label{classical}
m_{i}\frac{{\ud}^{2}\vec{x}}{{\ud}t^{2}} = - m_{g}\vec{\partial}\phi, 
\end{equation}     
where we left open the question of the eventual equality $m_{i} = m_{g}$, 
then we can eliminate the scalar term $\Lambda(\partial_{a}\partial_{b}\phi)$ from the 
wave equation applying the Ehrenfest equations (applying the short-wave limit -- the 
counterpart of the geometrical optics limit). This means that the Schr\"odinger equation
for \emph{one} particle fulfils the equivalence principle if the above assumption (\ref{classical}) for the classical
equations of motion are fulfilled. The assumption (\ref{classical}) is sufficient to derive the 
equivalence principle for the wave equation.
This is not so very surprising because the equivalence
principle is meaningless for the nonlocal laws 2) and the status of the equivalence principle (EP) in QM
remains still open, so that EP in QM does not follow from (\ref{classical}).   
But now the very strange fact comes in. One can think along the following
track. Because interactions do not change the mass in the nonrelativistic limit,
one can expect that from the equality of inertial and gravitational
masses for one quantum particle it follows the equality for many particle system. It is 
really the case if one assumes the standard many-particle QM  laws and that the many particle de Broglie wave of the 
system in question is \emph{testing}. That is, \emph{if the system does not  
influence the spacetime (the potential in our case)}. It can be shown,
if one uses the standard many-particle quantum formalism and the result obtained
for the system consisting of one quantum particle. The only additional
assumption is that the non-gravitational interactions between the particles are described by 
a potential which transforms as a scalar and depends on the distance of the interacting particles. 
After this, one can show, that 
there exists a degree of freedom belonging to the center of mass position.
The wave equation can be separated into the center of mass position
and the internal degrees of freedom. The wave equation for the center of mass
position is identical with that of one particle Schr\"odinger equation with
the inertial mass equal to the gravitational mass and equal to the sum
of all masses of constituent particles \cite{mass center}. But this is imposible,
because we cannot deduce the equality of inertial and gravitational masses
for many particle quantum system consisting of arbitrary many particles. 
This is because we believe that such a system of (appropriately) many particles
should be treated as a classical body, so that we were able to prove from
our assumptions (A) $\div$ (D) and (\ref{classical}) and many particle quantum formalism, that 
the inertial and gravitational masses for classical particle are equal -- which 
is the weak equivalence principle. Such a principle characterises the gravity
field and cannot be contained in our assumptions. Something is wrong, because 
there should be a place for two independent mass parameters -- inertial and gravitational. 
The equality of the two independent parameters should be a consequence of an independent extra 
assumption about the gravitational field. Where the gravity law comes from in the Quantum Mechanics?
We assume that this gravity law cannot be contained in the quantum mechanical principles and 
conclude from it 

\vspace{1ex}

THEOREM 7. \emph{(i) at least some of 
the orinary QM laws (partially assumed in (A) $\div$ (D)) of many particle system in the gravitational field are not valid 
when the system contains appropriately many particles or (ii) the system can influence
the spacetime geometry when it contains appropriately many particles}.

\vspace{1ex}

The system of many particles (in the Theorem 7) which come into play is still too small to 
influence significantly the gravitational field. By this the possibility (\emph{ii}) of Theorem 7 seems 
to be irrational. But it is qualitatively important if one takes into account the back reaction to 
the field or not -- which is a peculiar property of the gravitational field. We explain it now,
and show that also the possibility (\emph{ii}) leads to some difficulties. Consider first a quantum system of particles
which does not influence the gravitational field (which will be assumed classical). 
Then the Milne group (\ref{tra}) is the covariance group of the whole system: particles + field. But it is not
a symmetry group. The field breaks any symmetry -- it does not possess any symmetry in general. 
So, the representation $T_{r}$ transforming the particle wave function has time-dependent phase factor
$e^{i\xi}$, see {\bf V.A}, Appendix {\bf A} and {\bf V.C}. The factor of $T_{r}$ can be time-independent if and 
only if the mass $m_{i} = m_{g}$ of the system is zero. We assume the mass to be $\neq 0$. By this,
 the factor of the representation $T_{r}$ of the Milne group \emph{has to be time-dependent}.   
Consider now a system, but this time it can influence the gravitational field. The Milne group is now 
a symmetry group of the whole system: particles + field. If one assumes that the whole system 
can be described in terms of quantum mechanics of infinitely many degrees of freedom (Quantum Field
Theory), then some fundamental difficulties arise. Those difficulties are due to the fact that now a unitary projective
representation $U_{r}$ of the symmetry Milne group acts in the Hilbert space. By this the factor 
of the representation $U_{r}$ \emph{has to be time-independent}, see {\bf V.A}. This by no means can be reconciled 
with the fact that if only one particle is present in the field, then its wave function transforms according
to the representation $T_{r}$ with time-dependent factor, see the more detailed discussion below (Theorems 8 
and 9). 
\\ The situation is completely different for the non-gravitational fields -- there are no difficulties in this case. 
After the back reaction to the (non-gravitational) field is allowed for, the spacetime symmetry becomes 
the symmetry of the whole system. So, the symmetry group is a subgroup of the Galilean group at most
(we consider here the nonrelativistic case), and its projective unitary representation $U_{r}$ has time-independent
factor, of course. If the back reaction to the field is negligible, then the Galilean group becomes a covariance group
and not any symmetry group -- the field does not possess any symmetry in general. By this, the representation 
$T_{r}$ of the Galilean group as a covariance group has a time dependent factor in general.
But this time a representation $T'_{r}$ of the Galilean group (as a covariance group) exists which is
equivalent to $T_{r}$ and the factor of $T'_{r}$ is time-independent, see {\bf V.B}. So, there is no difficulty 
with non-gravitational field.   
         
It is probable that the two possibilities (\emph{i}) and (\emph{ii}) of Theorem 7 are deeply connected. 
Namely, note that (\emph{i}) as well as (\emph{ii}) is connected with the system subjected to 
the influence of the gravitational field, where in the case (\emph{ii}) the back reaction to the field 
is taken into consideration. By this one could suppose that this difficulty has its origin in the gravitational
field. We have to be careful, however, because it is \emph{a priori} possible 
that the QM laws for the system consisting of too many particles are not valid in general,
and the above difficulty is not connected with the gravitational field itself. Nonetheless, 
we will show in this section, that any quantum theoretical description of ineracting
quantum system with the gravitational field has some fundamental difficulties even in this
nonrelativistic limit -- the troubles connected with (\emph{ii}) of Theorem 7.

We shall consider the freely falling Schr\"odinger particle in the Newton-Cartan
spacetime. Then, we should describe the system in the canonical form and quantize it to
obtain the nonrelativistic limit of the quantum gravitational interactions between quantum 
particles. We ignore now the insurmountable difficulties of recasting such a system into the 
canonical form \cite{cangrav}. 
Any attempt to quantize the gravitational interactions with the Schr\"odiger particles rely 
on some fundamental \emph{postulates} of the quantum field theory. 
Namely, if such a theory possesses a symmetry
group $G$, then there exist a unitary projective (up to constant factor, of course) representation $U_{g}$
of the group $G$ in the Hilbert space $\mathfrak{H}$ of the system (see the discussion in 
{\bf V. A}, Theorem 1). The quantum fields $\widehat{\Phi}(X)$ are operator valued distributions,
the transformation laws of which cannot be uniquely deduced in general and have to be postulated.
However, there are always physical limitations for those transformation rules, for example
when we know what is the action of the group in question on the mean values of those operators.
Now we will analyse what are the postulates in our case. In contradiction to the General Relativity,
the Newton-Cartan theory possesses the absolute elements and the symmetry group -- the Milne group
(\ref{tra}) -- which is the symmetry group for those absolute elements \cite{symmetry}. It is therefore natural
to assume that the absolute elements remain to be absolute when the field is allowed to interact
with quantum particles. That is, we rely on the customary assumption of the quantum field theory, that
only dynamical objects can be subjected to the "second quantization". After this we have

\vspace{1ex} 
   
{\bf Postulate 1}. \emph{The Milne group is the symmetry group of the Newton-Cartan-Schr\"odinger
quantum field theory}.

\vspace{1ex}

Such a postulate was assumed for example in \cite{Chr}. Because of Postulate 1 it will be useful to
confine ourselves
to the galilean coordinate systems. The dynamical object consists now of the wave function $\psi$ 
and the potential $\phi$. But they are promoted to the status of operator-valued distributions $\widehat{\psi}(X),
\widehat{\Phi}(X)$ at the "second quanization" level. From the very general laws of the quantum
field theory it follows

\vspace{1ex} 
      
{\bf Postulate 2}. \emph{There exist a unitary projective (up to a constant factor) representation $U_{g}$ of 
the Milne group $G$ in the Hilbert space $\mathfrak{H}$ of the system}.

\vspace{1ex} 

{\bf Postulate 3}. \emph{There exist the vacuum state $\vert 0 \rangle \in \mathfrak{H}$ invariant under the 
Milne group $G$: $U_{g}\vert 0 \rangle = \vert 0 \rangle$ for $g \in G$}.

\vspace{1ex}

The meaning of the Postulate 3 is clear. Namely, it is well known that the canonical quantization
of a system with finite degrees of freedom is unique up to the unitary equivalence, which is the consequence
of the von Neumann-Stone uniqueness theorem. This is not the case if the system has infinite number of degrees 
of freedom. We require the representation to contain the vacuum state defined as in Postulate 3, and make  
the choice between inequivalent representations, see \cite{Chr} where this argument was used. 
After this the Hilbert space $\mathfrak{H}$ is the ordinary
Fock space and the numbers of particles are well defined.  
The transformation law for the fields $\widehat{\psi}(X)$ and  $\widehat{\Phi}(X)$ cannot be uniquely 
determined, but there is a natural condition which such a law should fulfil:

\vspace{1ex}

{\bf Postulate 4}. \emph{In the limit where Newton-Cartan gravity assumes classical spacetime, and the 
quantum system is a single particle, the Eqs}. (\ref{trapsi}), (\ref{ec1}) \emph{and} (\ref{ec2}) 
\emph{should naturally emerge}.

\vspace{1ex}

Let $n_{p}$ denote the total number of particles and let $n_{g}$ denote the quantum numbers 
characterizing the gravitational field \cite{graviton}.
The states $\Psi_{n_{p}n_{g}} \in \mathfrak{H}$ with a fixed number of particles $n_{p}$ have well determined 
transformation law obtained by the restriction of $U_{g}$, 
in accordance with Postulate 2, which is always a \emph{unitary} projective representation of $G$
independently of the number of particles $n_{p}$ . 
Because the \emph{vacuum} state is invariant the one-particle state
$\Psi_{1_{p}n_{g}}$ transforms under an element $g \in G$ into a one particle state $\Psi'_{1_{p}n'_{g}}$,
and in general the state $\Psi_{n_{p}n_{g}}$ transforms into a state $\Psi'_{n_{p}n'_{g}}$ with the same
number of particles $n_{p}$. 
So, the states $\Psi_{n_{p}n_{g}}$ compose a unitary projective representation of $G$.
Similarily the states $\Psi_{1_{p}0} = \widehat{\psi}^{\dagger}\vert 0 \rangle$ as well as $\Psi_{0_{p}n_{g}}$
compose a unitary projective representation of $G$. 
That means that the unitary character of the transformation law is achieved independently of the value
of the quantum numbers $n_{g}$ characterizing the gravitational field as well as independently of 
the total number of particles $n_{p}$. Suppose $n_{g}$ to be so large, that the gravitational
field is classical and one particle in the field, \emph{i.e.} $n_{p}=1$. According to the Postulate 4, the 
field evolves independently of the particle degrees of freedom (the field evolves as if the particle were absent)
and can be separated into the particle and field evolution. Then, the state 
$\Psi_{1_{p}n_{g}} = \psi(X) \Psi_{0_{p}n_{g}}$ is the product of the particle 
state and the field state. In accordance to our discussion $\Psi_{0_{p}n_{g}}$ as well as
$\Psi_{1_{p}n_{g}}$ composes a unitary projective representation of the Milne group $G$ with constant factor. 
By this $\psi(X)$ also composes a unitary projective representation of $G$ with constant factor.  
Because the transformation law for the one-particle state $\psi(X)$, when the field is classical
and the numbers $n_{g}$ are appropriately large, is given by (\ref{trapsi}) and (\ref{ec2}), then 
we have obtained a contradiction. Recall, that the transformation law $T_{g}$ as given by Eqs. (\ref{trapsi})
and (\ref{ec2}) does not compose any projective representation of the Milne group $G$ --
the factor of $T_{g}$ is time-dependent (see subsections {\bf V.A} and {\bf V.C}). So, we have 

\vspace{1ex}
 
THEOREM 8. \emph{The Postulates 2, 3 and 4 are self-contradictory}.

\vspace{1ex}

This is not so very surprising, because there are serious physical obstacles to the 
Postulate 3, even in the nonrelativistic case discussed here. Before we proceed further on,
let us turn our attention for the moment to the analog of the Postulate 3 for a nongravitational quantum field 
in the Minkowski spacetime (with
the Poincar\'e group instead of the Milne group). It is meaningful to say that the field is zero,
independently of the inertial observer. In the Newton-Cartan theory the situation is substantially 
different. In accordance to the equivalence principle the inertial forces are gravitational in origin.
By this, it is meaningless to say that the potential of the 
field is zero independently of the galilean observer \cite{curvature}.
This is because two galilean frames may be connected by an acceleration (element of the Milne group).
Such an acceleration creates the inertial field -- which is gravitational in origin. 

Nonetheless, even if we assume a much more weakened Postulate 3' instead of the Postulate 3,
the difficulty still remains. Namely, we assume 

\vspace{1ex}

{\bf Postulate 3'}. \emph{The number $n_{p}$ of particles is still well
defined, or the states $\vert 0_{p} \rangle \in \mathfrak{H}$ with no
particles in the gravitational field are well defined, and the space of such 
states is invariant under the action $U_{g}$ of the Milne group}.

\vspace{1ex}

This time the vacuum state $\vert 0 \rangle$ is not well defined, and the Hilbert space
$\mathfrak{H}$ is not the ordinary Fock space. The Postulate 3' seems to be 
physically justified. Of course,  we do not propose the Postulate 3' as a
postulate which can replace the Postulate 3 in the sense that we are able
to omit non-uniqueness property of the second quantization. We propose
only to assume that Postulate 3' is true. Using the Postulate 3' and the reasoning
such as in the proof of the Theorem 8, we can show  

\vspace{1ex}

THEOREM 9. \emph{The Postulates 2, 3' and 4 are self contradictory}.

\vspace{1ex}

The origin of all those difficulties is as follows. 1) The Milne group as a covariance
group has the time-dependent factor non-equivalent to any constant factor if the 
mass of the particle is not equal to zero, see {\bf V.A}, Appendix {\bf A}
and {\bf V.C}. 2) The Milne group becomes a symmetry group, when
the back-reaction of the particles to the gravitational field is considered.
But, as we know, if $G$ is a symmetry group in quantum mechanics or quantum field
theory, then the unitary projective representation of $G$ 
acts in the Hilbert space and by this the factor of such a representation has to be constant
(time-independent in our case), see {\bf V.A}. In particular the one-particle states
should compose irreducible unitary projective representations of the symmetry group,
with constant factor. This cannot be reconciled with 1), or strictly speaking with 
the transformation law (\ref{trapsi}) and (\ref{ec2}), with time-dependent factor.
Note, that the gauge freedom is essential for this difficulty -- the possibility of the existence
of the time-dependent factor is due to time-dependent gauge freedom, see {\bf V.A}.    
Note also, that this difficulty is deeply connected with the gravitational field, at least in this
nonrelativistic case. Indeed, the factors for the Galilean group treated as a symmetry group 
are the same as in the case when it is a covariance group, and only the interactions with
the gravitational field can "enlarge" the symmetry group to be a greater group then
the Galilean group (the maximal symmetry of the spacetime in the nonrelativistic case).
In the nonrelativistic theory, the symmetry group is "enlarged" to be the Milne group. 

It should be mentioned at this place that the troubles in quantum field theory generated
by the gauge freedom are of general character, and are well known. But -- one could argue -- 
we have learned what should be done to make the theory at least practically useful. 
For example, there do not exist vector particles with helicity = 1, which is a consequence
of the theory of unitary representations of the Poincar\'e group, as was shown by Jan {\L}opusza\'nski 
\cite{Lopuszanski}. This is apparently in contradiction with the existence of vector particles with
helicity = 1 in nature -- the photon, which is connected with the electromagnetic fourvector potential.
The connection of the problem with the gauge freedom is 
well known \cite{Lopuszanski}. We omit however the difficulty if we allow the inner product in
the Hilbert space to be not positively defined -- the inner product can be zero or even negative for
well defined states of the Hilbert space, see the formalism of Gupta and Bleuler \cite{Gupta}, 
or the so called BRST formalism \cite{BRST}. Then, we define the subset of the Hilbert space 
(the Lorentz set in the Gupta-Bleuler formalism, or the kernel of the BRST symmetry generator 
in the BRST formalism) on which the inner product is positively defined and can be normally interpreted,
which makes the theory at least practically useful. One could argue that the situation with the gravitational 
field could be the same -- \emph{i.e.} that there exists an analogous procedure, which makes the theory 
at least practically useful. Such a procedure, however, should be very radical and by this we 
think that the problem is fundamental rather. We explain it. The presented difficulty
has a deep connection with the gauge freedom difficulties indicated by J. {\L}opusza\'nski. Indeed,
due to \cite{Lopuszanski}, the vector potential (promoted to be an operator valued distribution in QED)
cannot be a vector field, if one wants to have the inner product positively defined -- together with
the coordinate transformation the gauge transformation has to be applied, which breaks the vector
character of the potential. Practically it means that any gauge condition which brings the theory into
the canonical form such that the quanization procedure can be consequently applied (with the positively
defined inner product in the Hilbert space) breaks the fourvector character of the electromagnetic potential, the 
Coulomb gauge condition is an example. To achieve the Poincar\'e symmetry of Maxwell
equations with such a gauge condition (the Coulomb gauge condition for example), 
it is impossible to preserve the vector character 
of the potential -- together with the coordinate transformation a well defined (by the coordinate transformation)
gauge transformtion $f$ has to be applied:
\begin{displaymath}
A_{\mu} \to A'_{\mu'} = \frac{\partial x^{\nu}}{\partial x^{\mu'}}(A_{\nu} + \partial_{\nu}f).
\end{displaymath}
This means that the electromagnetic potential has to form
a ray representation (in the sense of {\bf V.A } and Appendix {\bf A}) $T_{r}$ of the Poincar\'e group, with
the spacetime-dependent factor $e^{i\xi}$. Because it is imposible to form a projecive unitary representation
(of the Poincar\'e group in this case) with the spacetime-dependent factor we get a difficulty. There are, however, 
other realizations of the Poincar\'e group with the factor = 1 ($f = 0$), that is, with the potential which transforms as 
a vector, and such that the Maxwell equations have a canonical form and are invariant with 
respect to the Poincar\'e group (at the classical level) -- for example, the Maxwell equations with the Lorentz  
gauge condition. Thanks to that, the Gupta-Bleuler formalism can be built up. The situation with the gravitational
field is different. There do not exist any trivial realizations (with the factor = 1) of the Milne group formed by the 
geometrical structure of the spacetime (which is the analogue of the electromagnetic potential),
such that the Newton-Cartan theory is invariant with respect to the Milne group, see \cite{Waw}. By this there 
does not exist any construction analogous to that of Gupta and Bleuler.    
  
One could argue that the troubles indicated in Theorems 7 -- 8 are characteristic for the 
nonrelativistic case only, and they probably disappear in the relativistic theory. Such a reasoning
cannot be true. Indeed, any relativistic theory possesses a well defined nonrelativistic limit,
with the symmetries characteristic for the nonrelativistic theories investigated here. We believe
that such a limit should not lead to any selfcontradictory conclusions.  
 
Summing up the results, the many-particle quantum mechanics laws for the system in the gravitational field
are not self-consistent independently of the fact if the back reaction to the field is allowed for or not, 
at last in this nonrelativistic theory. There are two main possibilities, 1) the gravitational field is responsible
for those difficulties of many-particle quantum mechanics, or 2) the many-particle quantum mechanics itself
is not valid for systems consisting of appropriately many particles, independently of the gravitational field.
One could assume that the equivalence principle can be derived from QM laws, and by this the Theorem 7
would not be true -- there were no difficulties. But such an assumption would be strange and instead of the
Theorem 7 a new problem would arise: where the gravity law comes from in the QM laws? 

Some authors \cite{Pe1}, \cite{Ghi}, \cite{Dio}, \cite{kom}, \cite{Pe2}, \cite{Kib1}, \cite{Pen}, 
\cite{Kar} suppose the gravity field to be responsible
for the state vector reduction process. Those authors seem to agree with the possibility 1). 
Others \cite{Zur} treat the Quantum Mechanics as a selfconsistent theory and 
the  state vector reduction  as an apparent process which is an artefact of the 
interactions with environment. We do not agree with \cite{Zur} in this respect, 
the present paper can be treated as an argument in the discussion. 

There are experiments testing the quantum laws for "big" systems which are built of many quantum particles.
See for example the beautiful experiment with the superposition states of the fullerene  molecule C$_{60}$ \cite{Zei}. 
We expect that the Quantum Mechanical predictions fail in such experiments for the appropriately
big molecule.  

It seems necessary to make a comment on the work by J. Christian \cite{Chr}. This is an interesting
paper in which a recasting of the Newton-Cartan-Schr\"odinger system into a canonical form
is reached. However, a bit deeper analysis is needeed to derive some further consequences. 
Namely, 1) the investigations of the symmetry group of the Newton-Cartan theory is needed
and the role of the gauge freedom, 2) the investigation
of the quanization procedure of that canonical system is needed. 
We have attacked the problem 1)  in \cite{Waw}. The canonical formulation given in
\cite{Chr} cannot be reconciled with the invariance property of the Newton-Cartan theory
under the Milne group. The present paper shows, that even if it was possible
to recast the Newton-Cartan theory into a canonical form, then the problems 
indicated by Theorems 8 and 9 would arise.

\section{Summary}

The covariance condition in quantum mechanics is much stronger then the covariance condition
at the classical level. There are many possibilities for the generally covariant classical theory of classical 
particles in the gravitational field with the inertial mass $m_{i}$ not equal to the gravitational mass $m_{g}$.
This is imposible in quantum theory -- the covariance condition is so strong, that it almost uniquely 
determines the wave equation, and moreover, the equality $m_{i} = m_{g}$ has to hold in the quantum theory.
This is acceptable from the physical point of view, if the system of quantum particles is "small" --
the equivalence principle in QM cannot be derived from the equality $m_{i} = m_{g}$.
That is, the equivalence principle still remains an open question, 
and the equivalence principle cannot be derived, so that
the equality $m_{i} = m_{g}$ is not so strange. But the equality $m_{i} = m_{g}$ can be derived 
for arbitrary "big" quantum system, provided the back reaction of the system to the
gravitational field is negligible. We could derive, then, that 
$m_{i} = m_{g}$ for macroscopic systems, provided that the macroscopic system can be 
described by the many-particle Quantum Mechanics.
But we have assumed that this is impossible.  We conclude from this that 1) the 
macroscopic systems cannot be described by many-particle Quantum Mechanics, 
or 2) the back reaction to the gravitational field is not negligible.
Next, we have shown, that also the possibility 2) leads to some difficulties. So, in any case
the  QM laws of a many-particle system in the gravitational field are inconsistent. We have left 
open the question if the gravitational field is responsible for this inconsitency allowing the possibility
that the many-particle QM laws are not valid for "big" systems. We expect that some quantum
mechanical predictions fail for the superposition states of a suitably big molecule. According
to the beautiful experiment \cite{Zei} the molecule has to be bigger than the fullerene molecule C$_{60}$.

\section{Acknowledgments}

The author is indebted for helpful discussions to A. Staruszkiewicz, A. Herdegen and
J. Zejma. The paper was financially supported by the KBN grant no. 5 P03B 093 20. Without
this help the paper would have never been written.

\section{Appendix}

\subsection{Generalization of Bargmann's theory}

We investigate here the exponents $\xi$ of a ray representations $T_{r}$ in the space $\mathfrak{S}$ 
of waves functions $\psi$ fulfilling (\ref{rowt}). In the sequel any $\psi$ is considered as an element 
of $\mathfrak{S}$ and not as any element of the Hilbert space of wave functions in the "Schr\"odinger picture". 

It becomes clear in the further analysis that the group $G$ in question has to fulfil the cosistency condition that for 
any $r\in G$ $rt$ is a function of time only. Beside this we confine ourselves to the classical Lie groups. We follow
Bargmann's \cite{Bar} line of reasoning if only it is possible.

Consider two representatives $\psi_{i}, \, i=1,2$, of the unit rays $\boldsymbol{\psi_{i}}$. That is, 
$ \Vert \psi_{i} \Vert_{t}^{2} \equiv (\psi_{i},\psi_{i})_{t}= \Vert \psi_{i} \Vert^{2} = 1$, it is possible because 
$\psi_{i}$ are elements of rays and $\psi_{i}$ can be "normed" in such away. Note first, that 
\begin{displaymath}
(\psi_{1}-\psi_{2},\psi_{1}-\psi_{2})_{t}  \equiv \Vert \psi_{1} - \psi_{2} \Vert_{t}^{2} 
\end{displaymath}
does depend on the time in general. Indeed, we have (for unit $\psi_{i}$) 
\begin{displaymath}
2(1-\vert (\psi_{1}, \psi_{2})_{t} \vert ) \leq \Vert \psi_{1}-\psi_{2}\Vert_{t}^{2} = 
2(1-\textrm{Re}{(\psi_{1}, \psi_{2})_{t}}) \leq 
\end{displaymath}

\vspace{-0.5cm}

\begin{equation}\label{f-g}
\leq 2\vert 1- (\psi_{1},\psi_{2})_{t}\vert,
\end{equation}
where $\textrm{Re}$ stands for the real part. 

Second, note that Schwarz'-like inequality is fulfilled at each time $t$
\begin{displaymath}
\vert (\psi_{1}, \psi_{2})_{t}\vert \leq \sqrt{(\psi_{1},\psi_{1})_{t} (\psi_{2},\psi_{2})_{t}}
\end{displaymath}
and both sides are equal if and only if $\psi_{1}=\alpha(t)\psi_{2}$, in our
case $\vert\alpha(t) \vert =1$ because $\psi_{1}$ and $\psi_{2}$ are unit.

We define the \emph{inner product} -- a well defined time independent number
\begin{displaymath}
{\boldsymbol{\psi_{1} \centerdot \psi_{2}}}= \sup_{t \in {\mathcal{R}}}\vert (\psi_{1},\psi_{2})_{t}\vert  
\end{displaymath}
where $\psi_{1}$ and $\psi_{2}$ are any elements of ${\boldsymbol{\psi_{1}}}$ and ${\boldsymbol \psi_{1}}$ 
respectively. From the Schwarz'-like inequality $\boldsymbol{\psi_{1} \centerdot \psi_{2}} \leq 1$ and 
$\boldsymbol{\psi_{1} \centerdot \psi_{2}} =1$ if and only if $\boldsymbol{\psi_{1}}=
\boldsymbol{\psi_{2}}$. 
The \emph{distance} $d(\boldsymbol{\psi_{1}},\boldsymbol{\psi_{2}})$ between 
the two rays $\boldsymbol{\psi_{1}}$ and $\boldsymbol{\psi_{2}}$ is defined as follows
\begin{displaymath}
d(\boldsymbol{\psi_{1}},\boldsymbol{\psi_{2}}) = \inf_{\psi_{i} \in \boldsymbol{\psi_{i}}}
\sup_{t \in {\mathcal{R}}} \Vert \psi_{1} - \psi_{2} \Vert_{t}.
\end{displaymath} 
To show that $d$ is really a metric in the space of unit rays, we need a Lemma 
(we need this Lemma later on also). 

\vspace{1ex}

LEMMA 1. \emph{The inequality}
\begin{displaymath}
2(1- \vert (\psi_{1},\psi_{2})_{t} \vert ) \leq d(\boldsymbol{\psi_{1}}, \boldsymbol{\psi_{2}})^{2}
\end{displaymath}
\emph{is valid for any} $t \in \mathcal{R}$ \emph{and any} $\psi_{i} \in \boldsymbol{\psi_{i}}$. 

\vspace{1ex} 

PROOF. Let $\psi_{i}$ be any two representatives of $\boldsymbol{\psi_{i}}$ respectively. If
$\psi'_{i}\in \boldsymbol{\psi_{i}}$, then $\psi'_{i}= 
e^{i\sigma_{i}(t)}\psi_{i}$. After this one has $(\psi'_{1},\psi'_{2})_{t} = e^{i\sigma(t)}
(\psi_{1}, \psi_{2})_{t}$ with the relative phase $\sigma = \sigma_{2} - \sigma_{1}$. 
Denote the set of all differeniable real functions by $\mathcal{D}$. We have
\begin{displaymath}
\textrm{Re} (\psi'_{1}, \psi'_{2})_{t} = \textrm{Re} [ e^{i\sigma(t)}(\psi_{1},\psi_{2})_{t}] 
\leq \vert e^{i\sigma(t)}(\psi_{1},\psi_{2})_{t}\vert
\end{displaymath}

\vspace{-0.5cm}

\begin{displaymath}
 = \vert (\psi_{1},\psi_{2})_{t} \vert.
\end{displaymath}
By this one has
\begin{displaymath}
\inf_{t \in {\mathcal{R}}} \textrm{Re} [e^{i\sigma(t)}(\psi_{1}, \psi_{2})_{t}] 
\leq \inf_{t \in {\mathcal{R}}} \vert (\psi_{1},\psi_{2})_{t} \vert.
\end{displaymath}
So we have then
\begin{displaymath}
\sup_{\sigma \in {\mathcal{D}}} \inf_{t \in {\mathcal{R}}} \textrm{Re} [e^{i\sigma(t)}(\psi_{1},\psi_{2})_{t}] 
\leq \inf_{t \in {\mathcal{R}}} \vert f(t) \vert, 
\end{displaymath} 
and get the inequality 
\begin{equation}\label{f}
\sup_{\psi'_{i} \in \boldsymbol{\psi_{i}}} \inf_{t \in \mathcal{R}} \textrm{Re}(\psi'_{1}, 
\psi'_{2})_{t} \leq \inf_{t \in \mathcal{R}} \vert (\psi_{1}, \psi_{2})_{t} \vert.
\end{equation}
On the other hand one gets (for unit rays $\boldsymbol{\psi_{i}}$), using (\ref{f-g}) (where $d$ stands for 
$d(\boldsymbol{\psi_{1}}, \boldsymbol{\psi_{2}})$) 
\begin{displaymath}
d^{2} = \inf_{\psi'_{i} \in \boldsymbol{\psi_{i}}} \sup_{t \in \mathcal{R}} \Vert \psi'_{1} - \psi'_{2} \Vert_{t}^{2} =
\end{displaymath}

\vspace{-0.5cm}

\begin{displaymath}
 \inf_{\psi'_{i} \in \boldsymbol{\psi_{i}}} \sup_{t \in \mathcal{R}} 2(1 - \textrm{Re}(\psi'_{1}, \psi'_{2})_{t}) =
\end{displaymath}

\vspace{-0.5cm}

\begin{displaymath}
\inf_{\psi'_{i} \in \boldsymbol{\psi_{i}}} 2(1 - \inf_{t \in \mathcal{R}} \textrm{Re} (\psi'_{1}, \psi'_{2})_{t}) =
\end{displaymath}

\vspace{-0.5cm}

\begin{displaymath}
2(1 - \sup_{\psi'_{i} \in \boldsymbol{\psi_{i}}} \inf_{t \in \mathcal{R}} \textrm{Re} (\psi'_{1}, \psi'_{2})_{t}).
\end{displaymath}
From (\ref{f}) we get
\begin{displaymath}
d(\boldsymbol{\psi_{1}}, \boldsymbol{\psi_{2}})^{2} = 2(1- \sup_{\psi'_{i} \in \boldsymbol{\psi_{i}}} \inf_{t \in \mathcal{R}} 
\textrm{Re} (\psi'_{1}, \psi'_{2})_{t}) \geqslant 
\end{displaymath}

\vspace{-0.5cm}

\begin{displaymath}
\geqslant 2(1 - \inf_{t \in \mathcal{R}} \vert (\psi_{1}, \psi_{2})_{t} \vert \geqslant 2(1 - \vert (\psi_{1}, \psi_{2})_{t} \vert, 
\end{displaymath}
for any $t \in \mathcal{R}$. 

\vspace{1ex}

Note that the above proof is not valid for sets $\boldsymbol{S}_{i} = \{\tau(t)\psi_{i},
\vert \tau \vert =1 \} \subset \mathfrak{S}$ which are not equal to any rays. This is because the 
expression $\vert (\psi_{1},\psi_{2})_{t} \vert$ is time dependent in general if $\psi_{i} \in \boldsymbol{S_{i}}$,
and in general the elements $\psi_{i} \in \boldsymbol{S}_{i}$ cannot be normed, 
see the preceding subsection. 

From the Lemma 1 and the Schwarz'-like inequality follows that $d(\boldsymbol{\psi_{1}}, \boldsymbol{\psi_{2}}) = 0$
if and only if $\boldsymbol{\psi_{1}} = \boldsymbol{\psi_{2}}$. 

Making use of the triangle inequality
\begin{displaymath}
\Vert \psi_{1} - \psi_{3} \Vert_{t} \leq \Vert \psi_{1} - \psi_{2} \Vert_{t} + \Vert \psi_{2} - \psi_{3} \Vert_{t}
\end{displaymath}
one can show the triangle inequality for $d$. Indeed, taking the supremum of both sides of the above triangle inequality
with respect to $t \in \mathcal{R}$ and then the infimum with respect to $\psi_{1} \in \boldsymbol{\psi_{1}}$
and $\psi_{3} \in \boldsymbol{\psi_{3}}$ we get (recall that $\inf_{x \in A, y \in B}f(x,y) = \inf_{x \in A} \inf_{y \in B} f(x,y)$)
\begin{displaymath}
d(\boldsymbol{\psi_{1}}, \boldsymbol{\psi_{3}}) \leq 
\inf_{\psi_{1} \in \boldsymbol{\psi_{1}}}\sup_{t \in \mathcal{R}} \Vert \psi_{1} -\psi_{2} \Vert_{t} +
\end{displaymath}

\vspace{-0.5cm}

\begin{displaymath}
+\inf_{\psi_{3} \in \boldsymbol{\psi_{3}}}\sup_{t \in \mathcal{R}} \Vert \psi_{2} - \psi_{3} \Vert_{t}.  
\end{displaymath}
Now, taking the infimum of the first term on the right hand side with respect to $\psi_{2} \in \boldsymbol{\psi_{2}}$
and then the same infimum of the second term in the inequality (and making use of the inequality) we get
the triangle inequality: $d(\boldsymbol{\psi_{2}}, \boldsymbol{\psi_{3}}) \leq 
d(\boldsymbol{\psi_{1}}, \boldsymbol{\psi_{2}}) + d(\boldsymbol{\psi_{2}},\boldsymbol{\psi_{3}})$.  
We have proved in this way that the rays $\boldsymbol{\psi}$ with $d(\boldsymbol{\psi_{1}}, \boldsymbol{\psi_{2}})$
form a metric space.

\vspace{1ex}

LEMMA 2. \emph{The scalar product} $\boldsymbol{\psi_{1} \centerdot \psi_{2}}$ \emph{is continuous wih respect
to the metric} $d(\boldsymbol{\psi_{1}}, \boldsymbol{\psi_{2}})$.

\vspace{1ex}

PROOF. Consider three unit rays $\boldsymbol{\psi}, \boldsymbol{\psi_{1}}, \boldsymbol{\psi_{2}}$. Let 
$\psi$ and $\psi_{1}, \psi_{2}$ be any representatives of the rays.  Then, one has
\begin{displaymath}
\vert \boldsymbol{\psi \centerdot \psi_{1}} -\boldsymbol{\psi \centerdot \psi_{2}} \vert = 
\Big\vert \, \, \sup_{t \in \mathcal{R}}\vert (\psi, \psi_{1})_{t} \vert - 
\sup_{t \in \mathcal{R}}\vert (\psi, \psi_{2})_{t} \vert \, \, \Big\vert \leq
\end{displaymath}

\vspace{-0.5cm}

\begin{displaymath}
\Big\vert \sup_{t \in \mathcal{R}}\{\vert (\psi, \psi_{1})_{t} \vert - \vert (\psi, \psi_{2})_{t} \vert \} \Big\vert =
\sup_{t \in \mathcal{R}} \vert \, \, \vert (\psi, \psi_{1})_{t} \vert - \vert (\psi, \psi_{2})_{t} \vert \, \, \vert  
\end{displaymath} 

\vspace{-0.5cm}

\begin{displaymath}
\leq \sup_{t \in \mathcal{R}}\vert (\psi, \psi_{1})_{t} - (\psi, \psi_{2})_{t} \vert = 
\sup_{t \in \mathcal{R}}\vert (\psi, \psi_{1} - \psi_{2})_{t} \vert.
\end{displaymath}
But from Schwarz'-like inequality one has
\begin{displaymath}
\sup_{t \in \mathcal{R}}\vert (\psi, \psi_{1} - \psi_{2} )_{t} \vert \leq 
\sup_{t \in \mathcal{R}}\Vert \psi_{1} - \psi_{2} \Vert_{t}
\end{displaymath}
for any representatives $\psi_{i} \in \boldsymbol{\psi_{i}}$,
(note that $(\psi,\psi)_{t} = \Vert \psi \Vert_{t}$ is a well defined number constant in time and equal to 1,
because $\psi$ is a representative of a ray). Because the inequality is valid for all representatives $\psi_{i}$,
so taking the infimum of the right hand side with respect to $\psi_{i} \in \boldsymbol{\psi_{i}}$ we finally get 
\begin{displaymath}
\vert \boldsymbol{\psi \centerdot \psi_{2}} - \boldsymbol{\psi \centerdot \psi_{2}} \vert \leq
d(\boldsymbol{\psi_{1}}, \boldsymbol{\psi_{2}}).
\end{displaymath}
Repeating the same arguments one shows
\begin{displaymath}
\vert \boldsymbol{\psi_{1} \centerdot \psi_{2}} - \boldsymbol{\psi_{3} \centerdot \psi_{4}} \vert \leq
d(\boldsymbol{\psi_{1}}, \boldsymbol{\psi_{3}}) + d(\boldsymbol{\psi_{2}}, \boldsymbol{\psi_{4}}).
\end{displaymath}
 
\vspace{1ex}

Now we define the \emph{operator ray} 
\begin{displaymath}
{\boldsymbol{T}}=\{\tau T , \tau=\tau(t) \, and \, \vert\tau \vert = 1\}
\end{displaymath} 
which corresponds to the operator $T$. Any $T\in {\boldsymbol{T}}$ will be called a \emph{representative} of the 
ray ${\boldsymbol{T}}$. 

In the sequel only such operators $T$ (acting in the space $\mathfrak{S}$ 
of waves $\psi$) will be considered for which 
\begin{displaymath}
(T\psi_{1},T\psi_{2})_{t} = (\psi_{1},\psi_{2})_{t} \, \, \textrm{if} \, \, (\psi_{1},\psi_{2})_{t} = const = (\psi_{1}, \psi_{2}),
\end{displaymath}
for $\psi_{i} \in \mathfrak{S}$, and such that $T$ transforms rays into rays (note that our transformations $T_{r}$ in
(\ref{trapsi}) fulfils the assumptions). Of course any operator $T_{r}$ of a representation of a covariance group has to
transform rays into rays and "Schr\"odinger pictures" into "Schr\"odinger pictures".
A comment is necessary on the assumption. It could seem that
the more simple assumption can be applied, namely that $(T\psi_{1},T\psi_{2})_{t} = (\psi_{1}, \psi_{2})_{t}$ 
for all $\psi_{i} \in \mathfrak{S}$. But this is not the case. Such an assumption is too strong. 
Consider for example the transformations
(\ref{trapsi}), and the wave function $\psi$ and its transform $\psi'$ such that $(\psi,\psi')_{t}$ depends on $t$. Then
apply the time translation $T$ (according to (\ref{trapsi})), after this $(T\psi,T\psi')_{t} \ne (\psi,\psi')_{t}$ 
in general. This, however, poses no difficulty because this time $\psi$ and $\psi'$ do not belong to the same 
"Schr\"odinger picture" and $\vert (\psi,\psi')_{t} \vert$ cannot be interpreted as a transition probability. On the 
other hand if the two waves $\psi_{1}$ and $\psi_{2}$ both are members of the same "Schr\"odinger picture"
then $\vert(\psi_{1}, \psi_{2})_{t}\vert = \vert(\varphi_{1},\varphi_{2})\vert$ has the interpretation of the transition 
probability and is constant, but according to our assumption the transition probability is the same as
in the transformed frame: $(T\psi_{1},T\psi_{2})_{t} = (\psi_{1},\psi_{2})_{t}$. This is natural (and necessary
from a quite general point of view) that the transition probability between two physical states mesured 
by any observer should be the same. 
\\ Note, that in particular $(T\psi, T\psi)_{t} = (\psi,\psi)_{t} = 1$ for any element $\psi$ of any ray.
   
The product ${\boldsymbol{TV}}$ is defined as the set of all products $TV$ such that
$T\in {\boldsymbol{T}}$ and $V \in {\boldsymbol{V}}$. 

After this the condition (\ref{rowt}) can be written in the following equivalent form
\begin{displaymath}
\boldsymbol{T}_{r}\boldsymbol{T}_{s}=\boldsymbol{T}_{rs}.
\end{displaymath}
Any representation ${\boldsymbol{T}}_{r}$ (\emph{i.e.} a mapping $r \mapsto {\boldsymbol{T}}_{r}$ of $G$ into the 
operator rays) fulfilling the condition will be called a \emph{ray representation}. Because $T_{r}$ transforms rays 
into rays, we have $T_{r}(e^{i\xi(t)}\psi)=e^{i\xi_{r}(t)}T_{r}\psi$. In the sequel we assume that the the operators
$T_{r}$ are such that   $\xi_{r}(t)=\xi(r^{-1}t)$. Note again, that our operators $T_{r}$ in (\ref{trapsi}) fulfil it.    

Now we make the last assumption, namely the assumption that all transition probabilities vary continuously with
the continuous variation of the coordinate transformation $s \in G$:
\begin{enumerate}
\item[(D)] For any element $r$ in $G$, any ray ${\boldsymbol{\psi}}$ and any positive $\epsilon$ there exists a 
neighborhood $\mathfrak{N}$ of $r$ on $G$ such that 
$d(\boldsymbol{T}_{s}\boldsymbol{\psi},\boldsymbol{T}_{r}\boldsymbol{\psi}) < \epsilon$ if $s \in \mathfrak{N}$.
\end{enumerate}
We consider a Lie group $G$ in this subsection and the meaning of the word 'neighborhood' is clear. In the 
case of the group of transformations (\ref{tra}) the meaning has to be defined, but we will do it later on, see 
subsection {\bf V.E}.

Note that the assumption (D) is, in some sense, "natural" if the group $G$ is a symmetry group. In such a case 
$T_{r}\psi$ together with $\psi$ belongs to the same "Schr\"odinger picture" and the transition
from the state $\psi$ to the state $T_{r}\psi$ is possible such that the words 'transition probability
betwen $T_{r}\psi$ and $T_{s}\psi$' are meaningful. If the group $G$ is a covariance group,
then the above two wave fuctions do not belong to the same "Schr\"odinger picture" and in general the 
meaning of the above transition probability is not well defined in this case. 
But, (D) is still a "natural" assumption. This is because the operators $T$ of the representation
cannot depend on the dynamical objects of the spacetime geometry (in our case they cannot depend on 
the gravitational potential $\phi$). Indeed, $T = T(\phi,r)$ depending on $\phi$ would depend on
the coordinate frame in addition to the transformation $r$ contrary to the fact that $T(l,rl) = T_{r}$,
see \cite{Wigner}, because the potential has different 
numerical forms in different frames. So, the representation $T$ of
$G$ is universal in the sense that it has a common form for all spacetimes (all gravitational 
potentials in our case). So, one cannot exclude that for any fixed $r$ and $s$ there exists a potential
such that the two $T_{r}\psi$ and $T_{s}\psi$ do belong to the same "Schr\"odinger picture"  (of course
it can happen that it is not the case for any $s' \ne s$ and $r' \ne r$).
This justifies a "naturality" of (D). 

Basing on the assumption (D) one can prove the following 

\vspace{1ex}

THEOREM 2. \emph{Let ${\boldsymbol{T}}_{r}$ be a continuous ray representation of a group $G$. For all 
$r$ in a suitably chosen neighborhood $\mathfrak{N}_{0}$ of the unit element $e$ of $G$ one may select 
a strongly continuous set of representatives $T_{r}\in {\boldsymbol{T}}_{r}$. That is, for any compact set 
$\mathcal{C} \subset \mathcal{R}$, any wave function $\psi \in \mathfrak{S}$, any $r \in {\mathfrak{N}}_{0}$ 
and any positive $\epsilon$ there exists a neighborhood $\mathfrak{N}$ of $r$ such that $\Vert T_{s}\psi 
- T_{r}\psi \Vert_{t} < \epsilon$ if $s \in \mathfrak{N}$ and $t \in \mathcal{C}$}.     

\vspace{1ex}

PROOF. (We follow closely Bargmann's-Wigner's exposition \cite{Bar}, the very small modification
is due to the time dependence of the phase factor). For each ray $\boldsymbol{\psi}$ we select one and
only one representative $\psi$. Let $\boldsymbol{\psi_{o}}$ be a fixed ray and $\psi_{o}$ its representative
chosen in accordance with our selection. Set $\rho_{r} = \boldsymbol{\psi_{o} \centerdot T_{r}\psi_{o}}$. 
By (D) and Lemma 2 $\rho_{r}$ is a continuous function of $r$; note also that $\rho_{e} = 1$. Then one can
choose for a fixed positive $\alpha, \, 0< \alpha < 1$, a neighborhood $\mathfrak{N}_{0}$ of $e$ such that 
$\alpha <\rho_{r} \leq 1$ for every $r \in \mathfrak{N}_{0}$. Therefore we may select a uniquely determined 
representative $T_{r} \in \boldsymbol{T_{r}}$ for $r \in \mathfrak{N}_{0}$ which fulfils the equation 
\begin{equation}\label{1.7}
(\psi_{o}, T_{r}\psi_{o})_{t} =\rho_{r} = \boldsymbol{\psi_{o} \centerdot T}_{r}\boldsymbol{\psi_{o}},
\end{equation}   
at each time $t$ because $\rho_{r}$ is a number $\ne$ 0 for $r \in \mathfrak{N}_{0}$. From the Schwarz'-like 
inequality one has $T_{e}\psi_{o} = \psi_{o}$ and because $\boldsymbol{T}_{e}  = \boldsymbol{1}$, $T_{e} = 1$. 
Note, that (\ref{1.7}) is meaningful, because $T_{r}$ transforms rays into rays and 
$\boldsymbol{\psi_{o} \centerdot T}_{r}\boldsymbol{\psi_{o}}$ is a well defined number.    
This is the definition of $T_{r}$. Now we will show its strong continuity.

We need some auxiliary relations. Let $\psi \in \boldsymbol{\psi}$, and set
\begin{equation}\label{1.8}
d_{r,s}(\psi) \equiv d(\boldsymbol{T}_{r}\boldsymbol{\psi}, \boldsymbol{T}_{s}\boldsymbol{\psi}), \,
\sigma_{r,s}(\psi) \equiv (T_{r}\psi, T_{s}\psi)_{t}, 
\end{equation}

\vspace{-0.5cm}

\begin{equation}\label{1.8a}
z_{r,s}(\psi) \equiv T_{s}\psi - \sigma_{r,s}(\psi) T_{r} \psi. 
\end{equation}
Clearly $z_{r,s}(\psi)$ is orthogonal to $T_{r}\psi$ in the sense that $(z_{r,s}(\psi), T_{r}\psi)_{t} = 0$
for any $t$, so, we have, using Lemma 1
\begin{displaymath}
(z_{r,s}(\psi), z_{r,s}(\psi))_{t} = \Vert z_{r,s}(\psi) \Vert_{t}^{2} = (T_{s}\psi, T_{s}\psi)_{t} -
\end{displaymath}

\vspace{-0.5cm}

\begin{displaymath}
-(T_{s}\psi, T_{r}\psi)_{t} \sigma_{r,s}(\psi) =(\psi, \psi)_{t} - \sigma_{r,s}(\psi)^{*}\sigma_{r,s}(\psi)=
\end{displaymath}

\vspace{-0.5cm}

\begin{equation}\label{1.9}
= 1 - \vert \sigma_{r,s}(\psi) \vert^{2} \leq d_{r,s}(\psi)^{2}. 
\end{equation}

First, we show the continuity of $T_{r}\psi_{o}$. By (\ref{1.8a}) and (\ref{1.7}) $(\psi_{o}, z_{r,s}(\psi_{o})_{t} =
(\psi_{o}, T_{s}\psi_{o})_{t} - \sigma_{r,s}(\psi_{o})(\psi_{o}, T_{r}\psi_{o})_{t} = \rho_{s} - \sigma_{r,s}(\psi_{o}) \rho_{r}$, so that 
$1 - \sigma_{r,s}(\psi_{o}) = (1/\rho_{r}) \{ \rho_{r} - \rho_{s} + (\psi_{o}, z_{r,s}(\psi_{o}))_{t} \}$. From Lemma 2
$\vert \rho_{r} - \rho_{s} \vert \leq d_{r,s}(\psi_{o})$. Thus we have by (\ref{f-g}), (\ref{1.7}) and (\ref{1.9}), 
using the Schwarz'-like inequality
\begin{displaymath}
\Vert T_{s}\psi_{o} - T_{r}\psi_{o} \Vert_{t} \leq 2 \vert 1 - \sigma_{r,s}(\psi_{o}) \vert \leq (4/\alpha)d_{r,s}(\psi_{o}),
\end{displaymath}
for all $t$, so for any $t \in \mathcal{C}$ (with any compact $\mathcal{C}\subset \mathcal{R}$).  
But from (D) for every positive $\epsilon$ there exists a neighborhood $\mathfrak{N}$ of $r$ such that
$(4/\alpha)d_{r,s}(\psi_{o})< \epsilon^{2}$ if $s \in \mathfrak{N}$.

Second, consider a unit wave function $\psi$ (from the selected set) orthogonal to 
$\psi_{o}$ (\emph{i.e.} $(\psi,\psi_{o})_{t} = 0$ for each time $t$). Set $\psi_{1} = (1/\sqrt{2})(\psi_{o} + \psi)$, so that
$\Vert \psi_{1} \Vert_{t}^{2} = constant = 1$. By (\ref{1.8a}) $(T_{r}\psi_{o},z_{r,s}(\psi_{1}))_{t} = (T_{r}\psi_{o} - 
T_{s}\psi_{o}, T_{s}\psi_{1})_{t} +(T_{s}\psi_{o},T_{s}\psi_{1})_{t}  - \sigma_{r,s}(\psi_{1})(T_{r}\psi_{o},T_{r}\psi_{1})_{t}$.    
Since $(\psi_{o},\psi_{1})_{t} = 1/\sqrt{2} = const$ we have 
$(T_{r}\psi_{o},T_{r}\psi_{1})_{t} = (\psi_{o},\psi_{1})_{t} = 1/\sqrt{2}$,
according to our assumption about $T_{r}$. So, we obtain 
\begin{displaymath}
1- \sigma_{r,s}(\psi_{1}) = \sqrt{2}\{(T_{r}\psi_{o},z_{r,s}(\psi_{1}))_{t} + 
\end{displaymath}

\vspace{-0.5cm}

\begin{displaymath}
+ (T_{s}\psi_{o} - T_{r}\psi_{o}, T_{s}\psi_{1})_{t}\}
\end{displaymath}
and thus, by (\ref{1.9}),
\begin{displaymath}
\Vert T_{s}\psi_{1} - T_{r}\psi_{1} \Vert_{t} \leq 2 \vert 1 - \sigma_{r,s}(\psi_{1}) \vert \leq
\end{displaymath}

\vspace{-0.5cm}

\begin{displaymath} 
\leq 2^{3/2}\{d_{r,s}(\psi_{1}) + \Vert T_{s}\psi_{o} - T_{r}\psi_{o} \Vert_{t} \},
\end{displaymath}  
for any time $t$, so  for any $t \in \mathcal{C}$. The continuity of $T_{r}\psi_{o}$, proved above, implies the continuity
of $T_{r}\psi_{1}$ and by this that of $T_{r}\psi = \sqrt{2}T_{r}\psi_{1} - T_{r}\psi_{o}$. 

Third, let $\psi_{2}$ be arbitrary wave function from the selected set of representatives. Set $\psi_{2} = 
\lambda_{1}\psi_{o} + \lambda_{2}\psi_{1}$ where $\psi_{1}$ is a unit wave vector (from the
selected set) orthogonal to $\psi_{o}$. Then $T_{r}\psi_{2} = \lambda_{1}T_{r}\psi_{o} +
\lambda_{2} T_{r}\psi_{1}$ and the continuity of $T_{r}\psi_{2}$ for any element $\psi_{2}$ of the 
selected set follows. 

Fourth, let $\widetilde{\psi}$ be any element of any ray $\boldsymbol{\psi} \subset \mathfrak{S}$. So, we have
$\widetilde{\psi} = e^{i\xi(t)}\psi$ for some element $\psi$ of the selected set. We have
\begin{displaymath}
\Vert T_{s}\widetilde{\psi} - T_{r}\widetilde{\psi} \Vert_{t} \leq \vert e^{i\{\xi_{s}(t) - \xi_{r}(t)\}} -1 \vert + \Vert T_{s}\psi
- T_{r}\psi \Vert_{t}.
\end{displaymath}
So, according to our assumption about $T_{r}$
\begin{equation}\label{1.10}
\Vert T_{s}\widetilde{\psi} - T_{r}\widetilde{\psi} \Vert_{t} \leq \vert e^{i(\xi(s^{-1}t) - \xi(r^{-1}t)\}} \vert +
\Vert T_{s}\psi - T_{r}\psi \Vert_{t}.
\end{equation} 
Because $\xi(t)$ is a differentiable function (and by this continuous) and the group
$G$ is a Lie group (and by this a continuous group, and $rt$ is a continuous function of $r$ and $t$) then, 
for any compact $\mathcal{C}$ and any $\epsilon'$ there exists a neighborhood 
$\mathfrak{N}'$ of $r$ such that $\vert e^{i\{\xi(s^{-1}t) - \xi(r^{-1}t)\}} -1\vert < \epsilon'$
if $s \in \mathfrak{N}'$ and $t \in \mathcal{C}$. Indeed. The function $F= F(s,t) \equiv \vert e^{i\{\xi(s^{-1}t)
- \xi(r^{-1}t)\}} - 1 \vert \geqslant 0$ is a continuous function of $(s,t)$, such that $F(r,t) = 0$. Let $\epsilon'$ be any
positive number and $\mathcal{C}$ any compact subset of $\mathcal{R}$. For any $t \in \mathcal{C}$
there exist a neighborhood ${\mathfrak{N}}_{t}$ of $r$ and a neighborhood ${\mathcal{C}}_{t}$ of $t$,
such that $F(s,t) \leq \epsilon'$ if $(r,t) \in \mathfrak{U}_{t}={\mathfrak{N}}_{t} \times {\mathcal{C}}_{t}$. Because
$\mathcal{C}$ is compact and 
\begin{displaymath}
{\mathcal{C}} \subset \bigcup \limits_{t \in \mathcal{C}} {\mathcal{C}}_{t}, 
\end{displaymath}
then
\begin{displaymath}
{\mathcal{C}} \subset {\mathcal{C}}_{t_{1}} \cup \ldots \cup {\mathcal{C}}_{t_{n}}
\end{displaymath}
for a finite sequence $t_{1}, ... , \, t_{n}$. 
Now we define
\begin{displaymath}
{\mathfrak{N}}' ={\mathfrak{N}}_{t_{1}} \cap \ldots \cap {\mathfrak{N}}_{t_{n}}.
\end{displaymath} 
After this $0 \leq F(s,t) < \epsilon'$ for any $(s,t) \in \mathfrak{N}' \times \mathcal{C}$. Note that the compactness of 
$\mathcal{C}$ is essential. From this and from the continuity of $T_{r}\psi$ and the inequality (\ref{1.10})
the continuity of $T_{r}\widetilde{\psi}$ for any element $\widetilde{\psi}$ of any ray follows. 

Fifth, because any $\psi \in \mathfrak{S}$ is a linear combination of elements of rays and the sum of 
continuous functions is a continuous function, then the continuity of $T_{r}\psi$ for any $\psi \in \mathfrak{S}$
follows. 

\vspace{1ex}

REMARK. Note that the Theorem 2 is true for any continuous group $G$, provided (D) is assumed for $G$
(the proof remains unchanged). Also in the case when $\xi = \xi(X)$ the proof remains unchanged, of course the 
conception of the ray has to be consistent with (\ref{rowX}) and with the time evolution of $\psi$ (or $\Psi$ 
in the Quantum Electrodynamics) \emph{i.e.} with the gauge freedom of the time evolution. 

\vspace{1ex}

The representatives $T_{r} \in \boldsymbol{T}_{r}$ selected as in the Theorem 2 will be called \emph{admissible}
and the representation $T_{r}$ obtained in this way the \emph{admissible} representation. 
There are infinitely many possibilities of such a selection of admissible representation 
$T_{r}$ (or infinitely many $\xi$-s in (\ref{rowt})), and this is our aim now to classify them. 
We confine ourselves to the local \emph{admissible} representations 
defined on a fixed neighborhood $\mathfrak{N}_{o}$ of $e \in G$, as in the Theorem 2. 

Let $T_{r}$ be an \emph{admissible} representation. With the help of the phase
$e^{i\zeta(r,t)}$ with a real function $\zeta(r,t)$ differentiable in $t$ and continous in $r$ we can define
\begin{equation}\label{4}
T'_{r} = e^{i\zeta(r,t)}T_{r},
\end{equation}
which is a new \emph{admissible} representation. This is trivial, if one defines in the appropriate way 
the continuity of $\zeta(r,t)$ in $r$. Namely, from the Theorem 2 it follows that the continuity has to be defined
in the folowing way. The function $\zeta(r,t)$ \emph{will be called strongly continuous in r at $r_{0}$ if and 
only if for any compact set $\mathcal{C}$ and any positive $\epsilon$ there exist a 
neighborhood ${\mathfrak{N}}_{0}$ of $r_{0}$ such that } 
\begin{displaymath} 
\vert \zeta(r_{0},t) - \zeta(r,t) \vert < \epsilon,
\end{displaymath}
\emph{for all r $\in {\mathfrak{N}}_{0}$ and for all $t \in \mathcal{C}$. } 
 But the converse is also true. Indeed,
if $T'_{r}$ also is an \emph{admissible} representation, then (\ref{4}) has to be fulfilled for a real function $\zeta(r,t)$
differentiable in $t$ because $T'_{r}$ and $T_{r}$ belong to the same ray, and moreover, because both 
$T'_{r}\psi$ and $T_{r}\psi$ are strongly continuous (in $r$ for any $\psi$) then $\zeta(r,t)$ has to be 
\emph{strongly continuous} (in $r$).

Let $T_{r}$ be an \emph{admissible} representation, and by this continuous in the sense indicated in the 
Theorem 1. One can always choose the above $\zeta$ in such a way that $T_{e} = 1$ as will be assumed 
it in the sequel. 

Because $T_{r}T_{s}$ and $T_{rs}$ belong to the same ray one has
\begin{equation}\label{5}
T_{r}T_{s} = e^{i\xi(r,s,t)} T_{rs}
\end{equation}    
with a real  function $\xi(r,s,t)$  differentiable in $t$ (the motivation for the differentiability assumption 
has been given above). From the fact that $T_{e} = 1$ we have
\begin{equation}\label{9}
\xi(e,e,t) = 0.
\end{equation}  
From the associative law $(T_{r}T_{s})T_{g} = T_{r}(T_{s}T_{g})$ one gets
\begin{equation}\label{10}
\xi(r,s,t) + \xi(rs,g,t) = \xi(s,g,r^{-1}t) + \xi(r,sg,t).
\end{equation} 
The formula (\ref{10}) is very important and our analysis largely rests on this relation.
From the fact that the representation $T_{r}$ is \emph{admissible} follows that 
the exponent $\xi(r,s,t)$ is continuous in $r$ and $s$. Indeed, take a $\psi$ belonging
to a unit ray $\boldsymbol{\psi}$, then making use of (\ref{5}) we get
\begin{displaymath}
e^{i\xi(r,s,t)}(T_{rs} - T_{r's'})\psi + (T_{r'}(T_{s'} - T_{s})\psi + (T_{r'} - T_{r})T_{s}\psi 
\end{displaymath}  

\vspace{-0.5cm}

\begin{displaymath}
= (e^{i\xi(r',s',t)} - e^{i\xi(r,s,t)}) T_{r's'}\psi.
\end{displaymath}
Taking norms $\Vert \centerdot \Vert_{t}$ of both sides, we get 
\begin{displaymath}
\vert e^{i\xi(r',s',t)} - e^{i\xi(r,s,t)} \vert \leq \Vert (T_{r's'} - T_{rs})\psi \Vert_{t} + 
\end{displaymath}

\vspace{-0.5cm}

\begin{displaymath}
+ \Vert T_{r'}(T_{s'} - T_{s})\psi \Vert_{t} + \Vert (T_{r'} - T_{r})T_{s}\psi \Vert_{t}.
\end{displaymath} 
From this inequality and the continuity of $T_{r}\psi$, the continuity of $\xi(r,s,t)$ in $r$ and $s$ follows.
Moreover, from the Theorem 2 and the above inequality the \emph{strong continuity} of 
$\xi(t,s,t)$ in $r$ and $s$ follows.  

The formula (\ref{4}) suggests the following definition. Two \emph{admissible} representations
$T_{r}$ and $T'_{r}$ are called \emph{equivalent} if and only if $T'_{r} = e^{i\zeta(r,t)}T_{r}$ 
for some real function $\zeta(r,t)$ differentiable in $t$ and \emph{strongly continuous} in $r$. So, making use of 
(\ref{5}) we get $T'_{r}T'_{s} = e^{i\xi'(r,s,t)}T'_{rs}$, where
\begin{equation}\label{13}
\xi'(r,s,t) = \xi(r,s,t) + \zeta(r,t) + \zeta(s,r^{-1}t) - \zeta(rs,t).
\end{equation}
Then the two exponents $\xi$ and $\xi'$ are equivalent if and only if (\ref{13}) is fulfilled with
$\zeta(r,t)$ \emph{strongly continuous} in $r$ and differentiable in $t$. 

From (\ref{9}) and (\ref{10}) immediatelly follows that
\begin{equation}\label{11}
\xi(r,e,t) = 0 \, \, \, and \, \, \, \xi(e,g,t) = 0,
\end{equation}

\vspace{-0.5cm}

\begin{equation}\label{12}
\xi(r,r^{-1},t) = \xi(r^{-1},r,r^{-1}t).
\end{equation}
The relation (\ref{13}) between $\xi$ and $\xi'$ will be written in short by 
\begin{equation}\label{13'}
\xi' = \xi + \Delta[\zeta].
\end{equation}
The relation (\ref{13}) between exponents $\xi$ and $\xi'$ is reflexive, symmetric and transitive. Indeed, 
we have: $\xi = \xi + \Delta[\zeta]$ with $\zeta =0$. Moreover, if $\xi' = \xi + \Delta[\zeta]$ then 
$\xi = \xi' + \Delta[-\xi]$. At last if $\xi' = \xi +\Delta[\zeta]$ and $\xi'' = \xi' + \Delta[\zeta']$, then 
$\xi''= \xi + \Delta[\zeta + \zeta']$. So the relation is an equivalence relation, and will be sometimes
denoted by $\xi' \equiv \xi$. The equivalence relation preserves the linear structure, that is 
if $\xi_{i} \equiv \xi'_{i}$ (with the appropriate $\zeta_{i}$-s) then $\lambda_{1} \xi_{1} + \lambda_{2}\xi_{2}
\equiv \lambda_{1}\xi'_{1} + \lambda_{2}\xi'_{2}$ (with $\zeta = \lambda_{1}\zeta_{1} + \lambda_{2}\zeta_{2}$).

As was mentioned, our main task of this section is the classification of $\theta(t,X)$ in the 
transformation law $T_{r}$ given by (\ref{trapsi}) for the Milne transformations $r$ of the form (\ref{tra}). Before
we proceed further on, we show that the correspondence $\xi \to \theta$ between 
$\xi$ and $\theta$ defined globally on the group of transformations (\ref{tra}) is one-to-one,
if only the corresponding $\theta$ does exist (the correspondence is up
to an irrelevant equivalence transformation of $\theta$). That is, not for all $\xi$ there exists the transformation
$T_{r}$ of the form (\ref{trapsi}) with the exponent equal to $\xi$, but if such a transformation does exist, then it
is unique (up to an irrelevant equivalence transformation). By this, the classification of $\xi$-s gives 
the full classification of all possible $\theta$-s. 
Strictly speaking this is true for the Milne group (\ref{tra}) 
with differentiable $\vec{A}(t)$ (up to any order). But it is sufficient for us, that it is true for 
any Lie subgroup $G$  of (\ref{tra}), which contains the subgroup of space translations 
(see {\bf IV} and {\bf V. C}). Let $G$ be such a Lie subgroup of (\ref{tra}). 
Note, that instead of the wave functions $\psi(\vec{x},t)$ one can investigate the 
appropriately defined functions $f(g,t)$,
where $g \in G$. We make the definition below. 
Let us introduce the group coordinates on $G$, in general the whole atlas of coordinate maps
has to be used, of course. But taking into account the structure of the group (\ref{tra}),
$G$ is a semidirect product $G_{ST}\centerdot G_{1}$ of the space translation group 
$G_{ST}$ and a Lie group $G_{1}$. So, there exists an atlas of coordinates $(u, y_{a})$
 with the global coordinates $u \in {\mathcal{R}}^{3}$ for $G_{ST}$, and $y_{a}$ 
the coordinates of a map $(a)$ belonging to an atlas of $G_{1}$.
Assume the coordinates on $G$ to be defined by such an atlas, fixed once for all. 
Consider a \emph{galilean} reference frame in which the spacetime coordinates are $(\vec{x},t)$.
There is a biunique correspondence between $\vec{x}$ and the set $S_{\vec{x}}$ of such elements $g \in G$, 
that the space translation coordinates $u$ of $g$ are equal to $\vec{x}$.  
We define the function $f(g,t)$ to correspond uniquely to $\psi$
if and only if $f(S_{\vec{x}},t) = \psi(\vec{x},t)$, that is, for any $g \in S_{\vec{x}}$ $f(g,t) = \psi(\vec{x},t)$.
After this the transformation $T_{r}$ takes on
the following form
\begin{equation}\label{traf}
T_{r}f(u,t) = e^{- i\vartheta(r, u, t)}f(r^{-1}u, r^{-1}t),
\end{equation} 
where $\vartheta(r,S_{\vec{x}}, t) = \theta(t, \vec{x},t)$. Similarily,
\begin{displaymath}
(\psi_{1}, \psi_{2})_{t} = \int_{G_{ST}} f_{1}^{*}(g,t) f_{2}(g,t) \, {\ud}'g \equiv (f_{1},f_{2})_{t},
\end{displaymath}
where $G_{ST}$ stands for the space translation subgroup of $G$ and ${\ud}'g$ is the 
left invariant measure on $G_{ST}$ induced by the left invariant measure ${\ud}g$ on $G$.
So we have constructed another realization of $\mathfrak{S}$, which will be denoted by the same 
symbol $\mathfrak{S}$. 
We will show that there is a biunique correspondence between
equivalence classes of $\vartheta$ (or $\theta$) and equivalence classes of exponents $\xi$
for the group $G$. It is trivial, of course, that for any transformation $T_{r}$ defined by
$\vartheta$ there exist uniquely determined exponent $\xi$, which can be easily computed to be
equal
\begin{equation}\label{xivtheta}
\xi(r,s,t)  = \vartheta(rs,g,t) - \vartheta(s,r^{-1}g,r^{-1}t) - \vartheta(r,g,t),  
\end{equation}
and the dependence on $g$ cancels on the right hand side. 
But also conversely, if $\vartheta$ defines a strongly continuous representation $T_{r}$ with 
the exponent $\xi$, then $\vartheta$ is determined up to
an equivalence transformation. Indeed, because $T_{r}$ corresponding to $\vartheta$ is a ray representation
with the exponent equal to $\xi$, then $\vartheta$ has to fulfil Eq. (\ref{xivtheta}).
Substituting $s = r^{-1}g$ to the equation (\ref{xivtheta}) we get
\begin{displaymath}
\vartheta(r,g,t) = - \xi(r,r^{-1}g,t) + \vartheta(g,g,t) - \vartheta(r^{-1}g,r^{-1}g,r^{-1}t) 
\end{displaymath}

\vspace{-0.5cm}

\begin{equation}\label{xitheta}
\equiv - \xi(r,r^{-1}g,t) + \Delta\vartheta(r,g,t).
\end{equation}
Because $\xi$ is an exponent it fulfils (\ref{10}) and it can be shown that the function 
$\widetilde{\vartheta}(r,g,t) = - \xi(r,r^{-1}g,t)$ fulfils the equation (\ref{xivtheta}), so the function
$\Delta \vartheta(r,g,t) = \vartheta(g,g,t) - \vartheta(r^{-1}g,r^{-1}g,r^{-1}t)$ has to fulfil the equation
\begin{displaymath}
\Delta\vartheta(rs,g,t) - \Delta\vartheta(s,r^{-1}g, r^{-1}t) - \Delta\vartheta(r,g,t) = 0,
\end{displaymath}
which is indeed, identically fulfilled. We show that the two functions $\vartheta(r,g,t)$
and $\vartheta'(r,g,t) = \vartheta(r,g,t) +\vartheta(g,g,t) - \vartheta(r^{-1}g, r^{-1}g,r^{-1}t)$
$= \vartheta(r,g,t) + \Delta\vartheta(r,g,t)$
define two equivalent representations $T_{r}$ and $T_{r}'$, given by (\ref{traf}).
That is, from (\ref{xitheta}) it follows that $\vartheta(r,g,t) = - \xi(r,r^{-1}g,t)$ up to an irrelevat
equivalence transformation.  

One can simply "reparametrize" the space $\mathfrak{S}$, and this 
"reparametrization" will induce a trivial "reparemetrization" of the form of  $T_{r}$: $T_{r} \to T_{r}'$.
Namely, consider a linear isomorphism $L$ of the space $\mathfrak{S}$: $L: \mathfrak{S} \to
\mathfrak{S}$, such that 
\begin{enumerate}
\item[(i)] $(Lf_{1}, Lf_{2})_{t} = (f_{1},f_{2})_{t}$, if $(f_{1},f_{2})_{t}$ does not
depend on the time $t$,
\item[(ii)] $L(e^{\zeta(t)}f) = e^{i\zeta(t)}Lf$,
\end{enumerate}
where we consider the action of $L$ in the space $\mathfrak{S}$ of $f(g,t)$ constructed above. 
Then the isomorphism induces a new form $T_{r}' = LT_{r}L^{-1}$ of the ray representation $T_{r}$ 
with exactly the same exponent $\xi$:
\begin{displaymath}
T_{r}'T_{s}' = e^{i\xi(r,s,t)}T_{rs}'.
\end{displaymath}
The two representations $T_{r}$ and $T_{r}' = LT_{r}L^{-1}$ are equivalent, or loosely speaking
they are "unitarily" equivalent. Compare to the notion of equivalence of two unitary representations \cite{Gelfand}.  
Take for example $L: f(g,t) \to e^{- i\vartheta(g,g,t)}f(g,t) = f'(g,t)$, 
then $T_{r}' = LT_{r}L^{-1}$ has the form (\ref{traf}) with 
\begin{displaymath}
\vartheta'(r,g,t) = \vartheta(r,g,t) + \vartheta(g,g,t) - \vartheta(r^{-1}g,r^{-1}g,r^{-1}t)  
\end{displaymath}

\vspace{-0.5cm}

\begin{displaymath}
= \vartheta(r,g,t) + \Delta\vartheta(r,g,t). 
\end{displaymath}
The space $L(\mathfrak{S})$ is another realization of $\mathfrak{S}$ obtained by a trivial
"rephasing" of  its elements by a phase common for all elements. Similarily, the representation $T_{r}'$
is only a "reparametrized" form of the representation $T_{r}$ and should not be treated as a new
representation essentially different from $T_{r}$.
Note, that in this new realisation $L(\mathfrak{S})$
of $\mathfrak{S}$ the condition $f'(S_{\vec{x}},t) = \psi(\vec{x},t)$ is no longer true in general.
Summing up, the correspondence between equivalence classes of $\vartheta$-s and $\xi$-s is one-to-one,
that is, there always exists the transformation $T_{r}$ of the form (\ref{traf}) with the \emph{a priori}
given exponent $\xi$, defined by $\vartheta(r,g,t) = - \xi(r,r^{-1}g,t)$. However, such a $T_{r}$ acts in the
space $\mathfrak{S}$ of functions $f(g,t)$ which depends on all group coordinates of $g$ and the identification
$\psi(\vec{x},t)  = f(S_{\vec{x}},t)$ is impossible in general. But it will be sometimes possible 
to connect with this representation
(of the form (\ref{traf})) the representation of the form (\ref{trapsi}). It is readily seen from our discussion, 
that this is the case if there exists such a function $\phi(g,t)$ that $e^{-i\phi(g,t)}f(g,t) = f'(g,t)$
as well as $\vartheta(t,g,t) + \Delta \phi(r,g,t)$ are constant along $g \in S_{\vec{x}}$. Indeed, we can 
construct a representation $T'_{r}$ equivalent to the initial one obtained
by the rephasing $f \to e^{-i\phi}f$ of the space $\mathfrak{S}$. The new representation $T'_{r}$
fulfils the condition: $f'(g_{1},t) = f'(g_{2},t)$ for all $g_{i} \in S_{\vec{x}}$, and by this the counterpart
representation $T_{r}$ of the form (\ref{trapsi}) can be built up, or equivalently the $\theta(r,X)$ can be
built up. The construction of the "$\theta$-type" representation is unique up to the equivalence relation.
Indeed, the phase function $\phi(g,t)$ described above is not unique. One can always add an additive
term $\phi'(g,t)$ to $\phi(g,t)$, but then, the additive term $\phi'(g,t)$ has to be constant along $g \in S_{\vec{x}}$.
This means that $\theta(r,X)$ in (\ref{trapsi}) is determined up to an additive term 
$\Delta \varphi = \varphi(X) - \varphi(r^{-1}X)$. But $\theta' = \theta + \Delta \varphi$ defines the representation
$T'_{r}$ (of the form (\ref{trapsi})) equivalent to the representation $T_{r}$ (of the type (\ref{trapsi})) defined
by $\theta$. The representation defined by $\theta'$ is obtained from the representation defined by $\theta$
by the trivial rephasing $\psi(X) \to e^{-i\varphi(X)}\psi(X)$ of the space $\mathfrak{S}$ of wave functions $\psi(X)$. 
That is, if there corresponds a $\theta$ to a given $\xi$, then the correspondence is unique up to a trivial
equivalence relation -- the rephasing of $\mathfrak{S}$. 
So, we have explained the motivation: to classify all $\xi$'s and by this to get the full classification of $\theta$'s
in (\ref{trapsi}). The classification is full in the sense that no $\theta$ can be omitted in it, 
but of course, there is the problem that no $\theta$ can exists for a given $\xi$. We solve
the problem for the Galilean group and the Milne group, see {\bf V.B} and {\bf V.C}.  
 There is of course another problem, if a given local $\xi$ can be extended on the whole 
group, we will investigate this problem later on (see Theorems 4 and 5).   

Having given the motivation, let us pass to the classification problem of all local $\xi$. 
We introduce now the group $H$, the very important notion for the further investigations. It is evident
that all operators $T_{r}$ contained in all rays $\boldsymbol{T}_{r}$ form a group under multiplication.
Indeed, consider an \emph{admissible} representation $T_{r}$ with a well defined $\xi(r,s,t)$ in the formula (\ref{5}).
Because any $T_{r} \in \boldsymbol{T}_{r}$ has the form $e^{i\theta(t)}T_{r}$ (with a real and differentiable
$\theta$, which should not be confused with the $\theta(r,X)$ in (\ref{trapsi})) one has
\begin{displaymath}
\lgroup e^{i\theta(t)}T_{r}\rgroup \lgroup e^{i\theta'(t)}T_{s}\rgroup = e^{i\{\theta(t) + \theta'(r^{-1}t) + \xi(r,s,t)\}}T_{rs}.
\end{displaymath}
This important relation suggest the following definition of the group $H$ connected with the \emph{admissible}
representation or with the exponent $\xi(r,s,t)$.
Namely, $H$ consists of the 
pairs $\{\theta(t), r\}$ where $\theta(t)$ is a differentiable real function and $r \in G$. The multiplication
rule, suggested by the above relation, is defined as follows
\begin{equation}\label{15}
\{\theta(t),r\} \centerdot \{\theta'(t),r'\} = \{\theta(t) + \theta'(r^{-1}t) + \xi(r,r',t), \, rr' \}.
\end{equation}
The associative law for this multiplication rule is equivalent to (\ref{10}) (in a complete
analogy with the classical Bargmann's theory). The pair 
$\check{e} =\{0,e\}$ plays the role of the unit element in $H$. For any element $\{\theta(t), r\} \in H$ there 
exists the inverse $\{\theta(t),r\}^{-1} = \{-\theta(rt) - \xi(r,r^{-1},rt), \, r^{-1} \}$. Indeed, from (\ref{12}) it follows that 
$\{\theta, r\}^{-1} \centerdot \{\theta,r \} = \{\theta, r\} \centerdot \{\theta, r\}^{-1} = \check{e}$. The elements
$\{\theta(t), e\}$ form an abelian subgroup $T$ of $H$. Any $\{\theta,r\} \in H$ can be uniquely written as
$\{\theta(t),r\} = \{\theta(t),e\} \centerdot \{ 0, r\}$. Also the same element can be uniquely expressed in the form 
$\{\theta(t),r\} = \{0,r\} \centerdot \{\theta(rt),e\}$. So, we have $H = T \centerdot G = G \centerdot T$. 
The abelian subgroup $T$ is a normal factor subgroup of $H$. But this time $G$ does not form any normal 
factor subgroup of $H$ (contrary to the classical case investigated by Bargmann, when the exponents are 
time independent). So, this time $H$ is not direct product of $T$ and $G$, but it is a 
semidirect product of $T$ and $G$, see e.g. \cite{Nachbin} where the semidirect 
product of two continuous groups is investigated in detail. 
In this case however the theorem that $G$ is locally isomorphic
to the factor group $H/T$ is still valid, see \cite{Nachbin}. Then the group $H$ composes a
\emph{semicentral extension} of $G$ and not a central extension of $G$ as in the Bargmann's theory.
We introduce the explicit definition
of a topology in which the multiplication rule (\ref{15}) is continuous.
The topology is not arbitrary, and it has to be such a topology
which assures the  \emph{strong continuity} of $\xi(r,s,t)$ in $r$ and $s$. But on the other hand 
the topology cannot be more restrictive, because some exponents would be omitted. Such a topology
is uniquely determined. 
As we have seen $H$ is a semidirect product  $T\centerdot G$
of the abelian group $T$ and $G$. So, it is sufficient to introduce a topology in $T$ and in $G$ 
separately and define the topology of $H$ as the \emph{semi-cartesian} product of $T$ and $G$,
see \cite{Nachbin}.
It is sufficient to introduce such a topology in $T$ that the following three operations
are continuous: (1) addition of elements of $T$, (2) number multiplication: $(\lambda, \theta(t)) \to \lambda \theta(t)$,
(3) the translation: $(r,\theta(t)) \to \theta(r^{-1}t)$ (as we will see the continuity of 
$(\lambda,\theta(t)) \to \theta(t - \lambda)$ is sufficient). In addition, from the continuity of (\ref{15})
the \emph{strong continuity} of $\xi(r,s,t)$ should follow (recall, that \emph{strong} continuity of $\xi$
is a consequence of Theorem 2). But on the other hand the topology cannot be more restrictive  
-- no $\xi$ fulfilling the conditions of Theorem 2 can be ommited if one wants to get the full classification of $\xi$'s. 
That is $T$ should be a topological
linear space of differentiable functions $\theta(t)$ in which the translation is continuous and the 
convergence $\theta_{n} \to \theta$ is equivalent to the strong convergence:
\emph{for any compact set $\mathcal{C}$ and any positive $\epsilon$ there exist a number $n_{0}$
such that} $\vert \theta_{n}(t) - \theta(t) \vert < \epsilon$ \emph{for all $n > n_{0}$ and for all $t \in  \mathcal{C}$ }.
Such a topology is given by the metric 
\begin{equation}\label{topology}
d(\theta_{1}, \theta_{2}) = 
\max_{n \in {\mathcal{N}}} \frac{2^{-n}p_{n}(\theta_{2} - \theta_{1})}{1 + p_{n}(\theta_{2} - \theta_{1})},
\end{equation}
where 
\begin{displaymath}
p_{n}(\theta) = \sup_{t \in {\mathcal{C}}_{n}} \vert \theta(t) \vert,
\end{displaymath}
where ${\mathcal{C}}_{n}$ are compact sets such that ${\mathcal{C}}_{1} \subset {\mathcal{C}}_{2} \subset \ldots$
and $\bigcup_{n \in {\mathcal{N}}} {\mathcal{C}}_{n} = {\mathcal{R}}$   
(compare any handbook of Functional Analysis). 

The rest of this subsection is based on the following reasoning (the author was largely
inspired by the Bargmann's work \cite{Bar}). If the two exponents $\xi$ and $\xi'$ are \emph{equivalent},
that is $\xi' = \xi + \Delta[\zeta]$, then the \emph{semicentral extensions} $H$ and $H'$ connected
with $\xi$ and $\xi'$ are isomorphic. The isomorphism $h: \{\theta,r\} \mapsto \{\theta', r'\}$
is given by
\begin{equation}\label{izo}
\theta'(t) = \theta(t) - \zeta(r,t), \, \, r' = r.
\end{equation}
Indeed, $h$ is a homomorphism, because
\begin{displaymath} 
h(\{\theta_{1},r_{1}\}\{\theta_{2}, r_{2}\}) = h(\{\theta_{1},r_{1}\})h(\{\theta_{2},r_{2}\}),
\end{displaymath}
and $h$ is continuous with respect to the topology defined as above, because 
$\zeta(r,t)$ is \emph{strongly continuous} in $r$. Using an \emph{Iwasawa-type
construction} we show that any exponent $\xi(r,s,t)$ is equivalent to a differentiable one
(in $r$ and $s$). Next, we show that if $G$ is a Lie group then one can consider a finite-dimensional space $T$
of differentiable functions $\theta(t)$ such that the elements $\{\theta, r\}$ ($\theta \in T$)
form a group $H_{\mathfrak{H}}$ with the multiplication law given by (\ref{15}), such that 
$H_{\mathfrak{H}}$ is a Lie group itself. After this 
the above homomorfism $h$ given by (\ref{izo}) defines a local isomorphism between
Lie groups $H_{\mathfrak{H}}$ and $H'_{\mathfrak{H}}$. But from the classical theory of Lie groups 
follows that any local Lie group defines in a unique way a Lie algebra and \emph{vice versa}. So, in our case to any 
local exponent $\xi$ corresponds an equivalent differentiable exponent $\xi'$ which defines (by (\ref{15}))
a local Lie group $H_{\mathfrak{H}}$, and by this a Lie algebra $\mathfrak{H}$ of $H_{\mathfrak{H}}$, 
and \emph{vice versa}. 
As we will see the algebra defines a time dependent antilinear form $\Xi$ on the Lie algebra
$\mathfrak{G}$ of $G$, the so called \emph{infinitesimal exponent} . So, basing on the classical theory of Lie groups
we can see that the correspondence between local exponents $\xi$ and infinitesimal exponents $\Xi$ is one-to-one.
That is, we can translate the equivalence
of local exponents $\xi$ into the local isomorphism of Lie groups $H_{\mathfrak{H}}$ and by this, on the 
isomorphism of algebras $\mathfrak{H}$, \emph{i.e.} into the equivalence of infinitesimal exponents $\Xi$
(into a kind of a linear relation between infinitesimal exponents $\Xi$). So, we will simplify the 
problem of the classification to a largely linear problem.

\vspace{1ex}    

IWASAWA CONSTRUCTION. Denote by ${\ud}r$ and ${\ud}^{*}r$ the left and right invariant Haar
measure on $G$. Let $\nu(r)$ and $\nu^{*}(r)$ be two infinitely differentiable
functions on $G$ with compact supports contained in the fixed neighborhood $\mathfrak{N}_{0}$ 
of $e$. Multiplying them by the appropriate constants we can always reach:
$\int_{G} \nu(r) \, {\ud}r = \int_{G} \nu^{*}(r) \, {\ud}^{*}r =1$. Let $\xi(r,s,t)$ be any
\emph{admissible} local exponent defined on $\mathfrak{N}_{0}$. We will construct a
differentiable (in $r$ and $s$) exponent $\xi''(r,s,t)$ which is \emph{equivalent} to $\xi(r,s,t)$ and is defined
on $\mathfrak{N}_{0}$, in the following two steps:
\begin{displaymath}
\xi' = \xi + \Delta[\zeta], \, \, with \, \, \zeta(r,t) = - \int_{G} \xi(r,l,t) \nu(l) \, {\ud}l, 
\end{displaymath}

\vspace{-0.5cm}

\begin{displaymath}
\xi'' = \xi' + \Delta[\zeta'], \, \, with \, \, \zeta'(r,t) = - \int_{G} \xi'(u,r,ut)\nu^{*}(u) \, {\ud}^{*}u.
\end{displaymath}
Using first (\ref{13}) and (\ref{10}) and then the left invariance property of the integral, we get:
\begin{displaymath}
\xi'(r,s,t) = \int_{G} \{\xi(r,s,t) - \xi(r,l,t) - \xi(s,l,r^{-1}t) + 
\end{displaymath}

\vspace{-0.5cm}

\begin{displaymath}
+ \xi(rs,l,t)\}\nu(l) \, {\ud}l =
\end{displaymath}
  
\vspace{-0.5cm}

\begin{displaymath}
= \int_{G} \{ \xi(r,sl,t) - \xi(r,l,t)\}\nu(l) \, {\ud}l =
\end{displaymath}

\vspace{-0.5cm}

\begin{displaymath}
= \int_{G} \xi(r,sl,t)\nu(l) \, {\ud}l - \int_{G} \xi(r,l,t)\nu(l) \, {\ud}l =
\end{displaymath}

\vspace{-0.5cm}

\begin{displaymath}
= \int_{G} \xi(r,l,t) \nu(s^{-1}l) \, {\ud}l - \int_{G} \xi(r,l,t) \nu(l) \, {\ud}l =
\end{displaymath}

\vspace{-0.5cm}

\begin{displaymath}
= \int_{G} \xi(r,l,t) \{ \nu(s^{-1}) - \nu(l) \} \, {\ud}l. 
\end{displaymath}
So, we get
\begin{equation}\label{xi'}
\xi'(r,s,ur^{-1}t) = \int_{G} \xi(u,l,ur^{-1}t) \{ \nu(s^{-1}l) \nu(l)\} \, {\ud}l.
\end{equation}
In a similar way one gets
\begin{displaymath}
\xi''(r,s,t) = \int_{G} \{ \xi'(r,s,t) - \xi'(u,r,ut) - \xi'(u,s,ur^{-1}t) +
\end{displaymath}

\vspace{-0.5cm}
\begin{displaymath}
+\xi'(u,rs,ut) \} \nu^{*}(u) \, {\ud}^{*}u =
\end{displaymath}

\vspace{-0.5cm}

\begin{displaymath}
= \int_{G} \{ \xi'(ur,s,ut) - \xi'(u,s,ur^{-1}t) \} \nu^{*}(u) \, {\ud}^{*}u =
\end{displaymath}

\vspace{-0.5cm}

\begin{displaymath}
= \int_{G} \xi'(ur,s,ut) \nu^{*}(u) \, {\ud}^{*}u - \int_{G} \xi'(u,s,ur^{-1}t) \nu^{*}(u) \, {\ud}^{*}u 
\end{displaymath}

\vspace{-0.5cm}

\begin{displaymath}
= \int_{G} \xi'(u,s,ur^{-1}t) \nu^{*}(ur^{-1}) \, {\ud}^{*}u 
\end{displaymath}

\vspace{-0.5cm}

\begin{displaymath}
- \int_{G} \xi'(u,s,ur^{-1}t) \nu^{*}(u) \, {\ud}^{*}u 
\end{displaymath}

\vspace{-0.5cm}

\begin{displaymath}
= \int_{G} \xi'(u,s,ur^{-1}t)\{ \nu^{*}(ur^{-1}) - \nu^{*}(u) \} \, {\ud}^{*}u.
\end{displaymath}
Inserting (\ref{xi'}) to the formula we finally get
\begin{displaymath}
\xi''(r,s,t) = \int\!\!\int_{G} \xi(u,l,ur^{-1}t) \{\nu(s^{-1}l) - \nu(l) \} \times
\end{displaymath}

\vspace{-0.5cm}

\begin{displaymath}
\times \{ \nu^{*}(ur^{-1}) - \nu^{*}(u) \} \, {\ud}l \, {\ud}^{*}u.
\end{displaymath}
Because $\nu$ and $\nu^{*}$ are differentiable (up to any order) and $\xi(r,s,t)$ is a differentiable function of $t$ 
(up to any order, see {\bf V.A}) then $\xi''(r,s,t)$ is a differentiable (up to any order)
exponent in all variables $(r,s,t)$.

\vspace{1ex}

LEMMA 3. \emph{If two differentiable exponents} $\xi$ \emph{and} $\xi'$ \emph{are equivalent,
that is, if} $\xi' = \xi + \Delta[\zeta]$, \emph{then} $\zeta(r,t)$ is \emph{differentiable in} $r$.

\vspace{1ex}

PROOF. Clearly, the function $\chi(r,s,t) = \xi'(r,s,t) - \xi(r,s,t)$ is differentiable. Similarly the function
$\eta(r,t) = \int_{G} \chi(r,u,t) \nu(u) \, {\ud}u$, where $\nu$ is defined as in the Iwasawa construction,
is a differentiable function. But the difference $\zeta' = \eta - \zeta$ is equal 
\begin{displaymath}
\zeta'(r,t) = \int_{G} \{\zeta(u,r^{-1}t) - \zeta(ru,t) \} \nu(u) \, {\ud}u =
\end{displaymath}   

\vspace{-0.5cm}

\begin{displaymath}
= \int_{G} \{ \zeta(u,r^{-1}t)\nu(u) -\zeta(u,t)\nu(r^{-1}u) \} \, {\ud}u 
\end{displaymath}
and clearly it is a differentiable function. By this $\zeta = \eta - \zeta'$ also is a differentiable function
(recall that $\zeta(r,t)$ is differentiable function of $t$, see {\bf V.A}).

\vspace{1ex}

LEMMA 4. \emph{Every (local) exponent of one-parameter group is equivalent to zero}.

\vspace{1ex}

PROOF. We can map such a group $r = r(\tau) \rightleftarrows \tau$ on the real line 
($\tau \in \mathcal{R}$) in such a way that $r(\tau) r(\tau') = r(\tau + \tau')$. Set
\begin{displaymath}
\vartheta(\tau,\sigma,t) = \frac{\partial \xi(\tau,\sigma,t)}{\partial \sigma}.
\end{displaymath}
From (\ref{9}), (\ref{11}) and (\ref{10}) one gets 
\begin{equation}\label{0e}
\xi(0,0,t) = 0, \, \, \xi(\tau,0,t) = 0,
\end{equation}

\vspace{-0.5cm}

\begin{displaymath}
\xi(\tau, \tau',t) + \xi(\tau + \tau',t) = \xi(\tau', \tau'', r(-\tau)t) + 
\end{displaymath}

\vspace{-0.5cm}

\begin{equation}\label{1par}
+ \xi(\tau, \tau' + \tau'',t).
\end{equation}
Now we derive the expression with respect to $\tau''$ at $\tau'' = 0$. This yields
(with the above definition of $\vartheta$)
\begin{equation}\label{16}
\vartheta(\tau + \tau', 0, t) = \vartheta(\tau', 0,r(-\tau)t) + \vartheta(\tau,\tau',t).
\end{equation}
Let us define now
\begin{displaymath}
\zeta(\tau,t)= \int_{0}^{\tau} \vartheta(\sigma, 0,t) \, {\ud}\sigma =
\int_{0}^{1} \tau \vartheta(\mu\tau,0,t) \, {\ud}\mu.
\end{displaymath}
We have then
\begin{displaymath}
- \Delta[\zeta] = \zeta(\tau + \tau',t) - \zeta(\tau,t) - \zeta(\tau',r(-\tau)t) =
\end{displaymath}

\vspace{-0.5cm}

\begin{displaymath}
= \int_{0}^{\tau + \tau'} \vartheta(\sigma,0,t) \, {\ud}\sigma - \int_{0}^{\tau} \vartheta(\sigma,0,t) \, {\ud}\sigma -
\end{displaymath}

\vspace{-0.5cm}

\begin{displaymath}
- \int_{0}^{\tau'} \vartheta(\sigma,0,r(-\tau)t) \, {\ud}\sigma =
\end{displaymath}

\vspace{-0.5cm}

\begin{displaymath}
= \int_{\tau}^{\tau + \tau'} \vartheta(\sigma,0,t) \, {\ud}\sigma - \int_{0}^{\tau'} \vartheta(\sigma,0,r(-\tau)t) \, {\ud}\sigma =
\end{displaymath}

\vspace{-0.5cm}

\begin{displaymath}
= \int_{0}^{\tau'} \{ \vartheta(\tau + \sigma,0,t) - \vartheta(\sigma,0,r(-\tau)t) \} \, {\ud}\sigma.
\end{displaymath}
Using now the Eq. (\ref{16}) and (\ref{0e}) we get
\begin{displaymath}
- \Delta[\zeta] = \int_{0}^{\tau'} \vartheta(\tau,\sigma,t) \, {\ud}\sigma =
\end{displaymath}

\vspace{-0.5cm}

\begin{displaymath}
= \int_{0}^{\tau'} \frac{\partial \xi(\tau,\sigma,t)}{\partial \sigma} \, {\ud}\sigma = \xi(\tau, \tau',t)
\end{displaymath}
and $\xi$ is equivalent to 0. 

\vspace{1ex}

Let us recall that the continuous curve $r(\tau)$ in a Lie group $G$ is a one-parameter subgroup 
if and only if $r(\tau_{1})r(\tau_{2}) = r(\tau_{1} + \tau_{2})$ \emph{i.e.} $r(\tau) = (r_{0})^{\tau}$, for
some element $r_{0} \in G$, note that the real power $r^{\tau}$ is well defined on a Lie group (at least
on some neighborhood of $e$). The coordinates $\rho^{k}$ in $G$ 
are \emph{canonical} if and only if any curve of the form 
$r(\tau) = \tau \rho^{k}$ (where the coordinates $\rho^{k}$ are fixed) is a one-parameter subgroup 
(the curve $r(\tau) = \tau \rho^{k}$ will be denoted in short by $\tau a$, with the coordinates 
of $a$ equal to $\rho^{k}$. The "vector" $a$ is called by physicists the \emph{generator} of the one-parameter
subgroup $\tau a$. Denote the coordinates of $r$, $s$ and $rs$ by $\rho^{k}$, $\sigma^{k}$ and
$f^{k}= f^{k}(\rho^{i}, \sigma^{j})$, assume the coordinates of $e$ to be 0. 
Then in any differentiable coordinates (not necessarily \emph{canonical})
\begin{displaymath}
f^{k}(\rho^{i},e) = f^{k}(\rho^{i}, 0, \ldots , 0) = \rho^{k},
\end{displaymath}

\vspace{-0.5cm}

\begin{displaymath}
f^{k}(0, \ldots, 0, \sigma^{j}) = \sigma^{k},
\end{displaymath}

\vspace{-0.5cm}

\begin{displaymath}
f^{k}(\rho^{i}, \sigma^{j}) = \rho^{k} + \sigma^{k} + a_{ij}^{k}\rho^{i}\sigma^{j} + 
g_{ijl}^{k}\rho^{i} \rho^{j} \sigma^{l} + 
\end{displaymath}

\vspace{-0.5cm}

\begin{equation}\label{exp x}
+ h_{ijl}^{k}\rho^{i} \sigma^{j} \sigma^{l} + \varepsilon^{k},
\end{equation}
where $\varepsilon^{k}$ are of the fourth order of magnitude in the coordinates $\rho^{k}$ and $\sigma^{k}$
and $a_{ij}^{k}, g_{ijl}^{k}, h_{ijl}^{k}$ are some constants. The structure constants $c_{ij}^{k}$
are equal
\begin{displaymath}
c_{ij}^{k} = a_{ij}^{k} - a_{ji}^{k}
\end{displaymath} 
if the group coordinates are \emph{canonical}. Note that it is true in any "new" coordinates ${\rho'}^{k}$
defined as functions ${\rho'}^{i}(\rho^{j})$ of \emph{canonical} coordinates $\rho^{k}$ in such a way that
\begin{displaymath}
\frac{\partial{\rho'}^{i}}{\partial \rho^{j}} = \delta^{i}_{j}.
\end{displaymath}
Such new coordinates will be called \emph{admissible}. Note that the alternative definition of 
\emph{admissible} coordinates is possible: in those coordinates any curve $r(\tau) = \tau a$ is 
a one-parameter subgroup up to the second order in $\tau$. 

A local exponent $\xi$ of a Lie group $G$ is called \emph{canonical} if $\xi(r,s,t)$ is differentiable 
in all variables and $\xi(r,s,t) = 0$ if $r$ and $s$ are elements of the same one-parameter subgroup.

\vspace{1ex}

LEMMA 5. \emph{Every local exponent $\xi$ of a Lie group is equivalent to a canonical local exponent.}

\vspace{1ex}

PROOF. Set $\rho^{j}$ and $\sigma^{i}$ for the canonical coordinates of the two elements $r,s \in G$ respectively, and
define 
\begin{displaymath}
\vartheta_{k} = \frac{\partial \xi(r,s,t)}{\partial\sigma^{k}}.
\end{displaymath}
Let us define now
\begin{displaymath}
\zeta(r,t) = \int_{0}^{1} \sum_{k=1}^{n} \rho^{k} \vartheta_{k}(\mu r,0,t) \, {\ud}\mu.
\end{displaymath}
Consider a one-parameter subgroup $r(\tau)$ generated by $a$, \emph{i.e}
$r(\tau) = \tau a$. Because $\xi$ is a local exponent fulfilling (\ref{9}), (\ref{10}) and (\ref{11}) then
$\xi_{0}(\tau, \tau',t) \equiv \xi(\tau a, \tau' a,t)$ fulfils (\ref{0e}) and (\ref{1par}). Repeating now the same 
steps as in the proof of Lemma 4 one can show that 
\begin{displaymath}
\xi(\tau a, \tau' a, t) + \Delta[\zeta(\tau a,t)] = 0. 
\end{displaymath} 

\vspace{1ex}

LEMMA 6. \emph{Let $\xi$ and $\xi'$ be two differentiable and equivalent local exponents of a Lie group $G$, 
and assume $\xi$ to 
be canonical. Then $\xi'$ is canonical if and only if $\xi' = \xi + \Delta[\Lambda]$, where $\Lambda(r,t)$ is 
a linear form in the canonical coordinates of $r$ fulfilling the condition}
\begin{displaymath}
\frac{d\Lambda(a, (\tau a)t)}{d\tau} = 0,
\end{displaymath}
\emph{i.e. $\Lambda(a,(\tau a)t)$ is constant as a function of $\tau$}.

\vspace{1ex}
 
In the sequel we will use the simple notation 
\begin{displaymath}
\boldsymbol{a} f(X) = \frac{df((\tau a)X)}{d\tau}\Big\vert_{\tau = 0},
\end{displaymath}
such that $\boldsymbol{a}$ is the generator of the regular representation of $r(\tau) = \tau a$ and 
$\boldsymbol{a}f(X) = 0$ means that $f(X)$ is constant along the integral curves $X(\tau) = (\tau a) X_{0}$. After this
from the condition of Lemma 6 follows that 
\begin{displaymath}
\boldsymbol{a}\Lambda(a,t) =0.
\end{displaymath}
The notation seems to be innatural
in the presented context when the exponent $\xi$ depends on the time $t$ only. But it is very useful 
in general when $\xi$ depends on all spacetime coordinates $X$ (recall that our proofs are valid also in this general
case and we mark explicitly the important difference between the time and the spacetime dependent $\xi$). 
 
\vspace{1ex}

PROOF of Lemma 6. 1$^{o}$. Necessity of the condition. Because the exponents are equivalent we have 
$\xi'(r,s,t) = \xi(r,s,t) +\Delta[\zeta]$. Because both $\xi$ and $\xi'$ are differentiable then $\zeta(r,t)$ also is 
a differentiable function, which follows from Lemma 3. Suppose that $r=\tau a$ and $s= \tau' a$. Because
of both $\xi$ and $\xi'$ are \emph{canonical} we have $\xi(\tau a, \tau' a,t) = \xi'(\tau a, \tau' a,t) = 0$, such 
that $\Delta[\zeta](\tau a, \tau' a,t) = 0$, \emph{i.e.}
\begin{displaymath}
\zeta((\tau + \tau')a, t) = \zeta(\tau a, t) + \zeta(\tau' a,(-\tau a)t).
\end{displaymath}  
Applying recurrently this formula one gets
\begin{equation}\label{17}
\zeta(\tau a,t) = \sum_{k=0}^{n-1} \zeta(\frac{\tau}{n}a,(-\frac{k}{n}\tau a)t).
\end{equation}
$\zeta$ is differentiable (up to any order) and we can use the Taylor Theorem. Because in addition $\zeta(0,t) =0$ 
we get the following formula 
\begin{displaymath}
\zeta(\frac{\tau}{n}a,t) = \zeta'(0,t)\frac{\tau}{n} + 
\frac{1}{2}\zeta''(\theta_{\frac{\tau}{n}}\frac{\tau}{n}a,t)\Big(\frac{\tau}{n}\Big)^{2},
\end{displaymath}
where $\zeta'$ and $\zeta''$ stand for the first and the second derivative of $\zeta(xa,t)$ with
respect to $x$, and $0 \leq \theta_{\frac{\tau}{n}} \leq 1$. Recall that
in the Taylor formula
$f(x+h) = f(x) + f'(x)h + 1/2f''(x + \theta_{h} h) h^{2}$ the $\theta_{h} \in [0,1]$ depends on $h$, which is marked by 
the subscript $h$: $\theta_{h}$. Inserting $\tau =n =1$ to the formula and multiplying it by $\tau/n$ (provided
the coordinates $a$ of an element $r_{0} \in G$ are chosen in such a way that $r_{0}$
lies in the neighborhood ${\mathfrak{N}}_{0}$ on which the exponents $\xi$ and $\xi'$ are defined)
one gets
\begin{displaymath}
\frac{\tau}{n}\zeta(a,t) = \frac{\tau}{n}\{\zeta'(0,t) + \frac{1}{2} \zeta''(\theta_{1}a,t)\}.
\end{displaymath} 
Taking now the difference of the last two formulas we get
\begin{displaymath}
\zeta(\frac{\tau}{n}a,t) = \frac{\tau}{n}\Big\{ \zeta(a,t) - \frac{1}{2} \zeta''(\theta_{1}a,t) \Big\} + 
\frac{1}{2}\Big(\frac{\tau}{n} \Big)^{2} \zeta''(\theta_{\frac{\tau}{n}} \frac{\tau}{n}a,t).
\end{displaymath}
Inserting this to the formula (\ref{17}) we get
\begin{displaymath}
\zeta(\tau a,t) = \frac{\tau}{n} \sum_{k=0}^{n-1} \Big\{\zeta(a,(-\frac{k}{n}\tau a)t) - 
\frac{1}{2}\zeta''(\theta_{1}a, (-\frac{k}{n}\tau a)t) \Big\} + 
\end{displaymath}

\vspace{-0.5cm}

\begin{displaymath}
+ \frac{1}{2}\Big(\frac{\tau}{n}\Big)^{2} \sum_{k=0}^{n-1} \zeta''(\theta_{\frac{\tau}{n}}\frac{\tau}{n}a, (-\frac{k}{n}\tau a)t).
\end{displaymath}
Denote the supremum and the infimum of the function $\zeta''(xa, (-ya)t)$ in the square
$(0 \leq x \leq \tau, 0 \leq y \leq \tau)$ by $M$ and $N$ respectively. We have
\begin{displaymath}
\frac{1}{2}\Big(\frac{\tau}{n}\Big)^{2}nN + \frac{\tau}{n}\sum_{k=0}^{n-1} \Big\{\zeta(a,(-\frac{k}{n}a)t) - 
\frac{1}{2}\zeta''(\theta_{1} a,(-\frac{k}{n}\tau a)t) \Big\} 
\end{displaymath}

\vspace{-0.5cm}

\begin{displaymath}
\leq \zeta(\tau a,t) \leq \frac{1}{2}\Big(\frac{\tau}{n}\Big)^{2}nM + 
\end{displaymath}

\vspace{-0.5cm}

\begin{displaymath}
+ \frac{\tau}{n}\sum_{k=0}^{n-1} \Big\{\zeta(a,(-\frac{k}{n}\tau a)t) - \frac{1}{2}\zeta''(\theta_{1}a,(-\frac{k}{n}\tau a)t) \Big\}.
\end{displaymath}
Passing to the limit $n \rightarrow + \infty$ we get
\begin{displaymath}
\zeta(\tau a,t) = \int_{0}^{\tau} \Big\{\zeta(a,(-\sigma a)t) - \frac{1}{2}\zeta''(\theta_{1}a, (-\sigma a)t) \Big\} \, {\ud}\sigma.
\end{displaymath}
Taking into account that the functions $\zeta'$ and $\zeta''$ are independent the general solution $\zeta$
fulfilling $\Delta[\zeta](\tau a, \tau' a) = 0$ for any $\tau$ and $\tau'$ can be written in the following form
\begin{equation}\label{18}
\zeta(\tau a,t) = \int_{0}^{\tau} \varsigma(a,(-\sigma a)t) \, {\ud} \sigma,
\end{equation}
where $\varsigma = \varsigma(r,t)$ is any differentiable function. 
Differentiate now the expression (\ref{18}) with respect to $\tau$ at $\tau = 0$. After this one gets
\begin{equation}\label{19}
\varsigma(a, t) = \sum_{k=1}^{n}\lambda_{k}(t)a^{k}, \, \, \, with \, \, \, \lambda_{k}(t) = 
\frac{\partial \zeta(0,t)}{\partial a^{k}},
\end{equation}
where $a^{k}$ stands for the coordinates of $a$. So, the function $\varsigma(a,t)$ is linear with respect to $a$. 
Denote the time coordinate $t$ of the spacetime point $X$ by $t(X)$. Suppose in addition that the spacetime
coordinates $X$ are chosen in such a way that the integral curves $X(x) = (xa)X_{0}$ are coordinate lines, 
which is possible for appropriately small $\tau$. There are of course three remaining families of coordinate lines
beside $X(x)$, which can be chosen in arbitrary way, the parameters of which will be denoted by $y_{i}$.
After this, Eq. (\ref{18}) reads
\begin{displaymath}
\zeta(\tau a,t(x,y_{i})) = \int_{0}^{\tau} \varsigma(a,t(x - \sigma, y_{i})) \, {\ud}\sigma.
\end{displaymath} 
Inserting $\tau = 1$ to this formula one gets
\begin{displaymath}
\zeta(a,t(x,y_{i})) = \int_{0}^{1} \varsigma(a,t(x - \sigma, y_{k})) \, {\ud}\sigma.
\end{displaymath}
So, because $\varsigma(a,t)$ is linear with respect to $a$ the function $\zeta(a,t)$ is also linear in $a$. 
From the linearity of $\zeta(a,t)$ and from (\ref{18}) we have
\begin{displaymath} 
\zeta(a,t(x,y_{i})) = \frac{1}{\tau} \int_{0}^{\tau} \varsigma(a,t(x - \sigma, y_{i})) \, {\ud} \sigma =
\end{displaymath}

\vspace{-0.5cm}

\begin{displaymath}
= \frac{1}{\tau} \int_{x - \tau}^{x} \varsigma(a,t(z,y_{i})) \, {\ud}z,
\end{displaymath} 
for any $\tau$ (of course with appropriately small $\vert \tau \vert$, in our case $\vert \tau \vert \leq 1$) and for any 
(appropriately small) $x$.
But this is possible for the function $\varsigma(a,t(x,y_{k}))$ continuous in $x$  (in our case differentiable in $x$) 
if and only if $\varsigma(a,t(x,y_{k}))$ does not depend on $x$. This means that $\zeta(a,t(x,y_{k}))$ 
does not depend on $x$ and the condition of Lemma 6 is proved.
\\ $2^{o}$. Sufficiency of the condition. If $\zeta(r,t) = \Lambda(r,t)$ fulfils the condition of Lemma 6, then
$\xi' = \xi + \Delta[\Lambda]$ is differentiable and because $\Lambda(a, (\tau a)^{-1}t)$ does not depend 
on $\tau$ (and by this $\Lambda(a, (\tau a)^{-1}t) = \Lambda(a,t)$) and $\Lambda$ is linear in the first argument, 
we have
\begin{displaymath}
\Delta[\Lambda](\tau a, \tau' a) = \Lambda(\tau a, t) + \Lambda(\tau' a, (\tau a)^{-1}t) -  
\end{displaymath}

\vspace{-0.5cm}

\begin{displaymath}
- \Lambda((\tau + \tau')a, t) = \tau \Lambda(a,t) + \tau' \Lambda(a,t) - 
\end{displaymath}

\vspace{-0.5cm}

\begin{equation}\label{D}
- (\tau + \tau') \Lambda(a,t) = 0. 
\end{equation}
At last because $\xi$ is \emph{canonical} $\xi(\tau a, \tau' a, t) = 0$, then, from (\ref{D})
$\xi'(\tau a, \tau' a, t) = 0$ and $\xi'$ is canonical. 

\vspace{1ex}

REMARK. Up to now there is no difference between the time-dependent exponent $\xi$ 
and spacetime-dependent exponent. All proofs are constructed in such a way that it is
sufficient to replace the time $t$ by $X$. But now we proceed to construct
a finite dimensional Lie subgroup of the \emph{semicentral extension} $H$ of $G$. In general,
such a construction is impossible for $X$-dependent $\xi$. Because
the classification of $X$-dependent $\xi$-s also has interesting physical applications we sketch
briefly the machinery of the classification of $\xi(r,s,X)$ in the separate Appendix {\bf B}.
In the rest of this subsection we confine ourselves to the time-dependent $\xi$.

\vspace{1ex}

LIE SUBGROUP $H_{\mathfrak{H}}$ OF $H$. The topology of $H$ was defined in the comment
to the formula (\ref{15}). In this topology the multiplication operation as defined by Eq. (\ref{15})
is a continuous operation.  From Lemma 5 it follows that
you can confine yourself to the \emph{canonical} exponents $\xi$. Let us assume the coordinates on $G$ 
to be canonical (in these coordinates the multiplication function $f^{k}$ fulfils (\ref{exp x})).
We will investigate first the one-parameter curves $\check{r}(\tau) = \{\tau\theta(t),\tau a\} \equiv
\tau\{\theta(t),a\} \equiv \tau \check{a}$. Let us define the expression
\begin{equation}\label{20}
[\check{a}, \check{b}] = \lim_{\tau \to 0} \frac{(\tau\check{a})(\tau\check{b})(\tau\check{a})^{-1}(\tau\check{b})^{-1}}{\tau^{2}}, 
\end{equation}
where the limes is in the sense of the topology of $H$.
Inserting the multiplication rule (\ref{15}) and the expansion (\ref{exp x}) of the multiplication function $f^{k}$
in $G$ to this definition we get after simple computations
\begin{equation}\label{20'}
[\check{a}, \check{b}] = \{\boldsymbol{a}\beta - \boldsymbol{b}\alpha + \Xi(a,b, t), [a,b] \},
\end{equation}
  
\vspace{-0.5cm}

\begin{displaymath}
\Xi(a,b,t) = \lim_{\tau \to 0} \tau^{-2}\{\xi((\tau a)(\tau b), (\tau a)^{-1}(\tau b)^{-1},t) + 
\end{displaymath}

\vspace{-0.5cm}

\begin{equation}\label{20''}
+ \xi(\tau a, \tau b, t) + \xi((\tau a)^{-1},(\tau b)^{-1}, (\tau b)^{-1}(\tau a)^{-1}t) \},
\end{equation}
where $\check{a} =\{\alpha(t), a\}$, $\check{b} = \{\beta(t), b\}$, and $[a,b]$ is the Lie bracket in the 
Lie algebra $\mathfrak{G}$ of the Lie group $G$. Recall that $\boldsymbol{a}\alpha(t)$ was defined
in the comment to the Lemma 6. The limit in (\ref{20''}), induced by the topology of $H$, is defined
by the metric (\ref{topology}). That is, this is the almost uniform convergence limit. 
Note that because $\xi$ is differentiable and \emph{canonical}
the expression (\ref{20'}) is well defined. Indeed, because $\xi$ is differentiable we can expand
it up to the (say) fourth order in the canonical coordinates on $G$ around the point 0 (\emph{i.e.} around 
$e \in G$), and taking into account (\ref{9}) and (\ref{11}) we get (around 0)
\begin{equation}\label{exp xi}
\xi(\rho^{i},\sigma^{j},t) = a_{ij}\rho^{i}\sigma^{j} + b_{ijl}\rho^{i}\rho^{j}\sigma^{l} +d_{ijl}\rho^{i}\sigma^{j}\sigma^{l}+
\varepsilon,
\end{equation} 
where $a_{ij}, b_{ijl}$, $d_{ijl}$ and $\varepsilon$ are some differentiable functions of the time $t$,
and $\varepsilon$ is of the fourth order of magnitude
in $(\rho^{i}, \sigma^{j})$ that is, $\varepsilon$ is of a higher order than the second in $\rho^{i}$
and $\sigma^{i}$ separately. Because $\xi$ is \emph{canonical} $a_{ij} = - a_{ji}$. Inserting it 
to (\ref{20''}) one can easily see that (\ref{20''}) is well defined and because $a_{ij}$ is antisymmetric
the expression $[\check{a},\check{b}]$ also is antisymmetric. Inserting the expansion (\ref{exp xi})
to the formula (\ref{20''}) one can easily see that $\Xi(a,b,t)$ is an antilinear form in $a$ and $b$.   
\\ From the associative law in $H$ one gets
\begin{displaymath}
((\tau \check{a})(\tau \check{b}))(\tau \check{c}) = (\tau \check{a})((\tau \check{b})(\tau \check{c})).
\end{displaymath}
We divide now the above expression by $\tau^{3}$ and then pass to the limit $\tau \rightarrow 0$:
\begin{displaymath}
\lim_{\tau \to 0} \frac{((\tau \check{a})(\tau \check{b}))(\tau \check{c})}{\tau^{3}} =
\lim_{\tau \to 0} \frac{(\tau \check{a})((\tau \check{b})(\tau \check{c}))}{\tau^{3}}.
\end{displaymath}
Again both limits exist, which can be shown in the analysis similar to that used for the existence investigation
of $[\check{a},\check{b}]$. Inserting the explicit values of those limits one obtains
\begin{displaymath}
\Xi([a,a'],a'',t) + \Xi([a',a''],a,t) + \Xi([a'',a],a',t) =
\end{displaymath} 

\vspace{-0.5cm}

\begin{equation}\label{21}
= \boldsymbol{a}\Xi(a',a'',t) + \boldsymbol{a'}\Xi(a'',a,t) + \boldsymbol{a''}\Xi(a,a',t),
\end{equation}
which can be shown to be equivalent to the Jacobi identity
\begin{equation}\label{21'}
[[\check{a}, \check{a}'], \check{a}''] + [[\check{a}',\check{a}''],\check{a}] +
[[\check{a}'',\check{a}], \check{a}'] = 0. 
\end{equation}
So, we have constructed in this way a Lie algebra with $[\check{a}, \check{b}]$ (defined
for all $\check{a} = \{\alpha(t), a\}$ with differentiable $\alpha(t)$ and $a$ belonging to the 
Lie algebra $\mathfrak{G}$ of $G$) using the 
formula (\ref{20'}). We will show that in our case this algebra always possesses a finite dimensional
Lie subalgebra $\mathfrak{H}$ which contains all elements $\check{a} = \{0,a\}$ with $a \in \mathfrak{G}$
(which is not the case, in general, for the spacetime-depending $\xi$).
In our case, however, according to our assumption about $G$ (compare for example any finite dimensional subgroup
of (\ref{tra})) any $r \in G$ transforms simultaneity hyperplanes into simultaneity hyperplanes. So, there are two
possibilities for any $r \in G$. First, when $r$ does not change the time: $t(rX) = t(X)$ and the second in
which the time is changed, but in such a way that $t(rX) - t(X) = f(t)$. \emph{We assume in addition that the 
base generators $a_{k} \in \mathfrak{G}$ can be chosen in such away that only one acts on the time as the translation
and the remaining ones do not act on the time}. After this the Jacobi identity (\ref{21}) reads
\begin{equation}\label{Jac1}
\Xi([a,a'],a'') + \Xi([a', a''],a) + \Xi([a'',a],a') = \partial_{t} \Xi(a',a''),
\end{equation}
if one and only one among $a,a',a''$ is the time translation generator, namely $a$, and 
\begin{eqnarray}\label{Jac2}
\Xi([a,a'],a'') + \Xi([a',a''], a) + \Xi([a'',a],a') = 0,
\end{eqnarray} 
in all remaining cases. So, the elements $\check{a}_{ijk} = \{\Xi_{ij}(t), a_{k}\}$ where $\Xi_{ij}(t) = \Xi(a_{i}, a_{j},t)$
compose the base of a finite dimensional algebra $\mathfrak{H}$ with the Lie bracket defined by (\ref{20'}). Indeed,
any $\boldsymbol{a}_{k}\Xi_{ij}(t)$ is equal to 0 or $\partial_{t}\Xi_{ij}(t)$, but, as we have seen above, it follows
from the Jacobi identity that $\partial_{t}\Xi_{ij}$ is a linear combination of $\Xi_{ij}$.

We will show 

\vspace{1ex}

LEMMA 7. \emph{The Lie group $H_{\mathfrak{H}}$ generated by the Lie algebra $\mathfrak{H}$
is really a subgroup of the semicentral extension $H$, and $H_{\mathfrak{H}}$ is a semicentral 
extension of $G$}.

\vspace{1ex}

It means that $H_{\mathfrak{H}}$ contains all elements $\{0,r\}$, $r \in G$

\vspace{1ex}

PROOF. Consider the one-parameter subgroups $\check{r}_{\check{a}}(\tau)$ of $H$
\begin{displaymath}
\check{r}_{\check{a}}(\tau) = \Big\{\int_{0}^{\tau} \alpha((\sigma a)^{-1}t) \, {\ud} \sigma, \tau a \Big\}.
\end{displaymath} 
We define the correspondence $\check{r}(\tau) = \tau \check{a} = \tau \{\alpha(t), a\} \to \check{r}_{\check{a}}(\tau)$
between $\tau \check{a}$ and $\check{r}_{\check{a}}(\tau)$ in the following way 
\begin{displaymath}
\tau \check{a} = 
\tau\{ \alpha , a\} \to \check{r}_{\check{a}}(\tau) = \Big\{\int_{0}^{\tau} \alpha((\sigma a)^{-1}t) \, {\ud}\sigma, \tau a\Big\}.
\end{displaymath}
We have to show that $1^{o}$. the above correspondence is one-to-one (at least for appropriately small $\tau$),
and $2^{o}$. that the one-parameter subgroups also generate $\mathfrak{H}$. 

$1^{o}$. First of all, note that the linear correspondence
\begin{equation}\label{con}
\alpha(t) \leftrightarrow \frac{1}{\tau} \int_{0}^{\tau} \alpha((\sigma a)^{-1}t) \, {\ud}\sigma
\end{equation}
is one-to-one for $\{\alpha(t), a\} \in \mathfrak{H}$ when $\tau$ is appropriately small. Indeed,
it is trivial when $\tau a$ does not act on the time $t$. Suppose then, that $\tau a$ does act on $t$
and -- according to our assumption -- $\tau a$ acts on the time $t$ as the time translation. 
But 
\begin{displaymath}
\frac{1}{\tau} \int_{0}^{\tau} \alpha((\sigma a)^{-1}t) \, {\ud}\sigma = 
\frac{1}{\tau} \int_{0}^{\tau} \alpha(t - \sigma) \, {\ud} \sigma = 0 
\end{displaymath}
for all $t$ and differentiable $\alpha(t)$ if and only if $\alpha(t)$ is a periodic function with the 
period $T \leq \vert \tau \vert$ the integer-multiple of which is equal to 
$\tau$. But $\mathfrak{H}$ -- being a finite-dimensional
vector space -- cannot contain $\{\alpha(t),a\}$ with $\alpha$ having arbitrary small period $T$
(if $\mathfrak{H}$ does not contain any $\{\alpha,a\}$ with a periodic $\alpha$, then the correspondence
(\ref{con}) is one-to-one). So, there exists the infimum $T_{0} > 0$ of all periods $T$ of 
periodic $\alpha$-s such that $\{\alpha,a\} \in \mathfrak{H}$. Then, the correspondence as defined by (\ref{con})
is one-to-one if $0 <\vert \tau \vert < T_{0}$. From this immediately follows that the correspondence
$\tau \check{a} \leftrightarrow \check{r}_{\check{a}}(\tau)$ is one-to-one if $0 < \vert \tau \vert < T_{0}$.

$2^{o}$. As is well known from the classical theory of Lie groups the elements of the 
Lie algebra are defined by the one-parameter subgroups $\check{r}_{\check{a}}(\tau)$ in this way
(the limits are in the sense of the topology of $H$)
\begin{equation}\label{vecal}
\lim_{\tau \to 0} \Big(\check{r}_{\check{a}}(\tau)\Big)^{\frac{1}{\tau}} =
\lim_{\tau \to 0} \frac{\check{r}_{\check{a}}(\tau)}{\tau}.
\end{equation}
Note, that the raising to a real power $(\check{r}_{\check{a}})^{\lambda}$ is well defined if only
$\check{a} \in \mathfrak{H}$. It is almost trivial if $\tau a$ does not act on the time, so, let us assume
$\tau a$ to be the time translation: $t \to t + \tau$. But two elements $\check{r}_{1} = \{\theta_{1}(t), \tau a\}$
and $\check{r}_{2} = \{\theta_{2}(t), \tau a\}$ fulfil the condition $\check{r}_{1}\check{r}_{1} = \check{r}_{2}\check{r}_{2}
= \{ \theta_{i}(t) + \theta_{i}(t - \tau), 2\tau a\}$
if and only if the function $\theta(t) = \theta_{2}(t) - \theta_{1}(t)$ fulfils the condition
$\theta(t) = - \theta(t - \tau)$ (note that we confine ourselves to the canonical exponent: $\xi(\tau a, \tau' a,t) = 0$).
From this follows that $\theta(t) = \theta(t - 2\tau)$ and $\theta(t)$ is a periodic
function with the period $T = 2\tau$. Choosing an appropriately small neigbourhood $\check{{\mathfrak{N}}}_{0}$
of $\check{e} \in H$ we can see that $(\check{r}_{\check{a}})^{\frac{1}{2}}$ is well (uniquely) defined
if $\check{r}_{\check{a}} \in \check{{\mathfrak{N}}}_{0}$. It is sufficient to choose the neigbourhood 
$\check{{\mathfrak{N}}}_{0}$ in such a way that it does not contain any time translations $t \to t + \eta$ ($\eta  = 2\tau$)
with $T_{0} \leq \vert \eta \vert$. But applying recurrently our reasoning
one can see that for any $\check{r}_{\check{a}} \in \check{{\mathfrak{N}}}_{0}$ there exists the rational
power of the form $\Big(\check{r}_{a}\Big)^{\frac{1}{2^{k}}}$, for any natural $k$. Because,
for any elements $\check{r}_{1}, \check{r}_{2} \in \check{{\mathfrak{N}}}_{0}$, 
there exist their inverse and their product $\check{r}_{1}\check{r}_{2}$ we can see that for any
$\check{r}_{\check{a}} \in \check{{\mathfrak{N}}}_{0}$ there exists uniquely defined rational
power $(\check{r}_{\check{a}})^{\lambda}$, where $\lambda =  p_{1}/2^{k_{1}} + \ldots +
p_{m} / 2^{k_{m}}$ for any integers $p_{i}, k_{j}$ and any natural $m$. But such a set of
rational $\lambda$-s is dense in $\mathcal{R}$ (compare to the binary system), and by continuity
in $H$ the rational power is well defined for any $\lambda$ with appropriately small $\vert \lambda \vert$.    

It is easy to see that in our case the above limit given by Eq. (\ref{vecal}) is unique,
which is the consequence of the biuniquenees of (\ref{con}) and is really equal to $\check{a}$.
 
Consider two one-parameter subgroups
\begin{displaymath}
\check{r}_{\check{a}}(\tau) = \Big\{ \int_{0}^{\tau} \alpha((\sigma a)^{-1}t),\tau a \Big\}
\end{displaymath}
and
\begin{displaymath}
\check{r}_{\check{a}}(\tau) = \Big\{ \int_{0}^{\tau} \beta((\sigma b)^{-1}t),\tau b \Big\}.
\end{displaymath}  
The vector addition $\check{a} \oplus \check{b}$ in the Lie algebra is defined by
\begin{displaymath}
\check{a} \oplus \check{b} = \lim_{\tau \to 0} \Big(\check{r}_{\check{a}}(\tau)\check{r}_{\check{b}}(\tau)\Big)^{\frac{1}{\tau}}
= \lim_{\tau \to 0} \frac{\check{r}_{\check{a}}(\tau)\check{r}_{\check{b}}(\tau)}{\tau},
\end{displaymath} 
which in our case is equal to the ordinary addition $\check{a} \oplus \check{b} = \{\alpha,a\} \oplus \{\beta, b\} =
\{\alpha + \beta,a + b\}$. 
Finally, the Lie bracket in the standard theory is defined by the one-parameter subgroups in the way
\begin{displaymath}
[\check{a},\check{b}] = \lim_{\tau \to 0} \Big(\check{r}_{\check{a}}(\tau)\check{r}_{\check{b}}(\tau)
(\check{r}_{\check{a}}(\tau))^{-1}(\check{r}_{\check{b}}(\tau))^{-1}\Big)^{\frac{1}{\tau}}=
\end{displaymath} 

\vspace{-0.5cm}

\begin{displaymath}
 = \lim_{\tau \to 0}
\frac{\check{r}_{\check{a}}(\tau)\check{r}_{\check{b}}(\tau)(\check{r}_{\check{a}}(\tau))^{-1}
(\check{r}_{\check{b}}(\tau))^{-1}}{\tau^{2}},
\end{displaymath} 
which, after the computations similar to those used to derive the Eq. (\ref{20'}), can be seen to be equal 
\begin{displaymath}
[\check{a}, \check{b}] = \{\boldsymbol{a}\beta - \boldsymbol{b}\alpha + \Xi(a,b), [a,b]\}.
\end{displaymath}
That is, we get a formula identical to (\ref{20'}) obtained for the curves $\tau \check{a}$ and $\tau \check{b}$. 
This means that the local Lie group $H_{\mathfrak{H}}$ generated by the Lie algebra $\mathfrak{H}$
is a subgroup of $H$, that is the multiplication law in $H_{\mathfrak{H}}$ is given by (\ref{15}). 
Geometrically, it means that the curves $\tau \check{a}$ and $\check{r}_{\check{a}}(\tau)$
define exactly the same element $\check{a} \in \mathfrak{H}$, or so to speak, the curves
both have exactly the same tangent vector $\check{a}$ at $\check{e}$. 

\vspace{1ex}

From the classical theory of Lie groups it follows that the correspondence 
${\mathfrak{H}} \leftrightarrow H_{\mathfrak{H}}$
is biunique, (at least for the local Lie group $H_{\mathfrak{H}}$). By this we have a

\vspace{1ex}

COROLLARY. \emph{The correspondence $\xi \to \Xi$ between the local $\xi$ and the infinitesimal exponent 
$\Xi$ is one-to-one}.

\vspace{1ex}

Note that the words 'local $\xi = \xi(r,s,t)$' mean that $\xi(r,s,t)$ is defined for $r$ and $s$ belonging
to a fixed neighbourhood ${\mathfrak{N}}_{0} \subset G$ of $e \in G$, but \emph{in our case it is defined 
globally as a function of the time variable} $t \in \mathcal{R}$.

\vspace{1ex} 

Suppose the dimension of $G$ to be $n$. Let $a_{k}$ with $k \leq n$ be the base in the Lie algebra
$\mathfrak{G}$ of $G$. Let us introduce the base $\check{a}_{j}$ in $\mathfrak{H}$
in the following way: $\check{a}_{n+1} = \{\alpha_{1}(t), 0\}, \ldots , \check{a}_{n+p} = 
\{\alpha_{p}(t), 0\}$ and $\check{a}_{1} = \{0, a_{1}\},
\ldots , \check{a}_{n} = \{0, a_{n}\}$.
After this we have
\begin{equation}\label{raycom}
[\check{a}_{i},\check{a}_{j}] = c_{ij}^{k}\check{a}_{k} + \Xi(a_{i},a_{j}),
\end{equation}
for $i,j \leq n$. It means that,
in general, the commutation relations of a \emph{ray} representation of $G$ are not equal to
the commutation relations $[A_{i}, A_{j}] = c_{ij}^{k}A_{k}$ of $G$, but they are equal to
$[A_{i},A_{j}] = c_{ij}^{k}A_{k} + \Xi(a_{i}, a_{j},t) \centerdot {\bf 1}$. The generator $A_{i}$ 
corresponding to $a_{i}$ is defined in the following way \cite{Stone} 
\begin{displaymath}
A_{i}\psi = \lim_{\tau \to 0} \frac{(T_{\tau a_{i}} - {\boldsymbol{1}})\psi}{\tau}.
\end{displaymath}

Now, we pass to describe the relation between the infinitesimal exponents $\Xi$ and 
local exponents $\xi$. Let us compute first the infinitesimal exponents $\Xi$ and $\Xi'$ given by (\ref{20''})
which correspond to the two equivalent canonical local exponents $\xi$ and $\xi' = \xi + \Delta[\Lambda]$.
Inserting $\xi' = \xi + \Delta[\Lambda]$ to the formula (\ref{20''}) one gets
\begin{equation}\label{24}
\Xi'(a,b,t) = \Xi(a,b,t) +  \boldsymbol{a}\Lambda(b,t) - \boldsymbol{b}\Lambda(a,t) - \Lambda([a,b],t). 
\end{equation}     
Recall, that according to the Lemma 5, we can confine ourselves to the canonical exponents.
According to Lemma 6 $\Lambda = \Lambda(a, (\tau b)t)$ is a constant function of $\tau$ if 
$a = b$, and $\Lambda(a,t)$ is linear with respect to $a$ (we use the canonical coordinates on $G$). 
By this $\Xi'(a,b,t)$ is antisymmetric in $a$ and $b$ and fulfils (\ref{21}) if only $\Xi(a,b,t)$
is antisymmetric in $a$ and $b$ and fulfils (\ref{21}). 
This suggests the definition: \emph{two infinitesimal exponents $\Xi$ and $\Xi'$ will be called
equivalent if and only if the relation} (\ref{24}) \emph{holds}. For short we write the relation (\ref{24}) 
as follows: 
\begin{displaymath}
\Xi' = \Xi + d[\Lambda].
\end{displaymath}

\vspace{1ex}

LEMMA 8. \emph{Two canonical local exponents $\xi$ and $\xi'$ are equivalent if and only
if the corresponding infinitesimal exponents $\Xi$ and $\Xi'$ are equivalent}.

\vspace{1ex}

PROOF. (1) Assume $\xi$ and $\xi'$ to be equivalent. Then, by the definition of equivalence of infinitesimal
exponents $\Xi' = \Xi + d[\Lambda]$. (2) Assume $\Xi$ and $\Xi'$ to be equivalent: $\Xi' = \Xi + d[\Lambda]$
for some linear form $\Lambda(a,t)$ such that $\Lambda(a,(\tau a)t)$ does not depend on $\tau$. 
Then $\xi + \Delta[\Lambda] \to \Xi'$, and by the uniqueness of the correspondence $\xi \to \Xi$,
$\xi' = \xi + \Delta[\Lambda]$, \emph{i.e.} $\xi$ and $\xi'$ are equivalent. 

\vspace{1ex}

At last from lemma 5 every local exponent is equivalent to a canonical one and 
by the Corollary to every $\Xi$ corresponds uniquely a local exponent, so we can
summarise our results in the

\vspace{1ex}

THEOREM 3. (1) \emph{On a Lie group G, every local exponent $\xi(r,s,t)$ is equivalent to a canonical
local exponent $\xi'(r,s,t)$  which, on some canonical neighbourhood ${\mathfrak{N}}_{0}$, is analytic
in canonical coordinates 
of r and s and in t and vanishes if r and s belong to the same one-parameter subgroup. Two canonical
local exponents $\xi,\xi'$ are equivalent if and only if $\xi' = \xi + \Delta[\Lambda]$ on some canonical
neighbourhood, where $\Lambda(r,t)$ is a linear form in the canonical coordinates of $r$ such that
$\Lambda(r,st)$ does not depend on $s$
if $s$ belongs to the same one-parameter subgroup as} $r$.
 (2) \emph{To every canonical
local exponent of $G$ corresponds uniquely an infinitesimal exponent $\Xi(a,b,t)$ on the Lie
algebra $\mathfrak{G}$ of $G$, i.e. a bilinear antisymmetric form which satisfies  the identity 
$\Xi([a,a'],a'',t) +\Xi([a',a''],a,t)+ \Xi(a'',a],a',t) = \boldsymbol{a}\Xi(a',a'',t) + \boldsymbol{a'}\Xi(a'',a,t) +
\boldsymbol{a''}\Xi(a,a',t)$. The correspondence is linear}. (3) \emph{Two canonical local exponents 
$\xi,\xi'$ are equivalent if and only if the corresponding $\Xi,\Xi'$ are equivalent, i.e. 
$\Xi'(a,b,t) = \Xi(a,b,t) + \boldsymbol{a}\Lambda(b,t) - \boldsymbol{b}\Lambda(a,t) - \Lambda([a,b],t)$
where $\Lambda(a,t)$ is a linear form in $a$ on $\mathfrak{G}$ such that} ${\ud}\Lambda(a,(\tau b)t) / {\ud}\tau = 0$
\emph{if} $a = b$.
(4) \emph{There exist a one-to-one correspondence between the equivalence classes of local exponents
$\xi$ (global in $t$) of $G$ and the equivalence classes of infinitesimal exponents $\Xi$ of} $\mathfrak{G}$.

\vspace{1ex}

Theorem 3 provides the full classification of exponents $\xi(r,s,t)$ local in $r$ and $s$, defined for all $t$.
But we will show that if $G$ is connected and simply connected then we have the following theorems.
(1) If an extension $\xi'$ of 
a given local (in $r$ and $s$) exponent $\xi$ does exist, then it is uniquely determined (up to 
the equivalence transformation (\ref{13})) (Theorem 4). (2) There exists such an extension $\xi'$ (Theorem 5).

\vspace{1ex}

THEOREM 4. \emph{Let $\xi$ and $\xi'$ be two equivalent local exponents of a connected and simply 
connected group $G$, so that $\xi' = \xi + \Delta[\zeta]$ on some neighbourhood, and assume the exponents 
$\xi_{1}$ and $\xi_{1}'$ of $G$ to be extensions of $\xi$ and $\xi'$ respectively. Then, for all $r,s \in G$
$\xi_{1}'(r,s,t) = \xi_{1}(r,s,t) + \Delta[\zeta_{1}]$ where $\zeta_{1}(r,t)$ is strongly
continuous in $r$ and differentiable in $t$, and $\zeta_{1}(r,t) = \zeta(r,t)$, for all $t$ and for all 
$r$ belonging to some neighbourhood of $e \in G$. }
 
\vspace{1ex}  

PROOF. The two exponents $\xi_{1}$ and $\xi_{1}'$ being \emph{strongly continuous} (by assumption) 
define two semicentral extensions $H_{1} = T_{1} \centerdot G$ and 
$H_{1}'= T'_{1}\centerdot G$, which are continuous groups, 
with the topology  defined in the comment to the formula (\ref{15}). 
Note, that the linear groups $T_{1},T'_{1}$ are connected and simply connected.
Because $H_{1}$ and $H_{1}'$ both
are \emph{semi-cartesian} products of two connected
and simply connected groups they are both connected and simply connected. 
Eq. (\ref{izo}) defines a local isomorphism mapping $h: \check{r} \to \check{r}' = h(\check{r})$ 
of $H_{1}$ into $H_{1}'$
\begin{displaymath}
h(\check{r}) = h(\theta,r) = \{\theta(t) - \zeta(r,t),r\}
\end{displaymath}
on the appropriately small neighbourhood of $e$ in $G$, on which $\xi_{1} = \xi$
and $\xi_{1}' = \xi'$. Because $H_{1}$ and $H_{1}'$ are connected and simply connected
the isomorphism $h$ given by (\ref{izo}) can be uniquely extended to an isomorphism 
$h_{1}(\check{r}) = h(\theta,r) = \check{r}'$ of the entire groups $H_{1}$ and $H_{1}'$ such 
that $h_{1}(\check{r}) = h(\check{r})$ on some neighbourhood of $H_{1}$, see
\cite{Pontrjagin}, Theorem 80. The isomorphism $h_{1}$ defines
an isomorphism of the two abelian subgroups  $T_{1}$ and $h_{1}(T_{1})$. By (\ref{izo})
$h_{1}(\theta,e) = \{\theta, e\}$ locally in $H_{1}$, that is for $\theta$ lying approprately 
close to 0 (in the metric sense defined previously).  
Both $T_{1}$ and $h_{1}(T_{1})$ are connected, and $T_{1}$ is in addition simply connected,
so applying once again the Theorem 80 of \cite{Pontrjagin}, one can see that
$h_{1}(\theta,e) = \{\theta,e\}$ for all $\theta$. Set $h_{1}(0,r) = \{ - \zeta_{1}(t),g(r)\}$.
Note, that because $f_{1}$ is an isomorphism it is continuous in 
the topology of $H_{1}$ and $H_{1}'$. By this $\zeta_{1}(r,t)$ is \emph{strongly
continuous} in $r$ and $g(r)$ is a continuous function of $r$. 
The equation $\{\theta,r\} = \{\theta,e\}\{0,r\}$ implies that $h_{1}(\theta(t),r) = \{\theta(t) - \zeta_{1}(r,t), g(r)\}$.
Computing now $h_{1}(0,r)h_{1}(0,s)$ we find that $g(rs) = g(r)g(s)$. So, 
$g(r)$ is an authomorphism of a connected and simply connected $G$, for which
$g(r) = r$ locally, then applying once more the Theorem 80 of \cite{Pontrjagin} one shows that
$g(r) =r$ for all $r$. Thus
\begin{displaymath}
h_{1}(\check{r}) = h_{1}(\theta(t),r) = \{\theta(t) - \zeta_{1}(r,t),r\},
\end{displaymath}   
for all $\check{r} \in H_{1}$. Finally, $h_{1}(0,r)h_{1}(0,s) = h_{1}(\xi_{1}(r,s,t),rs)$. Hence
\begin{displaymath} 
\{\xi_{1}'(r,s,t) - \zeta_{1}(r,t) - \zeta_{1}(s,r^{-1}t), rs\} =
\end{displaymath}

\vspace{-0.5cm}

\begin{displaymath}
 \{\xi_{1}(r,s,t) - \zeta_{1}(rs,t), rs\},
\end{displaymath}
for all $r,s,t$. That is, $\xi_{1}'(r,s,t) = \xi_{1}(r,s,t) + \Delta[\zeta_{1}]$ for all $r,s,t$ and 
by (\ref{izo}) $\zeta_{1}(r,t) =\zeta(r,t)$ on some neighbourhood of $e$ on $G$. 

\vspace{1ex}

THEOREM 5. \emph{Let $G$ be connected and simply connected Lie group. Then to 
every exponent $\xi(r,s,t)$ of $G$ defined locally in $(r,s)$ there exists an exponent 
$\xi_{0}$ of $G$ defined on the whole group $G$ which is an extension of $\xi$. If $\xi$
is differentiable, $\xi_{0}$ may be chosen differentiable.}

\vspace{1ex}

Because the proof of Theorem 5 is identical as that of the Theorem 5.1 in \cite{Bar}, 
we do not present it explicitly \cite{proof}.    

We have obtained the full classification of $\xi$ defined on the whole group $G$ for Lie groups
 $G$ which are connected and simply connected. But for any Lie group $G$ there exists
the universal covering group $G^{*}$ which is connected and simply connected. So, for $G^{*}$
 the correspondence $\xi \to \Xi$ is one-to-one, that is, to every $\xi$ there exists the unique $\Xi$
and vice versa, to every $\Xi$ corresponds uniquely $\xi$ defined on the whole group $G^{*}$ 
and the correspondence preserves the equivalence relation. Because $G$ and $G^{*}$ are locally
isomorphic the infinitesimal exponents $\Xi$'s are exactly the same for $G$ and for $G^{*}$.
Because to every $\Xi$ there does exist exactly one $\xi$ on $G^{*}$, so, if there
exists the corresponding $\xi$ on the whole $G$ to a given $\Xi$, then such a $\xi$ is unique.
We have obtained in this way the full classification of $\xi$ defined
on a whole Lie group $G$ for any Lie group $G$, in the sense that no $\xi$
 can be omitted in the classification. The set of equivalence classes of 
$\xi$ is considerably smaller than the set of equivalence classes of $\Xi$, 
it may happen that to some local $\xi$ there does not exist any global extension.

Take, for example, a Lie subgroup $G$ of (\ref{tra}) and its ray representation $T_{r}$
given by (\ref{trapsi}). We have classified in this way all exponents in (\ref{5}) for this
$T_{r}$ and $r \in G$. By this, as was previously shown, we obtained the full classification
of possible $\theta(r,X)$ in (\ref{trapsi}). In general such a $\Xi$ may exists that there does not exist any 
$\theta(r,X)$ corresponding to this $\Xi$ if the group $G$ is not connected and simply connected.
But this not important for us, the important fact is that no $\theta(r,X)$ can be omitted in this 
classification. We will use this fact in the subsections {\bf V.B} and {\bf V.C}.

\subsection{A general remark concernig the gauge freedom in the Quantum Theory}

Consider a quantum system with a covariance group $G$. Let in addition the time evolution
law for the wave function possesses a gauge freedom: if $\psi$ is a solution to the time evolution
law, then $e^{i\zeta(t)} \psi$ is a solution to the appropriately gauged time evolution law.
The covariance means that the representation $T_{r}$ of $G$ is such that
if $\psi$ is a solution to the evolution law then $T_{r}\psi$ is a solution to the transformed evolution
law. Of course $T_{r}$ acts in the space $\mathfrak{S}$ constructed as in the subsection {\bf V.A} and
fulfils (\ref{5}) with some phase factor $\omega(r,s,t) = e^{i\xi(r,s,t)}$. In general it is not obvious 
if there exists a time evolution compatible with arbitrary chosen factor $\omega$. 
We can, however, consider the problem in a "kinematical"
way and consider abstract representations not necessarily connected with a time evolution
law. In such an abstract situation, however, one can not construct the space $\mathfrak{S}$ which
implicitly uses the notion of time evolution (see subsection {\bf V.A}). But we can
consider a space $\widetilde{\mathfrak{S}}$ in which $T_{r}$ acts. Namely, we can
always construct a space $\widetilde{\mathfrak{S}}$ with a time dependent form $(\psi,\psi')_{t}$ which becomes 
the Hilbert scalar product if only a compatible time evolution exists and $\psi, \psi'$
both are members of the same  "Schr\"odinger picture". That is, the representation fulfils the condition:
$(T_{r}\psi, T_{r}\psi')_{t} = (\psi,\psi')_{t}$ if $(\psi,\psi')_{t} = const$ and $T_{r}(e^{i\zeta(t)}\psi) =
e^{i\zeta(r^{-1}t)}T_{r}\psi$. Such a representation will be called \emph{kinematical}. We have the following

\vspace{1ex}

THEOREM 6. \emph{ For every factor} $\omega(r,s,t)$ \emph{of a locally compact and transitive group} $G$ 
\emph{there exist a continuous kinematical ray representation of} $G$ \emph{compatible with} $\omega(r,s,t)$.

\vspace{1ex}

The notion 'transitive' means that $G$ acts transitively on the time coordinate, that is, 
there exists a fixed $t_{o}$ such that for any $t \in \mathcal{R}$ there exists $r \in G$ fulfilling
the equality $t = rt_{o}$.

\vspace{1ex}

PROOF. Because $G$ acts transitively on $t$ we can consider the factor $\omega$ as a function of the 
three group variables $(r,s,u)$: $\omega = \omega(r,s,ut_{o})$ instead of $\omega = \omega(r,s,t)$.
Since $G$ is locally compact there exists a Haar left invariant measure ${\ud}u$.
We define the linear space $\widetilde{\mathfrak{S}}$ as the space of functions $f(u,t)$
(where $u \in G$ and $t$ stands for the time $t$) for which 
$(f,f)_{t} = \int_{G} f^{*}f \, {\ud}u$ is finite for each $t$. 
Next, we define $(f,g)_{t} = \int_{G} f^{*}(u,t)g(u,t) \, {\ud}u$ and 
\begin{equation}\label{defT}
g(u,t) = T_{r}f(u,t) = \omega(r, r^{-1}u, ut_{o})f(r^{-1}u,r^{-1}t).
\end{equation} 
(1) Clearly $(T_{r}f,T_{r}g)_{t} = (f,g)_{t}$ if only $(f,g)_{t} =const$. (2) From $\omega(e,u,t)=1$ (compare 
(\ref{11})) it follows that $T_{e} = 1$. (3) We show that the kinematical representation fulfils (\ref{5})
with $e^{i\xi}=\omega$.
Set $g = T_{s}f$ and $h =T_{r}g = T_{r}T_{s}f$. From (\ref{defT}) we get $h(r(su),t) = \omega(r,su,rsut_{o})
g(su,r^{-1}t) = \omega(r,su,rsut_{o})\omega(s,u,sut_{o})f(u,(rs)^{-1}t)$. From (\ref{10}) we get
$h(rsu) = \omega(r,s,rsut_{o})\omega(rs,u,rsut_{o})f(u,(rs)^{-1}t)$, such that 
$h(rsu,t) = \omega(r,s,rsut_{o}) T_{rs}f(rsu,t)$, or equivalently $h(u,t) = \omega(r,s,ut_{o}) T_{rs}f(u,t)$,
which means that $T_{r}T_{s} = \omega(r,s,ut_{o})T_{rs}$. (4) Because the regular representation
$f(u,t) \mapsto f(r^{-1}u,r^{-1}t)$ is continuous then our representation also is continuous, because 
of the strong continuity of $\omega$. 

One can consider, however, an abstract situation, in which a quantum system is investigated, such that
its configuration space is not the simultaneity hyperplane of the spacetime. As for example 
in the above Proof where the group $G$ plays the role of the configuration space. Another example is 
the configuration space for the wave function of photon. 
After this, basing on the same footing
as in the subsection {\bf V.A}, it is natural to consider the spacetime-dependent gauge freedom.     
Note that the case with the spacetime-dependent gauge freedom $\psi \mapsto e^{i\zeta(X)}\psi$ 
can be treated in the same way, the proof of the following Theorem 6' is exactly the same,
if only an invariant measure and topology can be introduced to the covariance group $G$. 
The replacements are trivial. For 
example the condition $T_{r}(e^{i\zeta(t)}\psi) =e^{i\zeta(r^{-1}t)}T_{r}\psi$ should be replaced by 
$T_{r}(e^{i\zeta(X)}\psi) = e^{i\zeta(r^{-1}X)}T_{r}\psi$, and the Eqs (\ref{9})$\div$(\ref{13}) are replaced by
\begin{displaymath}
\xi(e,e,X) = 0,
\end{displaymath}

\vspace{-0.5cm}

\begin{displaymath}
\xi(r,s,X) + \xi(rs,g,X) + \xi(s,g,r^{-1}X) + \xi(r,sg,X),
\end{displaymath}

\vspace{-0.5cm}

\begin{displaymath}
\xi(r,e,X) = 0 \, \, \, and \, \, \, \xi(e,g,X) = 0,
\end{displaymath}

\vspace{-0.5cm}

\begin{displaymath}
\xi(r,r^{-1}, X) = \xi(r^{-1}, r, r^{-1}X),
\end{displaymath}

\vspace{-0.5cm}

\begin{displaymath}
\xi'(r,s,X) = \xi(r,s,X) + \zeta(r,X) + \zeta(s,r^{-1}X) - \zeta(rs,X).
\end{displaymath}
The more subtle situation appears, if one tries to classify the exponents $\xi(r,s,X)$ in the analogous way as 
the time-dependent exponents $\xi(r,s,t)$, even if one considers the Lie group $G$ as a covariance group.
We briefly sketch the machinery of this classification.

\vspace{1ex}

CLASSIFICATION OF $\xi(r,s,X)$-S. The multiplication rule in $H$ is given by (\ref{15}),
but this time with $t$ replaced by $X$ . Of course, by the analogue of the Theorem 2
$\xi(r,s,X)$ is strongly continuous in $r$ and $s$ (note, that the compact set 
$\mathcal{C} \subset \mathcal{R}$ should be replaced by the compact set of the spacetime points).
The metric which defines the topology in $H$ is given by (\ref{topology}) 
with the compact sets ${\mathcal{C}}_{n}$ of $\mathcal{R}$ in $p_{n}$ replaced by a compact
sets of the spacetime. Note that all Lemmas 1$\div$6 of the preceding subsection
are valid in this case also. However the construction of the finite dimensional algebra
$\mathfrak{H}$ falls in this case, and we lost the Corollary that the connection $\xi \leftrightarrow \Xi$
is biunique (at least the author is not able to prove the uniqueness in this case).
More precisely, the biunique correspondence between $\Xi$ and local $\xi(r,s,X)$, where by
local $\xi$ we mean $\xi(r,s,X)$ defined for $r$ and $s$ belonging to a fixed neighbourhood 
${\mathfrak{N}}_{0}$ of $e$ in $G$ \emph{and defined globally with respect to the spacetime variables $X$}.    
But we can still construct the Lie algebra ${\mathfrak{H}}_{\infty}$ with the help of the analogue of (\ref{20}), (\ref{20'})
and (\ref{20''}). Unfortunatelly this algebra ${\mathfrak{H}}_{\infty}$ is infinite-dimensional. 
To a given local exponent $\xi$ there always corresponds uniquely the infinitesimal exponent $\Xi$ given by the 
analogue of (\ref{20''}) (in this analogue $t$ is replaced by $X$). In this spacetime-dependent case,
however, we are not able to prove that also conversely: to a given $\Xi$ there always exists uniquely
corresponding local $\xi(r,s,X)$, \emph{i.e} $\xi$ local in $r,s$ and defined globally in $X$.
But we are able to prove that \emph{if there exist the corresponding $\xi$
to an a priori given $\Xi$, then such $\xi$ is uniquely determined}. We will not
present the complete proof, but only explain the main reason for the difference as compared to
the time-dependent $\xi$. The reason lies in the fact that the topology of the one variable $t$
as compared with the spacetime topology of the variables $X$ is much simpler (real line,
or the circle). One can hope, that the correspondence $\xi \to \Xi$ becomes one-to-one
for $\xi(r,s,X)$ which is local \emph{in all three variables} $(r,s,X)$. 
But the author has encountered a difficulty with the analysis of $\xi(r,s,X)$, defined
locally in all three variables $(r,s,X)$, such that even in this case the uniquenees of $\xi \to \Xi$
is an open question. Beside this, we are not able to prove the existence
of an extension $\xi(r,s,X)$ defined globally in $X$ of a given $\xi(r,s,X)$ defined locally in all three variables
(possibly the problem possesses the affirmative solution, but the author is not able to prove it). By this
we do not confine ourselves to the classification of $\xi(r,s,X)$ local in all variables $(r,s,X)$.
Instead, we investigate $\xi(r,s,X)$ which is local in $r,s$ and defined \emph{globally} in $X$, because we are
able to prove the  existence of the unique global extension $\xi(r,s,X)$ defined \emph{globally} for all
three variables $(r,s,X)$ of a given local $\xi(r,s,X)$ \emph{defined globally in} $X$. The unique global extension exist,
if the group $G$ is connected and simply connected.       

Our analysis rests on the beautiful theory of infinite dimensional Lie groups due to Garrett Birkhoff \cite{Birkhoff}.
The algebra ${\mathfrak{H}}_{\infty}$ does not determine the group $H$ even locally, and by this we are not
able to prove the existence of a local $\xi$ (global in $X$) corresponding to a given $\Xi$, but if such a group $H$
does exist it is uniquely determined, and by this $\xi$ is uniquely determined. To explain it we recapitulate
the classical theory of Lie groups. The recasting of the theory of Lie groups into the topologico-algebraic form
depends on recognizing other operations that the two: (1) group multiplication and (2) passage to the inverse.
the first such operation is (3) raising to a scalar power. It is well known that, within restricted neighbourhoods
of the identity $e$ of any Lie group $G$, the equation $r^{n} = s$ has one and only one solution $r$ for a fixed $s$
and non-zero integer $n$. Hence $r^{m} = s^{\frac{m}{n}}$ is also uniquely determined. By rational approximation and 
passage to the limit, one sees that the operation of raising a given element $s$ to a given scalar power $s^{\lambda}$
is an operation uniquely determined by the algebra and the topology of $G$. Using this fact it is easy to state the 
definition
\begin{displaymath}
a \oplus b = \lim_{\lambda \to 0}(a^{\lambda}b^{\lambda})^{\frac{1}{\lambda}}.
\end{displaymath}
The existence of $a \oplus b$ -- for sufficiently small $a$ and $b$ -- is also well-known. The sum $a \oplus b$ simply
corresponds to the vector sum of $a$ and $b$ under canonical coordinates, \emph{i.e}
it defines the vector sum in the Lie algebra. Furthermore, if one defines 
\begin{displaymath}
\lambda a = a^{\lambda} \, \, \, and \, \, \,
[a,b] = \lim_{\lambda \to 0} (a^{\lambda}b^{\lambda}a^{-\lambda}b^{-\lambda})^{\frac{1}{\lambda}}, 
\end{displaymath}
then all usual rules for vector calculus hold, and together with the binary operation $[\centerdot,\centerdot]$ 
one constructs in this way the Lie algebra. But as is well known, the converse is also true:
one can construct (at least locally) the Lie group $G$ from the algebra, if one associates the 
one-parameter subgroup $a^{\lambda}$ with the element $a$ of the algebra. As was shown by
G. Birkhoff there is no problem if the algebra is finite or infinite-dimensional, however it is important
if there exists a restricted neighbourhood on which the operation of raising to a scalar power
is well defined, he calls such groups -- the \emph{analytical groups}. 

Now, let us pass to our case. Unfortunatelly, the raising a given element to real power is not well
defined in $H$. Indeed, compare the two elemets $\check{r}_{1} = \{\theta(x), r\}$ and 
$\check{r}_{2} = \{ \theta(x) + \sin(\frac{\pi x}{\sigma}),r\}$, where $r$ is a translation: $x \to x + \sigma$.
For the two elements one has $\check{r}_{1}\check{r}_{1} = \check{r}_{2}\check{r}_{2} = \check{s}$ 
and the square root $\check{s}^{\frac{1}{2}}$
of the element $\check{s}$ is not well defined (recall that we use the \emph{canonical} exponent: $\xi(r,r,X) = 0$).
It can be shown that the problem is essential, that is, in general with any "natural" topology in $H$ the 
raising to a scalar power $\check{s}^{\lambda}$ is not uniquely defined for $\check{s}$ lying arbitrarily close to 
the unit element $\check{e} \in H$. We will omit, however, this problem in the way explained below.  An
additional condition is needed to pick up a unique solution $\check{r}$ to the equation $\check{r}^{n} = \check{s}$,
where $\check{s}$ lies appropriately close to the unit element $\check{e} \in H$. Such an (additional) condition
really exists, and moreover, it is well defined on a set ${\mathcal{S}} \subset H$ of elements $\check{s}$ and
$\mathcal{S}$ is dense in $H$. By this we are able to represent the one-parameter subgroups 
\begin{displaymath}
\check{r}_{\check{a}}(\tau) = \Big\{ \int_{0}^{\tau} \alpha((\sigma a)^{-1}X) \, {\ud}\sigma, \tau a\Big\} = \check{a}^{\tau},
\end{displaymath} 
with the help of raising a given element $\check{a}$ to a scalar power $\check{a}^{\tau}$, 
if only $\check{a} \in \mathcal{S}$. The correspondence $ \check{a}^{\tau} \leftrightarrow \check{r}_{\check{a}}(\tau)$
is biunique for $\check{a} \in \check{\mathfrak{N}}_{0} \cap \mathcal{S}$ with a suitable chosen neighbourhood
$\check{\mathfrak{N}}_{0}$ of $\check{e}$. There is a problem, of course, if there exists a compatible
continuous (local) group $H$ containing the whole family of one-parameter curves $\check{r}_{\check{a}}(\tau)
= \check{a}^{\tau}$, $\check{a} \in \check{\mathfrak{N}}_{0} \cap {\mathcal{S}}$ -- the one-parameter subgroups. 
But because the family is dense in $H$ it is obvious that if such a group $H$ does exist it is \emph{uniquely} 
determined. The $H$ constructed in this way does not depend on the choice of the actually assumed condition 
by which one can pick up a unique solution $\check{r}$ to the equation $\check{r}^{n} = \check{s}$. It is because we 
require the set ${\mathcal{S}}$ to be dense in the local group $H$. On the other hand the local group 
$H$ with the multiplication law 
\begin{displaymath}
\{\theta_{1}(X),r_{1}\}\{\theta_{2}(X),r_{2}\} = 
\end{displaymath}

\vspace{-0.5cm}

\begin{displaymath}
= \{\theta_{1}(X) + \theta_{2}(r_{1}^{-1}X) + \xi(r_{1},r_{2}, X), r_{1}r_{2} \}
\end{displaymath}
(compare (\ref{15})) uniquely determines the local $\xi$. So, one can see that if there exists a compatible
$\xi$ to an \emph{a priori} given $\Xi$ (determined by the algebra ${\mathfrak{H}}_{\infty}$) then
such a $\xi$ is \emph{uniquely} determined. 

Let us briefly sketch how to construct the additional condition under which the uniqueness 
of the raising to a real scalar power $(\check{r}_{a})^{\lambda}$ is achieved on a set dense in 
a neighbourhood $\check{\mathfrak{N}}_{0}$ of $\check{e} \in H$. Consider two elements (compare the 
proof of Lemma 7) $\check{r}_{1} = \{\theta_{1}(X),\tau a\}$ and $\{\theta_{2}(X), \tau a\}$ fulfilling the condition
\begin{displaymath}
\check{r}_{1}^{2} = \check{r}_{1}\check{r}_{1} = \check{r}_{2}\check{r}_{2} = \check{r}_{2}^{2} = 
\end{displaymath}

\vspace{-0.5cm}

\begin{equation}\label{*}
= \{\theta_{i}(X) + \theta_{i}((\tau a)^{-1}X),2\tau a\} = \check{s},
\end{equation} 
where we have used the fact that $\xi$ can be assumed to be \emph{canonical}:
$\xi(\tau a, \tau' a,X) = 0$. Let us use the coordinate system in which the curves 
$X(x) = (x a)X_{0}$ are coordinate lines together with three remaining families of coordinate
lines $X(y_{i})$. After this the condition (\ref{*}) reads 
\begin{displaymath}
\check{r}_{1}^{2} =\check{r}_{2}^{2} = \{\theta_{i}(x,y_{k}) + \theta_{i}(x - \tau,y_{k}), 2\tau a\} = \check{s}.
\end{displaymath}
It is fulfilled if and only if 
\begin{equation}\label{**}
\theta(x,y_{k}) = - \theta(x - \tau, y_{k}),
\end{equation}
$i = 1,2$, where $\theta= \theta_{2} - \theta_{1}$. Applying once again the formula (\ref{**}) we see that 
$\theta(x,y_{k}) = \theta(x - 2\tau,y_{k})$ is a periodic function of $x$ with the period $T = 2\tau$.
Because $\theta(x,y_{k})$ is continuous in $x$ it follows from (\ref{**}) that there
exists a point $x_{1}$, such that $- T/2 = - \tau  \leq x_{1} \leq \tau = T/2$ and $\theta(x_{1},y_{k}) = 0$. 
From the Weierstrass' Theorem it follows the existence of a point $x_{2}$ ($-T/2  \leq x_{2} \leq T/2$)
such that $\vert\theta(x_{2},y_{k})\vert = \sup_{x \in [-T/2,T/2]} \vert\theta(x,y_{k})\vert$.
At last from the mean-value Theorem it follows that there exist a point $x_{3}$ ($-T/2 \leq x_{3} \leq T/2$)
fulfilling the condition
\begin{displaymath}
\frac{\theta(x_{2}, y_{k}) - \theta(x_{1}, y_{k})}{x_{2} - x_{1}} = \frac{{\ud}\theta}{{\ud}x}(x_{3},y_{k}).
\end{displaymath}
But $\vert x_{2} - x_{1}\vert \leq 2\tau = T$, so we have
\begin{displaymath}
\frac{1}{T}\sup_{x \in [-T/2,T/2]}\vert \theta(x,y_{k})\vert \leq 
\sup_{x \in [-T/2,T/2]} \Big\vert\frac{{\ud}\theta}{{\ud}x}(x,y_{k})\Big\vert.
\end{displaymath} 
Applying recurrently the above reasoning to ${\ud}\theta / {\ud}x, {\ud}^{2}\theta / {\ud}x^{2}, \ldots$, 
and so on (${\ud}\theta(x) / {\ud}x$ again fulfils (\ref{**}) and is
a periodic function with the same period $T = 2\tau$), one gets
\begin{displaymath}
\frac{1}{T^{n}} \sup_{x \in [-T/2,T/2]} \vert \theta(x,y_{k})\vert \leq 
\sup_{x \in [-T/2,T/2]} \Big\vert \frac{{\ud}^{n}\theta}{{\ud}x^{n}}(x,y_{k})\Big\vert.
\end{displaymath}
Note that the supremas can be taken over any $x \in [-b,b]$ with any $b \geq T/2$, or even
over the whole real line $\mathcal{R}$. Assuming $T < 1$ we can see that 
\begin{displaymath}
\sup_{x \in [-T/2,T/2]} \Big\vert\frac{{\ud}^{n}\theta}{{\ud}x^{n}}(x,y_{k})\Big\vert 
\end{displaymath}   
goes to infinity if $n \to + \infty$ for any fixed $y_{k}$. Now, let us go back to the two solutions 
$\check{r}_{1} = \{\theta_{1}, \tau a\}$ and $\check{r}_{2} = \{\theta_{2}, \tau a\}$ of the equation 
$\check{r}^{2} = \check{s}$ (compare (\ref{*})). Suppose that the set of numbers 
\begin{displaymath}
 \sup_{x \in [-T/2,T/2]}\Big\vert\frac{{\ud}^{n}\theta_{1}}{{\ud}x^{n}}(x,y_{k})\Big\vert, n \in {\mathcal{N}},
\end{displaymath}
is \emph{bounded} for any fixed $y_{k}$. Then, the set of numbers
\begin{displaymath}
 \sup_{x \in [-T/2,T/2]} \Big\vert \frac{{\ud}^{n}\theta_{2}}{{\ud}x^{n}}(x,y_{k})\Big\vert, n \in {\mathcal{N}}, 
\end{displaymath}
with $\theta_{2} = \theta_{1} + \theta$ will be \emph{unbounded}. This suggest the following definition.
Suppose that the one-parameter subgroups $a^{\tau} \equiv \tau a$ of the local Lie group $G$ are well
defined for any $\tau$, $0 \leq \vert\tau \vert \leq \epsilon$. We define then
the set $\mathcal{S}$
of elements $\{\theta(X), \tau a\}$ for which the set of numbers
\begin{displaymath}
\sup_{\tau \in [- b, b]} \Big\vert\frac{{\ud}^{n}\theta}{{\ud}x^{n}}((\tau a)X) \Big\vert, n \in {\mathcal{N}},
\end{displaymath}
is bounded for any fixed $X$ and any fixed $a \in \mathfrak{G}$; in this formula $b$ stands
for $\min(\epsilon, 1/2)$. After this, we define the unique solution $\check{r}$ to the equation
$\check{r}^{2} = \check{s}$, with $\check{s} \in \mathcal{S}$, as that $\check{r}$ which also 
belongs to the set $\mathcal{S}$. It is really unique solution, if 
$\check{s} \in \check{\mathfrak{N}}_{0} \cap {\mathcal{S}} = T \times {\mathfrak{N}}_{0} \cap \mathcal{S}$,
such that ${\mathfrak{N}}_{0} \subset G$ is appropriately small neighbourhood of $e$ which does not contain
any $\tau a$ with $2\vert\tau\vert \geq 1$ ($T$ is the abelian 
group of elements $\{\theta,e\}$).
Applying the reasoning recurrently one can see that the equation 
$(\check{r})^{2^{n}} = \check{s}$ has a unique solution $\check{r} \in \mathcal{S}$ if 
$\check{s} \in \check{\mathfrak{N}}_{0} \cap \mathcal{S}$. Repeating the arguments presented in the proof of Lemma 7,
we show that the raising to any (appropriately small) real power is well defined for elements $\check{s}$ belonging
to the above mentioned neighbourhood.
It remains to show that $\mathcal{S}$ is dense in
the topology of $H$ , because there is no essential difficulty with
this point we leave the discussion of it.

Because we are not able to prove the existence of a local $\xi$ (global in $X$) corresponding to 
a given $\Xi$ we are not able to prove the complete analogue of the Theorem 3. 
Instead of Lemma 8, however, the following assertion is true:
 \emph{ If for a given infinitesimal $\Xi$ and $\Xi'$
there exist corresponding local canonical $\xi$ and $\xi'$ (global in $X$), 
then} (a) \emph{$\xi$ and $\xi'$ are unique and}
(b) \emph{the exponents $\xi$ and $\xi'$ are equivalent if and only if the infinitesimal exponents $\Xi$ and $\Xi'$  
are equivalent}. So, we have a slightly weaker Theorem which differs from the Theorem 3 in the point (4) only
(the additional replacement is trivial, namely, one should replace $t$ by $X$).
This allows us to classify \emph{all} exponents $\xi(r,s,X)$ local in $r$ and $s$ and global in $X$, in the sense
that no such an exponent $\xi$ is omitted in the classification, The set of infinitesimal exponents $\Xi$
is too rich this time, but to every local $\xi$ corresponds a unique $\Xi$ 
and if a local $\xi$ corresponds to a given $\Xi$ then such a $\xi$ is unique and the correspondence 
preserves the equivalence relation.

Note that the analog of the Theorem 4 is true in this case when $\xi$ depends on $X$, and
the proof of it is exactly the same, with the trivial replacement of $t$ by $X$. 
The more subtle situation appears in the proof of the analog of the Theorem 5. But, note
that indepedently of the fact if the analog of the Theorem 5 is true or not, we have obtained
in this way the full classification of $\xi(r,s,X)$ in the sense that no $\xi(r,s,X)$ is omitted
in this case also.


\begin{references}
\bibitem{Herdegen} Note that in particular the integral $\int \psi^{*}\psi \, {\ud}^{3}x$ has to be time-independent.
But this is in agreement with the Born interpretation of $\psi$.
\bibitem{mass center} Strictly speaking, we assume that the interactions between particles are attractive,
such that the wave function is practically zero outside some "small" bounded region.
The wave equation separates with the same accuracy as the classical equations of motion. Because this
result is rather trivial and expected we do not present explicitly the computations, which are rather long.
\bibitem{symmetry} Because of the degeneracy of the "metric" structure of the Newton-Cartan
spacetime, there were some troubles with the definition of the symmetry group for the Newton-Cartan
theory, see e.g. C. Duval, Class. Quantum Grav. {\bf 10}, 2217, (1993).
However there were rather clear physical arguments that the 
Milne group is the symmetry group of that theory (and by this plays the same role as the Poincar\'e group
in the Maxwell theory), see A. Milne, Q. J. Math. {\bf 5}, 64, (1934).
The strict proof of this was given by the author in Ref. \cite{Waw}.
\bibitem{graviton} The quantum numbers $n_{g}$ are the counterparts of the photon numbers in QED.
However, one should not think of them as of the graviton numbers, for the graviton is not a well defined object 
in the nonrelativistic theory even if the Postulate 3 was true. The numbers $n_{g}$ play the role of the 
longitudial or the scalar photons rather, compare \cite{Chr}.
\bibitem{curvature} The presence of the gravitational field reflects in the nonzero curvature tensor, independently
of the coordinate system of course. But the dynamical equations of the field cannot be obtained 
from the curvature tensor, because the action of the field depends in a purely algebraical way on
the curvature components. By this a potential is needed to get the field equations in a variational form,
exactly as in the electrodynamics. It seems to be imposible to find such a potential which could achieve
the zero value independently of the galilean frame -- compare for example the Newtonian potential.
\bibitem{Stone} The transformation $T_{r}$ does not act in the ordinary Hilbert space but in the space $\mathfrak{S}$,
by this we cannot refer to the Stone and G{\aa}rding Theorems. Nonetheless, it is still justified to tell about the 
generators $A$ of $T_{r}$. However, we omit the argumentation, because it is rather clear what we mean in 
concrete application discussed in this paper and moreover, we do not use this fact.
\bibitem{proof} In the proof we consider the Lie group $H_{\mathfrak{H}}$ instead of the Lie group
$H$ in the proof presented in \cite{Bar}. The remaining replacements are rather trivial, but we 
mark them here explicitly to simplify the reading. (1) insted of the formula 
$\bar{r}' = \bar{t}(\theta)\bar{r} = \bar{r}\bar{t}(\theta)$ of  (5.3) in \cite{Bar}
we have $\check{r}' = \check{t}(\theta(r^{-1}t))\check{r} = \check{r}\check{t}(\theta(t))$. 
By this, from the formula 
$(\check{h}_{1}(r)\check{h}_{1}(s))\check{h}_{1}(g)$ $ = \check{h}_{1}(r)(\check{h}_{1}(s)\check{h}_{1}(g))$
(see \cite{Bar}) follows $\xi(r,s,t) + \xi(rs,g,t)$ $= \xi(s,g,r^{-1}t) + \xi(r,sg,t)$ instead of (5.8) in \cite{Bar}.
(2) Instead of (4.9), (4.10) and (4.11) we use the Iwasawa-type construction presented in this paper.
(3) Instead of Lemma 4.2 in \cite{Bar} we use the Lemma 3.
\end{references}
\end{document}